\documentclass[english,aps,prl,twocolumn,longbibliography,reprint,final]{revtex4-2}
\usepackage[T1]{fontenc}
\usepackage[latin9]{inputenc}
\usepackage{babel}
\usepackage{units}
\usepackage{amsmath}
\usepackage{amssymb}
\usepackage{graphicx}
\usepackage[bookmarks=false,
 breaklinks=false,pdfborder={0 0 1},backref=false,colorlinks=false]
 {hyperref}

\makeatletter

\newcommand{\mathcircumflex}[0]{\mbox{\^{}}}

\providecommand{\tabularnewline}{\\}

\DeclareMathOperator{\Tr}{Tr}
\usepackage{physics}

\makeatother

\begin{document}
\global\long\def\k#1{\ket{#1}}%

\global\long\def\b#1{\bra{#1}}%

\global\long\def\kb#1#2{\dyad{#1}{#2}}%

\global\long\def\p#1{\dyad{#1}}%

\global\long\def\bk#1#2{\braket{#1}{#2}}%

\global\long\def\n#1{\braket{#1}}%

\global\long\def\ns#1{\left|#1\right|^{2}}%

\global\long\def\e#1{\expval{#1}}%

\global\long\def\ef#1#2{\ev**{#1}{#2}}%

\global\long\def\me#1#2#3{\mel**{#1}{#2}{#3}}%

\title{TR-ARPES Signal in Pumped Semiconductors\\
within Dynamical Projective Operatorial Approach (DPOA)}
\author{Amir Eskandari-asl$^{1}$ and Adolfo Avella$^{1,2,3}$}
\address{$^{1}$Dipartimento di Fisica ``E.R. Caianiello'', Università degli
Studi di Salerno, I-84084 Fisciano (SA), Italy}
\address{$^{2}$CNR-SPIN, Unità di Salerno, I-84084 Fisciano (SA), Italy}
\address{$^{3}$CNISM, Unità di Salerno, Università degli Studi di Salerno,
I-84084 Fisciano (SA), Italy}
\begin{abstract}
In this manuscript, after discussing in detail the internals of our
recently developed method, the dynamical projective operatorial approach
(DPOA), we provide the framework to apply this method to pumped semiconductor
lattice systems and, in particular, to study and analyze their electronic
excitations and TR-ARPES signal. The expressions for relevant out-of-equilibrium
Green's functions and TR-ARPES signal are given within the DPOA framework
and, defining a retarded TR-ARPES signal, it is shown that it is possible
to obtain an out-of-equilibrium version of the fluctuation-dissipation
theorem. We clarify how single- and multi-photon resonances, rigid
shifts, band dressings, and different types of sidebands emerge in
the TR-ARPES signal. We also propose protocols for evaluating the
strength of single- and multi-photon resonances and for assigning
the residual excited electronic population at each crystal momentum
and band to a specific excitation process. Hamiltonians, where intra-
and inter-band transitions are selectively inhibited, are defined
and used to analyze the effects on the TR-ARPES signal and the residual
electronic excited population. Three relevant cases of light-matter
coupling are examined within the dipole gauge: only a local dipole,
only the Peierls substitution in the hopping term, and both terms
at once. The transient and residual pump effects are studied in detail,
including the consequences of the lattice symmetries at different
crystal momenta on the TR-ARPES signal. A detailed study of the dependence
of the TR-ARPES signal on the probe-pulse characteristics is also
reported. To provide a guideline for understanding the complex effects
and interplays and the variety of possible physical phenomena without
being limited by the characteristics of a single particular \emph{real}
material, we have chosen to study a prototypical pumped two-band semiconductor
lattice system.
\end{abstract}
\maketitle

\section{Introduction}

The modern developments in technology made it possible to study condensed
matter systems in the attosecond regime and investigate their real-time
dynamics upon perturbation by ultra-short and intense electromagnetic
pulses, the so called pump-probe setups \citep{brabec2000intense,krausz2009attosecond,krausz2014attosecond,calegari2016advances,borrego2022attosecond,aoki2014nonequilibrium}.
Investigating the real-time behavior of electronic excitations induced
by the laser pulse reveals the fundamental processes that govern the
physics of the system under study \citep{Zurch_17,doi:10.1063/1.4985056,PhysRevB.97.205202,perfetti2008femtosecond,schmitt2008transient}.
One avenue is to investigate the response of the solid by reading
out the high-harmonic generation upon irradiation \citep{von1995generation,norreys1996efficient,chin2001extreme,ghimire2011observation,luu2015extreme,han2016high,Borja:16,you2017anisotropic,liu2017high,jiang2019crystal}.
In other pump-probe setups, the system is pumped with an intense laser
pulse, usually in the IR regime and with a duration ranging from few
to hundreds of $\unit{fs}$, and analyzed using a positively or negatively
delayed probe pulse by measuring either the transient change in the
optical properties \citep{schultze2013controlling,stojchevska2014ultrafast,schultze2014attosecond,lucchini2016attosecond,mashiko2016petahertz,Borja:16,zurch2017ultrafast,Zurch_17,schlaepfer2018attosecond,Kaplan:19,geneaux2019,inzani2022field,inzani2023photoinduced,neufeld2023attosecond}
or the time-resolved angle-resolved photoemission spectroscopy (TR-ARPES)
signal \citep{schmitt2008transient,rohwer2011collapse,smallwood2012tracking,hellmann2012time,papalazarou2012coherent,wang2013observation,johannsen2013direct,rameau2016energy,reimann2018subcycle}.
Even though the theoretical method we introduce in this manuscript
is in principle capable of dealing with any time-dependent system
response, as optical properties \citep{eskandari2024generalized},
in this paper, we will mainly focus on its description of the TR-ARPES
signal.

ARPES investigates the electronic band structure of materials by analyzing
the energy and momentum distribution of the electrons ejected from
a solid via photoelectric effect \citep{smith1975angular,himpsel1978experimental,kampf1990spectral,smith1991electronic,damascelli2003angle,hufner2013photoelectron,sobota2021angle}.
Instead, in pump-probe setups, TR-ARPES is exploited to determine
the out-of-equilibrium electronic properties of materials by measuring
the signal as a function of the time delay between the pump and probe
pulses \citep{sobota2021angle,bovensiepen2012elementary,smallwood2016ultrafast,zhou2018new,boschini2024time}.
TR-ARPES measurements in pump-probe setups can reveal the different
dynamical processes taking place in the system \citep{sobota2021angle},
which are of fundamental importance for understanding the underlying
physics and eventually engineering materials for practical purposes.
Thanks to the capability of monitoring the dynamics of the electronic
excitations, TR-ARPES can give valuable information about the bands
above the Fermi energy, well beyond what one can achieve by measuring
thermal excitations at equilibrium \citep{weinelt2002time,echenique2004decay,sobota2013direct}.
Moreover, TR-ARPES can measure and study the dressing of the main
bands and the emergence of side-bands due to the pump pulse \citep{wang2013observation,broers2022detecting},
opening a pathway to novel applications in ultra-fast engineering
of materials. TR-ARPES measurements can be used to investigate many
other complex effects induced by the pump pulse such as the perturbation
(melting, switching, emergence, etc.) of ordered states in materials
\citep{schmitt2008transient,rohwer2011collapse,smallwood2012tracking}
and the dynamical excitation of collective modes \citep{papalazarou2012coherent,avigo2013coherent,yang2015inequivalence},
just to give a few examples.

To understand the underlying physical phenomena and microscopic processes
induced by the pumping of materials, such advanced experimental studies
require their theoretical description and numerical simulation. The
standard approach to the numerical study of the out-of-equilibrium
behavior of a material pumped with an intense laser pulse is the time-dependent
density functional theory (TD-DFT) \citep{de2013inside,de2016monitoring,wopperer2017efficient,de2017first,pemmaraju2018velocity,tancogne2018atomic,schlaepfer2018attosecond,neufeld2023attosecond,de2018real},
which is unfortunately rather time-consuming and computationally expensive
\citep{armstrong2021dialogue}. Moreover, it is not easy to get deep
insights into the underlying physics through TD-DFT simulations just
using (without \emph{tampering}) currently available software packages,
while, in a model-Hamiltonian approach, it is possible to switch on
and off terms and investigate their relative relevance and interplay
\citep{armstrong2021dialogue}.

Model-Hamiltonian approaches, for both matter and light-matter interaction
terms, rely on parameters supplied by DFT calculations at equilibrium
for \emph{real} materials \citep{schuler2021gauge}. If the material
is strongly correlated, one can use the dynamical mean-field theory
to compute its out-of-equilibrium properties if the number of degrees
of freedom involved (spin, bands, atoms in the basis, etc.) is limited
\citep{freericks2006nonequilibrium,aoki2014nonequilibrium,golevz2019dynamics,eckstein2021time}.
On the other hand, for weakly correlated materials, such as most of
the semiconductors, the Hamiltonian can be mapped to an effective
quadratic form, for which one can in principle compute the time-dependent
single-particle density-matrix and/or higher-order correlation functions
according to the probing scheme \citep{broers2022detecting,d2022time}.
Another approach that is suitable for effective few-band models is
the so called Houston method in which one expands electronic single-particle
wave functions in terms of the instantaneous eigenstates of the time-dependent
Hamiltonian and solves the equations of motion for the expansion coefficients
within some approximations \citep{schlaepfer2018attosecond,sato2018role}.
The relevance of Houston method is to provide a framework to disentangle
the effects of different processes, in particular those related to
the inter-band and to the intra-band transitions, and their interplay.

At any rate, model-Hamiltonian approaches can not be applied so easily
to \emph{real} materials as one either run the risk to use oversimplified
models that could lose some important features or has to find an efficient
way to deal with the actual very complicated Hamiltonians describing
many degrees of freedom at once \citep{armstrong2021dialogue}. Even
for quadratic Hamiltonians, one needs to numerically solve the equations
of motion of the multi-particle density matrices or multi-time correlation
functions, which are needed to describe response functions, such as
the optical conductivity, or for computing the TR-ARPES signal. Unfortunately,
without a proper framework, such calculations can be computationally
quite heavy and eventually unaffordable. Recently, we designed and
developed a novel method, the dynamical projective operatorial approach
(DPOA), and used it to analyze the transient and residual electronic
photo-excitations in ultrafast (attosecond) pumped germanium \citep{inzani2022field,inzani2023photoinduced}.
We benchmarked our results with those obtained through TD-DFT calculations,
which were in turn validated by direct comparison to the experimental
results for the differential transient reflectivity. Moreover, DPOA
allowed us to use much finer momentum grids than the ones affordable
by TD-DFT, and unravel the actual relevant pumping processes, as well
as the individual roles of different mechanisms and their interplay.

DPOA is a quite versatile model-Hamiltonian approach that deals with
the time evolution of composite operators \citep{mancini2004,avella2007under,avella2012COM,avella2014hubbard,avella2014,DiCiolo2018}
and is capable of simulating \emph{real} materials, and the time-dependent
transitions among their actual numerous bands \citep{inzani2022field,inzani2023photoinduced}.
DPOA is, in principle, capable of tackling strongly correlated systems
as well \citep{eskandari2024out}. In this paper, we delve into DPOA
by reporting its detailed derivation and its quadratic-Hamiltonian
version, which is particularly fast and efficient. Such a version
is very useful for semiconductors where one can usually safely discard
the \emph{dynamical} Coloumb interaction. We also report how to compute
all single/multi-particle single/multi-time observables and correlation
functions within this approach. Moreover, we provide an efficient
way to implement the Peierls substitution through a numerically exact
expansion and to compute the $m$-th partial derivative in momentum
space of the hopping and of the dipole terms appearing in such expansion.
This allows to analyze and characterize the terms in such an expansion
defining the related characteristic frequencies, timescales, bandwidths,
and relative phases that explain the emergence and the features of
different kinds of sidebands (multi-photon resonant, non resonant,
envelope, ...) considering also the effects of the finite width of
the envelope of the pump pulse. The presence of the envelope modifies/generalizes
also the Rabi-like phenomenology that takes place when some of the
band gaps are in resonance with integer multiples of the central frequency
of the pump pulse. Such single/multi-photon resonances determine the
accumulation of (residual) electronic excited populations after the
pump pulse turns off. Then, we propose a procedure to determine the
strength of a multi-photon \emph{non-exact} resonance and through
this to assign residual electronic excited populations per momentum
and band to specific single/multi-photon resonant processes. Furthermore,
we use our approach to reproduce the Houston method, generalize it
to second quantization, and to obtain numerically exact expectation
values of the Houston coefficients overcoming its limitations and
drawbacks. We also show that the separation of inter-band and intra-band
transition effects can be obtained in DPOA without any ambiguity,
while such separation is questionable within the Houston approach.
Moreover, we show how to compute Green's functions (GFs) using DPOA,
and hence the TR-ARPES signal. As the standard spectral functions
can become negative out of equilibrium, there is no out-of-equilibrium
counter-part of the fluctuation-dissipation theorem \citep{eckstein2021time}.
Indeed, by defining the retarded TR-ARPES signal, we generalize the
fluctuation-dissipation theorem and find its equivalent out of equilibrium
for TR-ARPES signal, which can be useful to better understand and
compute the out-of-equilibrium energy bands of pumped systems.

As already mentioned above, very recently, we exploited DPOA to unveil
the various charge-injection mechanisms active in germanium \citep{inzani2022field,inzani2023photoinduced}.
In this work, to analyze and discuss a larger variety of fundamental
physical processes without the limitations imposed by the peculiarities
of a specific \emph{real} material, we apply DPOA to a non-trivial
toy model. We analyze a two-band (valence-conduction) model and consider
three relevant cases by switching on and off the Peierls substitution
in the hopping term (relevant to bulk systems) and a local dipole
term (relevant to systems such as quantum dots and molecules and low-dimensional
systems with transverse pumps). We discuss the main effects of the
two terms separately as well as the relevance of their interplay.
In particular, we analyze how the first-order (in the pumping field)
terms of the two types of light-matter couplings assist the higher-order
ones and how their decomposition in terms of intra- and inter-band
components can help understanding the actual phenomenology. We compute
and analyze, in connection to the symmetries of the system, the lesser
and the retarded TR-ARPES signals as well as the residual excited
population. We discuss the broadening of the out-of-equilibrium TR-ARPES
bands and their relationship to the equilibrium bands (the rigid shift
due to the even terms starting from the inverse-mass one) and the
instantaneous eigenstates. Moreover, we discuss the suppression of
the different kinds of sidebands and, in particular, of the resonant
ones in connection to the vanishing of velocity (one-photon sideband)
and inverse-mass (two-photon sideband) terms due to band symmetries
and how such \emph{symmetry protection} is lost in the presence of
the dipole term. We also introduce another type of side bands induced
by the envelope/even-terms. Additionally, we investigate the accumulation
of residual electronic excited population (clearly visible also in
the lesser TR-ARPES signal) induced by Rabi-like oscillations at single/multi-photon
resonant \emph{non-symmetry-protected} \textbf{k} points and the characteristics
of such oscillations in terms of the pump-pulse features. The effects
of inhibiting selectively intra- and inter-band transitions are also
studied on the TR-ARPES signal and on the residual electronic excited
population. Moreover, we study the changes in the characteristics
of TR-ARPES signal on varying the pump-probe delay and the width of
the probe pulse.

In addition, we report a detailed derivation of the dipole-gauge second-quantization
Hamiltonian for light-matter interaction from the velocity-gauge first-quantization
one within the minimal coupling. The expressions of Hamiltonian, electronic
current, and charge density operators are derived requesting charge
conservation and cast in real and momentum space and in Bloch and
Wannier basis. Such expressions are fundamental for the current study
(residual excited electronic population and TR-ARPES signal) and for
the determination of optical response functions.

The manuscript is organized as follow. In Sec.~\ref{sec:Theory},
we introduce DPOA and its quadratic-Hamiltonian version for a pumped
lattice system within the dipole gauge as well as its relation to
the single-particle density-matrix and Houston approaches, and discuss
in detail how to analyze different phenomena and their emergent effects.
Moreover, we provide the formulation to obtain the out-of-equilibrium
Green's function of a system within DPOA as well as the TR-ARPES signal
and an out-of-equilibrium version of the fluctuation-dissipation theorem.
In Sec.~\ref{sec:Two-band_system}, in order to show how DPOA works
in a fundamental and prototypical case, we present and discuss in
detail the DPOA results for the TR-ARPES signal and the residual electronic
excited population of a pumped two-band (valence-conduction) system
considering different cases of the case of light-matter interaction,
and conclude with a study of the TR-ARPES signal dependence on the
probe-pulse characteristics. Sec.~\ref{sec:Summary}, summarizes
this work and gives an outlook. Finally, we included four appendices
regarding the derivation of the velocity and the dipole gauges in
second quantization (App.~\ref{app:gauges}), the effects of the
oscillations of the diagonal elements on the multi-photon resonances
(App.~\ref{app:Oscillations-of-diag}), the Houston approach in first
quantization (App.~\ref{app:Houston}) and the out-of-equilibrium
spectral functions (App.~\ref{app:SF}).

\section{Theory\protect\label{sec:Theory}}

\subsection{Dynamical Projective Operatorial Approach (DPOA)\protect\label{subsec:DPOA}}

For any system at equilibrium, described by a general time-independent
Hamiltonian $\mathcal{H}$ in second quantization and Heisenberg picture,
one can find as many sets of \emph{composite} operators $\mathcal{C}_{\alpha}^{\dagger}=\left(\mathcal{C}_{\alpha,1}^{\dagger},\ldots,\mathcal{C}_{\alpha,a}^{\dagger},\ldots\right)$,
as many degrees of freedom characterizing the system (spin, orbital,
momentum, etc.), which close their hierarchy of the equations of motion
\citep{mancini2004,avella2007under,avella2012COM,avella2014hubbard,avella2014,DiCiolo2018}.
A very effective measure of the degree of correlation in the system
is the ratio between the number of independent (disjoint) sets and
the number of degrees of freedom: for a non-correlated system this
ratio is 1, and it tends to 0 (1) according to how much the system
is strongly (weakly) correlated.

To study the properties of a solid-state system and its linear response,
two types of sets are essential. One is the set stemming from the
canonical electronic (fermionic) operators of the system under study,
$c_{\nu}\left(\mathbf{r},t\right)$, where, for instance, $\mathbf{r}$
can be the site in a Bravais lattice and $\nu$ collects all possible
degrees of freedom (spin, orbital, atom in a basis, etc.). The other
is the set stemming from the canonical charge, spin, orbital, ...
number and ladder (bosonic-like) operators of the system under study
that allow to obtain the related susceptibilities.

Now, let us consider a general time-dependent external perturbation
applied to the system: $\mathcal{H}\rightarrow\mathcal{H}\left(t\right)$.
For instance, it can be an electromagnetic pump pulse whose interaction
with the system is usually described via the minimal coupling. Such
a perturbation preserves the \emph{closure} of the hierarchy of the
equations of motion of $\mathcal{C}_{\alpha}$ as it usually changes
only the single-particle term of the Hamiltonian~\citep{schuler2021gauge},
therefore

\begin{equation}
\mathrm{i}\hbar\partial_{t}\mathcal{C}_{\alpha}\left(t\right)=\left[\mathcal{C}_{\alpha}\left(t\right),\mathcal{H}\left(t\right)\right]=\Xi_{\alpha}\left(t\right)\cdot\mathcal{C}_{\alpha}\left(t\right).\label{eq:EOM_psi_ooe}
\end{equation}
where $\cdot$ is the matrix product in the space of the operators
in a specific set $\alpha$, while $\Xi_{\alpha}\left(t\right)$ and
$\mathcal{C}_{\alpha,a}\left(t\right)$ are the time-dependent energy
matrix and eigenoperators in the Heisenberg picture, respectively.
These considerations guided us to design and devise the Dynamical
Projective Operatorial Approach (DPOA) according to which we have

\begin{equation}
\mathcal{C}_{\alpha}\left(t\right)=P_{\alpha}\left(t,t_{0}\right)\cdot\mathcal{C}_{\alpha}\left(t_{0}\right)\quad\forall t\geq t_{0},\label{eq:P_proj}
\end{equation}
where $P_{\alpha}\left(t,t_{0}\right)$ are called dynamical projection
matrices. Eq.~\ref{eq:P_proj} can be verified using mathematical
induction as follows. Basis: At time $t=t_{0}$, Eq.~\ref{eq:P_proj}
obviously holds with $P_{\alpha}\left(t_{0},t_{0}\right)=\mathbf{1}$.
Induction step: Let us discretize the time axis in terms of an infinitesimal
time step $\Delta t\to0$ ($t_{n}=n\ \Delta t+t_{0}$) and let us
assume that Eq.~\ref{eq:P_proj} holds for time $t_{n}$, i.e.,
\begin{equation}
\mathcal{C}_{\alpha}\left(t_{n}\right)=P_{\alpha}\left(t_{n},t_{0}\right)\cdot\mathcal{C}_{\alpha}\left(t_{0}\right),\label{eq:P_proj_base}
\end{equation}
Then, for time $t_{n+1}=t_{n}+\Delta t$, we have
\begin{multline}
\mathcal{C}_{\alpha}\left(t_{n+1}\right)=\mathcal{C}_{\alpha}\left(t_{n}\right)+\Delta t\partial_{t}\mathcal{C}_{\alpha}\left(t_{n}\right)\\
=\left[P_{\alpha}\left(t_{n},t_{0}\right)-\Delta t\frac{\mathrm{i}}{\hbar}\Xi_{\alpha}\left(t_{n}\right)\cdot P_{\alpha}\left(t_{n},t_{0}\right)\right]\cdot\mathcal{C}_{\alpha}\left(t_{0}\right),\label{eq:P_proj_step}
\end{multline}
that closes the proof and suggests the following relation
\begin{equation}
P_{\alpha}\left(t_{n+1},t_{0}\right)=P_{\alpha}\left(t_{n},t_{0}\right)-\Delta t\frac{\mathrm{i}}{\hbar}\Xi_{\alpha}\left(t_{n}\right)\cdot P_{\alpha}\left(t_{n},t_{0}\right).\label{eq:EOM_P_disc}
\end{equation}
In the following, we choose as initial time $t_{0}$ any time before
the application of the pump pulse (e.g., $t_{0}\rightarrow-\infty$)
and, for the sake of simplicity, we indicate the dynamical projection
matrices using just one time argument $P_{\alpha}\left(t,t_{0}\right)\rightarrow P_{\alpha}\left(t\right)$
. Then, $\mathcal{C}_{\alpha}\left(t_{0}\right)$ simply stands for
the operatorial basis describing the system at equilibrium.

Applying the limit $\Delta t\to0$ to Eq.~\ref{eq:EOM_P_disc}, one
obtains the equation of motion for the dynamical projection matrix
as

\begin{equation}
\mathrm{i}\hbar\partial_{t}P_{\alpha}\left(t\right)=\Xi_{\alpha}\left(t\right)\cdot P_{\alpha}\left(t\right),\label{eq:EOM_P}
\end{equation}
with initial condition $P_{\alpha}\left(t_{0}\right)=\mathbf{1}$.
For stationary Hamiltonians, i.e., when $\Xi_{\alpha}\left(t\right)\rightarrow\Xi_{\alpha}^{\left(0\right)}$,
the solution of Eq.~\ref{eq:EOM_P} is simply $P_{\alpha}^{\left(0\right)}\left(t\right)=\mathrm{e}^{-\frac{\mathrm{i}}{\hbar}\left(t-t_{0}\right)\Xi_{\alpha}^{\left(0\right)}}$.
However, for a general perturbed system, where $\Xi_{\alpha}\left(t\right)=\Xi_{\alpha}^{\left(0\right)}+\Xi_{\alpha}^{\left(1\right)}\left(t\right)$,
one needs to numerically compute the dynamical projection matrix $P_{\alpha}\left(t\right)$
from which it is possible to obtain all out-of-equilibrium properties
and response functions of the system.

Finally, it is worth noting that rewriting $P_{\alpha}\left(t\right)=P_{\alpha}^{\left(0\right)}\left(t\right)\cdot P_{\alpha}^{\mathrm{int}}\left(t\right)=\mathrm{e}^{-\frac{\mathrm{i}}{\hbar}\left(t-t_{0}\right)\Xi_{\alpha}^{\left(0\right)}}\cdot P_{\alpha}^{\mathrm{int}}\left(t\right)$,
we can deduce the following reduced equation of motion
\begin{equation}
\mathrm{i}\hbar\partial_{t}P_{\alpha}^{\mathrm{int}}\left(t\right)=\Xi_{\alpha}^{\left(1\right)\mathrm{int}}\left(t\right)\cdot P_{\alpha}^{\mathrm{int}}\left(t\right),\label{eq:EOM_P_tilde}
\end{equation}
where $\Xi_{\alpha}^{\left(1\right)\mathrm{int}}\left(t\right)=\mathrm{e}^{\frac{\mathrm{i}}{\hbar}\left(t-t_{0}\right)\Xi_{\alpha}^{\left(0\right)}}\cdot\Xi_{\alpha}^{\left(1\right)}\left(t\right)\cdot\mathrm{e}^{-\frac{\mathrm{i}}{\hbar}\left(t-t_{0}\right)\Xi_{\alpha}^{\left(0\right)}}$.
Eq.~\ref{eq:EOM_P_tilde} can be helpful (i) to stabilize the numerical
solution when high frequencies are involved and (ii) to apply any
approximation only to the time-dependent component of the Hamiltonian
and preserve intact the equilibrium dynamics. The equivalent (iterative)
integro-differential equation reads as
\begin{equation}
P_{\alpha}^{\mathrm{int}}\left(t\right)=\mathbf{1}-\frac{\mathrm{i}}{\hbar}\int_{t_{0}}^{t}dt_{1}\,\Xi_{\alpha}^{\left(1\right)\mathrm{int}}\left(t_{1}\right)\cdot P_{\alpha}^{\mathrm{int}}\left(t_{1}\right).\label{eq:EOM_P_tilde_intdiff}
\end{equation}

\subsection{Quadratic Hamiltonians\protect\label{subsec:Quadratic}}

Quadratic Hamiltonians play a fundamental role in many fields of physics
as they retain the full complexity of a system in terms of its degrees
of freedom as well as the possibility to describe to full extent the
effects of applying a (time-dependent) external field or gradient
to the system. Obviously, one cannot describe strong correlations,
that is a deep and intense interplay between degrees of freedom, but
this is not essential in many cases.

As it specifically regards solid-state systems, the most relevant
quadratic Hamiltonians are the tight-binding ones that can be built
for \emph{real} materials through wannierization (for example, by
exploiting Wannier90 code \citep{MOSTOFI20142309}) of the basic standard
results of almost any DFT code available. This procedure preserves
the \emph{static} Coloumb interaction among the electrons (appearing
in the exchange integral within DFT), which usually results in the
opening of gaps and in band repulsion. In presence of a time-dependent
perturbation, e.g., a pump pulse, TD-DFT is usually applied although
it results in very lengthy and very resource-consuming calculations.
DPOA for time-dependent quadratic Hamiltonians is instead very fast
and efficient although it neglects the \emph{dynamical} Coloumb interaction,
which can be safely discarded in many cases. Even excitonic effects
can be easily described in DPOA by choosing the proper effective terms
in the Hamiltonian under analysis and working with effective excitonic
creation and annihilation operators. It is worth noticing that DPOA
allows to retain and to catch the physics of all time-dependent \emph{complications}
and all transitions among the actual, although very numerous, bands
of \emph{real} materials \citep{inzani2022field}.

Let us consider a system described by the following completely general
time-dependent quadratic Hamiltonian in second quantization and Heisenberg
picture
\begin{equation}
\mathcal{H}\left(t\right)=a^{\dagger}\left(t\right)\cdot\Xi\left(t\right)\cdot a\left(t\right),\label{eq:ham_quad}
\end{equation}
where $a^{\dagger}\left(t\right)=\left(a_{1}^{\dagger}\left(t\right),\ldots,a_{n}^{\dagger}\left(t\right),\ldots\right)$
is the creation operator of a general quasi particle (either fermionic
or bosonic including free electrons, electrons in lattice, electrons
in molecules, phonons, magnons, photons, excitons, plasmons, polarons,
polaritons, etc.) in Heisenberg picture and we use a vectorial notation
with respect to the set of quantum numbers $n=\left(n_{1},n_{2},\ldots\right)$
that label all degrees of freedom of the system under analysis. $\Xi\left(t\right)=\Xi^{\left(0\right)}+\Xi^{\left(1\right)}\left(t\right)$
is the energy matrix in which $\Xi^{\left(0\right)}$ gives the equilibrium
Hamiltonian, $\mathcal{H}^{\left(0\right)}$, and matrix $\Xi^{\left(1\right)}\left(t\right)$
describes the coupling of the system to the time-dependent external
pump pulse and gives $\mathcal{H}^{\left(1\right)}\left(t\right)$
. As the main simplification comes from the Hamiltonian being quadratic,
the eigenoperators of the system are just the $a_{n}\left(t\right)$,
and according to the general theory discussed above, we have
\begin{align}
 & a\left(t\right)=P\left(t\right)\cdot a\left(t_{0}\right),\label{eq:P_a}\\
 & \mathrm{i}\hbar\partial_{t}P\left(t\right)=\Xi\left(t\right)\cdot P\left(t\right),\label{eq:EOM_P_quad}
\end{align}
with the initial condition $P\left(t_{0}\right)=\mathbf{1}$. At each
instant of time, the canonical commutation relations obeyed by $a_{n}\left(t\right)$
lead to $P\left(t\right)\cdot P^{\dagger}\left(t\right)=\mathbf{1}$,
which is a useful relation to check the stability and the precision
over time of any numerical approach used to compute $P\left(t\right)$.

\subsection{Single-Particle Density Matrix (SPDM)\protect\label{subsec:SPDM}}

To show how to obtain the dynamical properties of the system using
the dynamical projection matrices $P\left(t\right)$, we consider
first the single-particle density matrix (SPDM) $\rho\left(t\right)=\left\langle a\left(t\right)\otimes a^{\dagger}\left(t\right)\right\rangle $,
whose equation of motion reads as
\begin{equation}
\mathrm{i}\hbar\partial_{t}\rho\left(t\right)=\left[\Xi\left(t\right),\rho\left(t\right)\right].\label{eq:EOM_rho}
\end{equation}
Once the time evolution of $\rho\left(t\right)$ is known, it is possible
to compute the average of any single-particle single-time operator
$X\left(t\right)=a^{\dagger}\left(t\right)\cdot\mathcal{X}\left(t\right)\cdot a\left(t\right)$
and, therefore, of the corresponding physical quantity as follows,
\begin{equation}
\left\langle X\left(t\right)\right\rangle =\Tr\left(\mathcal{X}\left(t\right)\cdot\left[\boldsymbol{1}-\rho\left(t\right)\right]\right).\label{eq:ave_rho}
\end{equation}

Given that $\rho\left(t\right)=P\left(t\right)\cdot\left\langle a\left(t_{0}\right)\otimes a^{\dagger}\left(t_{0}\right)\right\rangle \cdot P^{\dagger}\left(t\right)=P\left(t\right)\cdot\rho\left(t_{0}\right)\cdot P^{\dagger}\left(t\right)$,
we have
\begin{equation}
\left\langle X\left(t\right)\right\rangle =\Tr\left[P^{\dagger}\left(t\right)\cdot\mathcal{X}\left(t\right)\cdot P\left(t\right)\cdot\left[\boldsymbol{1}-\rho\left(t_{0}\right)\right]\right].\label{eq:ave_P}
\end{equation}
If we choose the quantum numbers $n$ such that the corresponding
operators diagonalize the equilibrium Hamiltonian $H^{\left(0\right)}$,
that is $\Xi_{nm}^{\left(0\right)}=\delta_{nm}\varepsilon_{n}$, we
simply have $\rho_{nm}\left(t_{0}\right)=\delta_{nm}\left(1-\eta f_{\eta}\left(\varepsilon_{n}\right)\right)$
where $\eta=\pm1$ correspond to fermionic (bosonic) system and $f_{\eta}\left(\varepsilon\right)=\frac{1}{\mathrm{e}^{\beta\varepsilon}+\eta}$
is the related equilibrium distribution function. Once the dynamical
projection matrices $P\left(t\right)$ are known at all times, it
is possible to recover all the results of the SPDM approach, and more
importantly, go beyond them.

Indeed, we are not limited to single-particle properties and even
within these latter not to single-time ones. For instance, given a
general single-particle two-time operator $Y\left(t,t^{\prime}\right)=a^{\dagger}\left(t\right)\cdot\mathcal{Y}\left(t,t^{\prime}\right)\cdot a\left(t^{\prime}\right)$,
the time evolution of its average $\left\langle Y\left(t,t^{\prime}\right)\right\rangle $
is simply given by
\begin{equation}
\left\langle Y\left(t,t^{\prime}\right)\right\rangle =\Tr\left[P^{\dagger}\left(t\right)\cdot\mathcal{Y}\left(t,t^{\prime}\right)\cdot P\left(t^{\prime}\right)\cdot\left[\boldsymbol{1}-\rho\left(t_{0}\right)\right]\right].\label{eq:ave_P_2t}
\end{equation}
The extension to multi-particle multi-time operators is straightforward
and requires only the knowledge of equilibrium averages, which for
quadratic Hamiltonians can be easily calculated thanks to the Wick's
theorem.

\subsection{Pumped lattice systems, Peierls expansion and multi-photon resonances\protect\label{subsec:Pumped_Peierls}}

Let us consider an electromagnetic pump pulse applied to a lattice
system after time $t_{0}$, and described by the vector potential
$\boldsymbol{A}\left(t\right)$ and the electric field $\boldsymbol{E}\left(t\right)=-\partial_{t}\boldsymbol{A}\left(t\right)$
in the Coulomb gauge. Accordingly, in the dipole gauge, the dynamics
is governed by the Hamiltonian $\mathcal{H}\left(t\right)=\sum_{\mathbf{k},\nu,\nu{}^{\prime}}\tilde{c}_{\mathbf{k},\nu}^{\dagger}\left(t\right)\tilde{\Xi}_{\mathbf{k},\nu,\nu{}^{\prime}}\left(t\right)\tilde{c}_{\mathbf{k},\nu{}^{\prime}}\left(t\right)$,
where $\tilde{c}_{\mathbf{k},\nu}\left(t\right)$ is the annihilation
operator of an electron with momentum $\mathbf{k}$ (in the first
Brillouin zone) in the localized state $\nu$ (e.g., a maximally localized
Wannier state) and~\citep{schuler2021gauge} (see App.~\ref{app:gauges}
for derivation)
\begin{equation}
\tilde{\Xi}_{\mathbf{k},\nu,\nu{}^{\prime}}\left(t\right)=\tilde{T}_{\mathbf{k}+\frac{\mathrm{e}}{\hbar}\boldsymbol{A}\left(t\right),\nu,\nu{}^{\prime}}+\mathrm{e}\boldsymbol{E}\left(t\right)\bullet\tilde{\boldsymbol{D}}_{\mathbf{k}+\frac{\mathrm{e}}{\hbar}\boldsymbol{A}\left(t\right),\nu,\nu{}^{\prime}}.\label{eq:ham_dg}
\end{equation}
$\tilde{T}_{\mathbf{k},\nu,\nu{}^{\prime}}$ and $\tilde{\boldsymbol{D}}_{\mathbf{k},\nu,\nu{}^{\prime}}$
are the hopping and dipole matrix elements in the reciprocal space,
respectively, and the over-script $\sim$ indicates that they are
expressed in the basis of the localized states, $\mathrm{e}>0$ is
the value of electronic charge, and $\bullet$ is the scalar product
between two vectors in the Cartesian space. The momentum shift by
the vector potential, $\mathbf{k}+\frac{\mathrm{e}}{\hbar}\boldsymbol{A}\left(t\right)$,
resembles the Peierls substitution~\citep{peierls1933theorie,ismail2001coupling}
and Eq.~\ref{eq:ham_dg} can be considered as its generalization
to multi-band systems~\citep{schuler2021gauge}. Eq.~\ref{eq:ham_dg}
shows that the coupling to the pump pulse is two fold: the Peierls
substitution (in both $\tilde{T}_{\mathbf{k},\nu,\nu{}^{\prime}}$
and $\tilde{\boldsymbol{D}}_{\mathbf{k},\nu,\nu{}^{\prime}}$) and
the dipole term $\boldsymbol{E}\left(t\right)\bullet\boldsymbol{D}$.
It is worth noting that, for $\left(d<3\right)$-dimensional systems
with transverse pump-pulse polarization there is no coupling through
the Peierls substitution and the dipole term is the only coupling
to the external field.

The equilibrium Hamiltonian reduces to $\tilde{\Xi}_{\mathbf{k},\nu,\nu{}^{\prime}}\left(t\leq t_{0}\right)=\tilde{T}_{\mathbf{k},\nu,\nu{}^{\prime}}$,
which can be diagonalized through the matrix $\Omega_{\mathbf{k},\nu,n}$
as follows,

\begin{equation}
\delta_{n,n^{\prime}}\varepsilon_{\mathbf{k},n}=\sum_{\nu,\nu{}^{\prime}}\Omega_{\mathbf{k},n,\nu}^{\dagger}\tilde{T}_{\mathbf{k},\nu,\nu{}^{\prime}}\Omega_{\mathbf{k},\nu{}^{\prime},n{}^{\prime}},\label{eq:w2b}
\end{equation}
where $n$ indicates the energy band. Being diagonal at equilibrium,
the band basis provides a great advantage in computations. The transformation
to the band basis is performed as

\begin{align}
 & \boldsymbol{D}_{\mathbf{k},n,n{}^{\prime}}=\sum_{\nu,\nu{}^{\prime}}\Omega_{\mathbf{k},n,\nu}^{\dagger}\tilde{\boldsymbol{D}}_{\mathbf{k},\nu,\nu{}^{\prime}}\Omega_{\mathbf{k},\nu{}^{\prime},n{}^{\prime}},\label{eq:dip_b}\\
 & \Xi_{\mathbf{k},n,n{}^{\prime}}\left(t\right)=\sum_{\nu,\nu{}^{\prime}}\Omega_{\mathbf{k},n,\nu}^{\dagger}\tilde{\Xi}_{\mathbf{k},\nu,\nu{}^{\prime}}\left(t\right)\Omega_{\mathbf{k},\nu{}^{\prime},n{}^{\prime}},\label{eq:ham_b}
\end{align}
and
\begin{equation}
c_{\mathbf{k},n}\left(t\right)=\sum_{\nu}\Omega_{\mathbf{k},n,\nu}^{\dagger}\tilde{c}_{\mathbf{k},\nu}\left(t\right).\label{eq:c_b}
\end{equation}
It is worth recalling that $c_{\mathbf{k}}\left(t\right)=P_{\mathbf{k}}\left(t\right)\cdot c_{\mathbf{k}}\left(t_{0}\right)$,
where $P_{\mathbf{k}}\left(t_{0}\right)=\boldsymbol{1}$ and $\mathrm{i}\hbar\partial_{t}P_{\mathbf{k}}\left(t\right)=\Xi_{\mathbf{k}}\left(t\right)\cdot P_{\mathbf{k}}\left(t\right)$.
Moreover, $N_{\mathbf{k},n}\left(t\right)=\left\langle c_{\mathbf{k},n}^{\dagger}\left(t\right)c_{\mathbf{k},n}\left(t\right)\right\rangle $,
the time-dependent number of electrons in band $n$ with momentum
$\mathbf{k}$, is given by

\begin{equation}
N_{\mathbf{k},n}\left(t\right)=\sum_{n^{\prime}}P_{\mathbf{k},n,n{}^{\prime}}\left(t\right)f_{+}\left(\varepsilon_{\mathbf{k},n^{\prime}}\right)P_{\mathbf{k},n{}^{\prime},n}^{\dagger}\left(t\right).\label{eq:N_P}
\end{equation}

For \emph{real} materials (our recent work on germanium being an example
\citep{inzani2022field}), with many bands involved in the dynamics
and hopping and dipole parameters obtained in real space through wannerization,
the presence of the Peierls substitution, $\mathbf{k}+\frac{\mathrm{e}}{\hbar}\boldsymbol{A}\left(t\right)$,
in Eq.~\ref{eq:ham_dg} makes any time-dependent measure extremely
time-consuming, as it is necessary, at each time step in the numerical
time grid, to Fourier transform again and again, because of the shift,
the hopping and dipole matrices to momentum space on the numerical
momentum grid and, finally, perform the rotation to the band space.
A very efficient way to deal with this problem, which makes it possible
to study systems with many bands without overheads in terms of time
consumption and numerical precision, exploits the expansion of the
hopping matrix and of the dipole matrix with respect to the vector
potential, to sufficiently high order (determined by the maximum strength
of the vector potential and the bandwidth of the system) and uses
the expansion coefficients, computed once for all, at all times:
\begin{align}
 & T_{\mathbf{k}+\frac{\mathrm{e}}{\hbar}\boldsymbol{A}\left(t\right)}\left(t\right)=\sum_{m=0}^{\infty}\frac{1}{m!}\Omega_{\mathbf{k}}^{\dagger}\cdot\left[\partial_{k_{A}}^{\left(m\right)}\tilde{T}_{\mathbf{k}}\right]\cdot\Omega_{\mathbf{k}}\left(\frac{\mathrm{e}}{\hbar}A\left(t\right)\right)^{m},\label{eq:Peierls_exp_T}\\
 & \boldsymbol{D}_{\mathbf{k}+\frac{\mathrm{e}}{\hbar}\boldsymbol{A}\left(t\right)}=\sum_{m=0}^{\infty}\frac{1}{m!}\Omega_{\mathbf{k}}^{\dagger}\cdot\left[\partial_{k_{A}}^{\left(m\right)}\tilde{\boldsymbol{D}}_{\mathbf{k}}\right]\cdot\Omega_{\mathbf{k}}\left(\frac{\mathrm{e}}{\hbar}A\left(t\right)\right)^{m},\label{eq:eq:Peierls_exp_D}
\end{align}
where $\partial_{k_{A}}^{\left(m\right)}$ is the $m$-th partial
derivative in momentum space in the direction of the pump-pulse polarization,
$\hat{A}$, and $A\left(t\right)$ is the magnitude of the vector
potential: $\boldsymbol{A}\left(t\right)=A\left(t\right)\hat{A}$.
We call this procedure \emph{Peierls expansion} hereafter. It is noteworthy
that $T_{\mathbf{k}}$ and $\boldsymbol{D}_{\mathbf{k}}$ are, by
construction, analytic functions of the momentum $\mathbf{k}$ (see
Eqs.~\ref{eq:Tk_from_TR} and~\ref{eq:Dk_from_DR}), and, therefore,
the \textit{Peierls expansion} always converges. The expansion coefficients,
that is, the $m$-th partial derivatives, tend to zero by increasing
$m$ since the hopping and dipole matrices in direct space are not
infinitely long range thanks to the localization of the Wannier states.
These coefficients can be efficiently computed by means of the Fourier
transformation as,
\begin{align}
 & \partial_{k_{A}}^{\left(m\right)}\tilde{T}_{\mathbf{k}}=\sum_{\mathbf{i}}\left(-\mathrm{i}\hat{A}\bullet\mathbf{R}_{\mathbf{i}}\right)^{m}\mathrm{e}^{-\mathrm{i}\mathbf{k}\bullet\mathbf{R}_{\mathbf{i}}}\tilde{T}_{\mathbf{R}_{\mathbf{i}}},\label{eq:der_T}\\
 & \partial_{k_{A}}^{\left(m\right)}\tilde{\boldsymbol{D}}_{\mathbf{k}}=\sum_{\mathbf{i}}\left(-\mathrm{i}\hat{A}\bullet\mathbf{R}_{\mathbf{i}}\right)^{m}\mathrm{e}^{-\mathrm{i}\mathbf{k}\bullet\mathbf{R}_{\mathbf{i}}}\tilde{\boldsymbol{D}}_{\mathbf{R}_{\mathbf{i}}},\label{eq:der_D}
\end{align}
where $\tilde{T}_{\mathbf{R}_{\mathbf{i}}}$ and $\tilde{\boldsymbol{D}}_{\mathbf{R}_{\mathbf{i}}}$
are the hopping and dipole matrices, respectively, in the direct space,
as outputted, for example, by the wannierization procedure. Another
important point to mention is that even for very high intensities
of the pump pulse, $\frac{\mathrm{e}}{\hbar}A\left(t\right)$ is at
maximum just a few percent of the extension of the Brillouin zone
in \emph{real} materials \citep{inzani2022field}. This fixes an upper
bound for the maximum value of $m$ to be actually used in the numerical
evaluation of the \textit{Peierls expansions}.

Such an expansion, even if not used in the actual numerical calculations,
is of fundamental relevance as it gives insight into the actual excitation
processes active in the system and connects them to the symmetries
of the band structure and of the dipole couplings. According to a
well-established practice, we call the coefficient of the first-(second-)order
term of the \emph{Peierls expansion}, Eq.~\ref{eq:Peierls_exp_T},
of the hopping term $T$ as the velocity (inverse-mass) term.

The pump pulse $A\left(t\right)$ can be usually represented as $A\left(t\right)=A_{0}S\left(t\right)\cos\left(\omega_{\mathrm{pu}}t+\phi\right)$
where $\omega_{\mathrm{pu}}$ is the central frequency of the pulse,
$\phi$ is its phase, and $S\left(t\right)$ is an envelope function
that vanishes at $t\rightarrow\pm\infty$. A usual expression for
the envelope function is a Gaussian, $S\left(t\right)=e^{-4\ln2t^{2}/\tau_{\mathrm{pu}}^{2}}$,
where $\tau_{\mathrm{pu}}$ is its full-width at half maximum (FWHM)
and, for the sake of simplicity, its center is just at $t=0$. Such
an envelope gives a finite bandwidth to the pulse of the order $2\pi\hbar\tau_{\mathrm{pu}}^{-1}$,
where $\tau_{\mathrm{pu}}^{-1}$ is FWHM of the corresponding Gaussian
in frequency domain.

Given the above expression for the pump pulse, $A\left(t\right)$,
we can expand its $m$-th power, $A^{m}\left(t\right)$, and get
\begin{align}
 & \Lambda_{\mathbf{k}+\frac{\mathrm{e}}{\hbar}\boldsymbol{A}\left(t\right)}\left(t\right)=\Omega_{\mathbf{k}}^{\dagger}\cdot\tilde{\Lambda}_{\mathbf{k}}\cdot\Omega_{\mathbf{k}}+\sum_{m=1}^{\infty}\Theta_{0,m}\left(t\right)\nonumber \\
 & +2\sum_{l=1}^{\infty}\left[\sum_{m=0}^{\infty}\Theta_{l,m}\left(t\right)\right]\cos\left(l\omega_{\mathrm{pu}}t+l\phi\right),\label{eq:Peierls_exp_L}\\
 & \Theta_{l,m}\left(t\right)=\frac{\left(\frac{\mathrm{e}A_{0}S\left(t\right)}{2\hbar}\right)^{2m+l}}{m!\left(m+l\right)!}\Omega_{\mathbf{k}}^{\dagger}\cdot\left[\partial_{k_{A}}^{\left(2m+l\right)}\tilde{\Lambda}_{\mathbf{k}}\right]\cdot\Omega_{\mathbf{k}}\label{eq:Peierls_exp_Theta}
\end{align}
where $\Lambda$ can be either the hopping matrix $T$ or the dipole
matrix $D$. Such an expression allows us to understand the excitation
processes. The first term on the right-hand side is just the pristine
(time-independent) hopping/dipole matrix. The second term would result
in a $\mathbf{k}$-dependent energy shift coming from the even derivatives
(mainly from the inverse-mass coefficient of the hopping term) if
there would be no envelope function $S\left(t\right)$. Actually,
it is time-dependent because of the envelope function $S\left(t\right)$,
but not periodic, and will lead to the emergence of non-resonant side
bands, as we will show in Sec.~\ref{sec:Two-band_system}, on a timescale
of the order $\frac{\tau_{\mathrm{pu}}}{\sqrt{2}}$ around the envelope
center provided that the energy-band symmetries do not require the
inverse-mass term (and higher-order even terms) to be zero. The third
term leads to Rabi-like $l$-photon resonances whenever the energy
gap between any two bands in the system, not both empty or full at
a certain instant of time, is close to $l\hbar\omega_{\mathrm{pu}}$
within a bandwidth of order $2\pi\hbar\sqrt{l}\tau_{\mathrm{pu}}^{-1}$.
In realistic pump-probe setups, given the bandwidths of the valence
and conduction portion of the band structure involved in the dynamical
processes, usually the maximum number of relevant $l$ (i.e., the
maximum relevant number of photon processes) is low, which is connected
to the fact that only the first few terms of the \emph{Peierls expansions}
are needed in actual numerical calculations, as discussed above. Each
$l$-component of this term is active on a timescale of the order
$\frac{\tau_{\mathrm{pu}}}{\sqrt{l}}$ around the envelope center
and has a phase shift of $(l-1)\phi$ with respect to the $l=1$ component.
For very short values of $\tau_{\mathrm{pu}}$ with respect to $2\pi\omega_{\mathrm{pu}}^{-1}$,
that is, when we have so few cycles of the pump pulse within the envelope
to hardly recognize any oscillation, we end up in an impulsive regime.
Actually, given that the oscillation period decreases with $l^{-1}$
while the FWHM decreases with $l^{-\frac{1}{2}}$, even in the case
where lower-$l$ terms are impulsive, sufficiently-higher-$l$ terms
are anyway oscillatory, although these latter can have a negligible
effect on the dynamics.

Consequently, in lattice systems, one origin of multi-photon resonances
are the non-linear terms in the Peierls expansion. Another origin
is the oscillatory behavior of the diagonal terms of the coupling
Hamiltonian. This is discussed in App.~\ref{app:Oscillations-of-diag}
where we consider a simple two-level system and show the emergence
of multi-photon resonances.

\subsection{Resonances and residual electronic excited population\protect\label{subsec:Resonances_residual}}

At resonance, the dynamics of the electronic population has a Rabi-like
behavior which is completely different from the off-resonance behavior.
In particular, the residual electronic population $N_{\mathbf{k},n}^{\mathrm{res}}=N_{\mathbf{k},n}\left(t\rightarrow\infty\right)$,
that is the electronic population in band $n$ at momentum $\mathbf{k}$
after the application of the pump pulse, becomes a very relevant quantity
to measure and analyze. For a perfectly periodic pump pulse, that
is, with infinite extension in time and no envelope, checking the
$l$-photon resonance condition requires just the comparison of the
energy gaps to $l\hbar\omega_{\mathrm{pu}}$. Instead, the presence
of an envelope broadens the range of frequencies appearing in the
Fourier transform of the pump pulse and hence increases the range
of resonant energy gaps. To quantify this occurrence and on the basis
of what reported in the previous section, we define the normalized
strength of a $l$-photon resonance with respect to an energy gap
$\varepsilon_{\mathrm{gap}}$, $w_{l}\left(\varepsilon_{\mathrm{gap}}\right)$
as

\begin{align}
 & w_{l}\left(\varepsilon_{\mathrm{gap}}\right)=e^{-\frac{\tau_{\mathrm{pu}}^{2}}{8\ln2\hbar^{2}l}\left(\varepsilon_{\mathrm{gap}}-l\hbar\omega_{\mathrm{pu}}\right)^{2}},\label{eq:strenght_res_l}
\end{align}
where $\omega_{\mathrm{pu}}$ is the pump-pulse frequency and $\tau_{\mathrm{pu}}$
is the FWHM of its Gaussian envelope. The expression resembles the
square of the amplitude of the $\varepsilon_{\mathrm{gap}}/\hbar$
component in the spectrum of the $n_{\mathrm{ph}}$-th power of the
pump pulse centered at $n_{\mathrm{ph}}\omega_{\mathrm{pu}}$. Then,
to measure the total number of effective $l$-photon resonant energy
gaps, $W_{l}$, it is sufficient to sum up all normalized strengths
over all points $\mathbf{k}$ of the numerical momentum grid for all
possible pairs of valence-conduction bands

\begin{equation}
W_{l}=\sum_{\mathbf{k},n_{C},n_{V}}w_{l}\left(\varepsilon_{\mathbf{k},n_{C}}-\varepsilon_{\mathbf{k},n_{V}}\right),\label{eq:eff_res_l}
\end{equation}
where $n_{C}$($n_{V}$) runs over all conduction(valence) bands.

$N_{\mathbf{k},n_{C}}^{\mathrm{res}}$, the residual electronic population
in one specific conduction band $n_{C}$ at momentum $\mathbf{k}$,
is the result of resonant processes originating in different valence
bands at the same momentum $\mathbf{k}$. Each of these valence bands
will contribute to $N_{\mathbf{k},n_{C}}^{\mathrm{res}}$ with an
undetermined portion of its residual hole population $N_{\mathbf{k},n_{V}}^{\left(h\right)\mathrm{res}}=1-N_{\mathbf{k},n_{V}}^{\mathrm{res}}$:
$\sum_{n_{V}}N_{\mathbf{k},n_{V}}^{\left(h\right)\mathrm{res}}=\sum_{n_{C}}N_{\mathbf{k},n_{C}}^{\mathrm{res}}$.
Here, we suggest a procedure that allows to determine the contribution
$N_{\mathbf{k},n_{C},n_{V}}^{\mathrm{res}\left(l\right)}$ of the
residual hole population of the valence band $n_{V}$ due to a $l$-photon
resonant process to $N_{\mathbf{k},n_{C}}^{\mathrm{res}}$: $N_{\mathbf{k},n_{C}}^{\mathrm{res}}=\sum_{l,n_{V}}N_{\mathbf{k},n_{C},n_{V}}^{\mathrm{res}\left(l\right)}$.
The rationale is to assign to each valence band $n_{V}$ such a contribution,
$N_{\mathbf{k},n_{C},n_{V}}^{\mathrm{res}\left(l\right)}$, according
to the strength of the $l$-photon resonant process involved, $w_{l}\left(\varepsilon_{\mathbf{k},n_{C}}-\varepsilon_{\mathbf{k},n_{V}}\right),$
and to the actual value of $N_{\mathbf{k},n_{V}}^{\left(h\right)\mathrm{res}}$
with respect to those of all other valence bands:
\begin{equation}
N_{\mathbf{k},n_{C},n_{V}}^{\mathrm{res}\left(l\right)}=\frac{N_{\mathbf{k},n_{V}}^{\left(h\right)\mathrm{res}}w_{l}\left(\varepsilon_{\mathbf{k},n_{C}}-\varepsilon_{\mathbf{k},n_{V}}\right)}{\sum_{n'_{V}}N_{\mathbf{k},n'_{V}}^{\left(h\right)\mathrm{res}}\sum_{l^{\prime}}w_{l^{\prime}}\left(\varepsilon_{\mathbf{k},n_{C}}-\varepsilon_{\mathbf{k},n'_{V}}\right)}N_{\mathbf{k},n_{C}}^{\mathrm{res}}.\label{eq:N_k_nc_nv_l}
\end{equation}
Given these ingredients, it is now possible to compute (i) the contribution
to $N_{\mathbf{k},n_{C}}^{\mathrm{res}}$ coming from all $l$-photon
resonant processes, $N_{\mathbf{k},n_{C}}^{\mathrm{res}\left(l\right)}$,
\begin{equation}
N_{\mathbf{k},n_{C}}^{\mathrm{res}\left(l\right)}=\sum_{n_{V}}N_{\mathbf{k},n_{C},n_{V}}^{\mathrm{res}\left(l\right)},\label{eq:N_k_nc_l}
\end{equation}
(ii) the contribution to $N_{\mathbf{k},n_{C}}^{\mathrm{res}}$ coming
from each valence band $n_{V}$, $N_{\mathbf{k},n_{C},n_{V}}^{\mathrm{res}}$,
\begin{equation}
N_{\mathbf{k},n_{C},n_{V}}^{\mathrm{res}}=\sum_{l}N_{\mathbf{k},n_{C},n_{V}}^{\mathrm{res}\left(l\right)},\label{eq:N_k_nc_nv}
\end{equation}
(iii) the total residual electronic population at momentum $\mathbf{k}$
coming from all $l$-photon resonant processes, $N_{\mathbf{k}}^{\mathrm{res}\left(l\right)}$,
\begin{equation}
N_{\mathbf{k}}^{\mathrm{res}\left(l\right)}=\sum_{n_{V},n_{C}}N_{\mathbf{k},n_{C},n_{V}}^{\mathrm{res}\left(l\right)},\label{eq:N_k_l}
\end{equation}
(iv) the average residual electronic population per momentum point
coming from all $l$-photon resonant processes, $N^{\mathrm{res}\left(l\right)}$,
\begin{equation}
N^{\mathrm{res}\left(l\right)}=\frac{1}{M_{\mathrm{grid}}}\sum_{\mathbf{k},n_{V},n_{C}}N_{\mathbf{k},n_{C},n_{V}}^{\mathrm{res}\left(l\right)},\label{eq:N_l}
\end{equation}
where $M_{\mathrm{grid}}$ is the total number of momentum points
in the numerical grid, and, finally, we can be interested in (v) the
average residual excited electronic population per momentum point,
$N^{\mathrm{res}}$, which is actually the residual excitation population
per unit cell,
\begin{equation}
N^{\mathrm{res}}=\frac{1}{M_{\mathrm{grid}}}\sum_{\mathbf{k},n_{C}}N_{\mathbf{k},n_{C}}^{\mathrm{res}}.\label{eq:N_}
\end{equation}

\subsection{The generalized Houston approach\protect\label{subsec:Houston}}

One of the model-Hamiltonian methods to simulate the behavior of pumped
semiconductors is the Houston approach \citep{schlaepfer2018attosecond,sato2018role},
which has been formulated and is generally used in first quantization
and in the velocity gauge (see App.~\ref{app:Houston}). Here, we
reformulate this approach in second quantization within the DPOA framework,
highlighting its limitations and drawbacks.

We have seen that the Hamiltonian of a pumped quadratic lattice system
has the general form $H\left(t\right)=\sum_{\mathbf{k}}H_{\mathbf{k}}\left(t\right)$
where $H_{\mathbf{k}}\left(t\right)=c_{\mathbf{k}}^{\dagger}\left(t\right)\cdot\Xi_{\mathbf{k}}\left(t\right)\cdot c_{\mathbf{k}}\left(t\right)$
and $c_{\mathbf{k}}\left(t_{0}\right)=\left(c_{\mathbf{k},1}\left(t_{0}\right),\ldots,c_{\mathbf{k},\nu}\left(t_{0}\right),\ldots\right)$
is the canonical operatorial basis at equilibrium in vectorial notation
for an electron with momentum $\mathbf{k}$ and with $\nu$ denoting
all possible degrees of freedom of the system. Let us consider the
time-dependent transformation matrix $U_{\mathbf{k}}\left(t\right)$
that diagonalizes $\Xi_{\mathbf{k}}\left(t\right)$ at each instant
of time, i.e., $\Xi_{\mathbf{k}}^{S}\left(t\right)=U_{\mathbf{k}}^{\dagger}\left(t\right)\cdot\Xi_{\mathbf{k}}\left(t\right)\cdot U_{\mathbf{k}}\left(t\right)$
has only diagonal elements that are usually called instantaneous bands.
Then, we can define a new operatorial basis for the system, the Houston
basis $c_{\mathbf{k}}^{S}\left(t\right)$, given by $c_{\mathbf{k}}^{S}\left(t\right)=U_{\mathbf{k}}^{\dagger}\left(t\right)\cdot c_{\mathbf{k}}\left(t\right)$.
Within the DPOA framework, we can write $c_{\mathbf{k}}^{S}\left(t\right)=P_{\mathbf{k}}^{S}\left(t\right)\cdot c_{\mathbf{k}}\left(t_{0}\right)$
where $P_{\mathbf{k}}^{S}\left(t\right)$ is the Houston projection
matrix that satisfies the following equation of motion

\begin{equation}
\mathrm{i}\hbar\partial_{t}P_{\mathbf{k}}^{S}\left(t\right)=\left[\Xi_{\mathbf{k}}^{S}\left(t\right)+\Pi_{\mathbf{k}}\left(t\right)\right]\cdot P_{\mathbf{k}}^{S}\left(t\right),\label{eq:EOM_P_Hou}
\end{equation}
where $\Pi_{\mathbf{k}}\left(t\right)=\mathrm{i}\hbar\partial_{t}U_{\mathbf{k}}^{\dagger}\left(t\right)\cdot U_{\mathbf{k}}\left(t\right)$.
Another variant of the Houston method can be obtained, within second
quantization, by the following transformation 
\begin{equation}
P_{\mathbf{k}}^{\prime S}\left(t\right)=\mathrm{e}^{\frac{\mathrm{i}}{\hbar}\int_{t_{0}}^{t}dt^{\prime}\Xi_{\mathbf{k}}^{S}\left(t^{\prime}\right)}\cdot P_{\mathbf{k}}^{S}\left(t\right),\label{eq:P_Hou}
\end{equation}
which results in the following equation of motion
\begin{equation}
\mathrm{i}\hbar\partial_{t}P_{\mathbf{k}}^{\prime S}\left(t\right)=\Pi_{\mathbf{k}}^{\prime}\left(t\right)\cdot P_{\mathbf{k}}^{\prime S}\left(t\right),\label{eq:EOM_P_Hou_tilde}
\end{equation}
where
\begin{equation}
\Pi_{\mathbf{k}}^{\prime}\left(t\right)=\mathrm{e}^{\frac{\mathrm{i}}{\hbar}\int_{t_{0}}^{t}dt^{\prime}\Xi_{\mathbf{k}}^{S}\left(t^{\prime}\right)}\cdot\Pi_{\mathbf{k}}\left(t\right)\cdot\mathrm{e}^{-\frac{\mathrm{i}}{\hbar}\int_{t_{0}}^{t}dt^{\prime}\Xi_{\mathbf{k}}^{S}\left(t^{\prime}\right)}.\label{eq:Ham_P_Hou_tilde}
\end{equation}

Computing $\Xi_{\mathbf{k}}^{S}\left(t\right)$ and $\Pi_{\mathbf{k}}\left(t\right)$,
or equivalently $\Pi_{\mathbf{k}}^{\prime}\left(t\right)$, is not
only more time-consuming when many bands are involved as in \emph{real}
materials than just using $\Xi_{\mathbf{k}}\left(t\right)$, as in
DPOA, because of the numerical diagonalizations necessary to obtain
$U_{\mathbf{k}}\left(t\right)$ and $\partial_{t}U_{\mathbf{k}}^{\dagger}\left(t\right)$
at each instant of time, but it can be extremely difficult to calculate
it numerically, because of the well-known difficulty of tracking the
phase of eigenvectors between different instants of time in particular
in the presence of instantaneous-band crossing (dynamical degeneracy)
\citep{yue2022introduction}. This usually leads to implementing the
Houston method only for very few \emph{effective} bands and to use
\emph{approximate} $\mathbf{k}$-independent matrix elements. Actually,
DPOA can yield, if ever needed, the exact Houston-method results just
computing $P_{\mathbf{k}}^{S}\left(t\right)$ as
\begin{equation}
P_{\mathbf{k}}^{S}\left(t\right)=U_{\mathbf{k}}^{\dagger}\left(t\right)\cdot P_{\mathbf{k}}\left(t\right),
\end{equation}
where $P_{\mathbf{k}}\left(t\right)$ is the usual DPOA dynamical
projection matrix.

\subsection{Inter- and intra-band transitions\protect\label{subsec:Inter_intra}}

Within DPOA, it is straightforward to separate the effects of the
so-called inter-/intra-band transitions. In order to have only intra-band
transitions in the dynamics, in the basis of equilibrium bands, those
indexed by $n$, one needs to keep only the diagonal elements of $\Xi_{\mathbf{k}}\left(t\right)$
and remove all off-diagonal ones which cause transitions among the
bands:

\begin{equation}
\mathrm{i}\hbar\partial_{t}P_{\mathbf{k},n,n{}^{\prime}}^{\mathrm{intra}}\left(t\right)=\varepsilon_{\mathbf{k},n}^{\mathrm{intra}}\left(t\right)P_{\mathbf{k},n,n{}^{\prime}}^{\mathrm{intra}}\left(t\right),\label{eq:EOM_P_intra}
\end{equation}
where $\varepsilon_{\mathbf{k},n}^{\mathrm{intra}}\left(t\right)=\Xi_{\mathbf{k},n,n}\left(t\right)=\sum_{\nu,\nu{}^{\prime}}\Omega_{\mathbf{k},n,\nu}^{\dagger}\tilde{\Xi}_{\mathbf{k},\nu,\nu{}^{\prime}}\left(t\right)\Omega_{\mathbf{k},\nu{}^{\prime},n}$.
Eq.~\ref{eq:EOM_P_intra} has the formal solution $P_{\mathbf{k},n,n{}^{\prime}}^{\mathrm{intra}}\left(t\right)=\delta_{n,n^{\prime}}e^{-\frac{\mathrm{i}}{\hbar}\int_{t_{0}}^{t}\varepsilon_{\mathbf{k},n}^{\mathrm{intra}}\left(t^{\prime}\right)dt{}^{\prime}}$.
Even though here we use the standard term \emph{intra-band transition},
one should keep in mind that performing the calculation for a specific
$\mathbf{k}$ point, one does not need to take into account the equilibrium
band structure or electronic distribution at any other adjacent $\mathbf{k}$
point.

On the other hand, in order to keep only inter-band transitions, it
is needed to keep the off-diagonal elements of $\Xi_{\mathbf{k}}\left(t\right)$
and discard the Peierls substitution in its diagonal elements: $\varepsilon_{\mathbf{k},n}^{\mathrm{inter}}\left(t\right)=\varepsilon_{\mathbf{k},n}+\mathrm{e}\boldsymbol{E}\left(t\right)\bullet\boldsymbol{D}_{\mathbf{k},n,n}$.
Accordingly, we have

\begin{multline}
\mathrm{i}\hbar\partial_{t}P_{\mathbf{k},n,n{}^{\prime}}^{\mathrm{inter}}\left(t\right)=\varepsilon_{\mathbf{k},n}^{\mathrm{inter}}\left(t\right)P_{\mathbf{k},n,n{}^{\prime}}^{\mathrm{inter}}\left(t\right)\\
+\sum_{\bar{n}\neq n}\Xi_{\mathbf{k},n,\bar{n}}\left(t\right)P_{\mathbf{k},\bar{n},n{}^{\prime}}^{\mathrm{inter}}\left(t\right).\label{eq:EOM_P_inter}
\end{multline}
Usually, the diagonal elements of the dipole matrix are negligible,
$\boldsymbol{D}_{\mathbf{k},n,n}\simeq0$, and therefore $\varepsilon_{\mathbf{k},n}^{\mathrm{inter}}\left(t\right)$
is almost equal to the equilibrium band energy $\varepsilon_{\mathbf{k},n}$.

The Houston method is often used to perform the same kind of analysis.
Within the velocity gauge, to remove the intra-band dynamics and define
an only inter-band one, one sets $\mathbf{k}+\frac{\mathrm{e}}{\hbar}\boldsymbol{A}\left(t\right)\rightarrow\mathbf{k}$
in the instantaneous eigenenergies and eigenvectors reducing them
to the equilibrium ones, but one still computes the projection coefficients
(see Eq.~\ref{eq:Houston_coe}) through the full equation of motion
whose inter-band term just comes from the differentiation of the very
same Peierls-like term. This is somehow questionable and ambiguous.
Moreover, defining inter- and intra-band dynamics in the Houston basis
is ambiguous as the instantaneous bands are superpositions of equilibrium
bands and therefore any interpretation becomes very cumbersome.

\subsection{Green's functions and TR-ARPES signal\protect\label{subsec:GF_TR-ARPES}}

Green's functions (GFs) are extremely important tools as they allow
to compute many interesting properties of a system. The most relevant
single-particle two-time electronic GFs are the retarded, $G^{R}$,
and the lesser, $G^{<}$, GFs, defined in the vectorial notation as
follows
\begin{align}
 & G_{\mathbf{k},n,n^{\prime}}^{R}\left(t,t^{\prime}\right)=-\mathrm{i}\theta\left(t-t^{\prime}\right)\left\langle \left\{ c_{\mathbf{k},n}\left(t\right),c_{\mathbf{k},n^{\prime}}^{\dagger}\left(t^{\prime}\right)\right\} \right\rangle ,\label{eq:GF_R}\\
 & G_{\mathbf{k},n,n^{\prime}}^{<}\left(t,t^{\prime}\right)=\mathrm{i}\left\langle c_{\mathbf{k},n^{\prime}}^{\dagger}\left(t^{\prime}\right)c_{\mathbf{k},n}\left(t\right)\right\rangle .\label{eq:eq:GF_L}
\end{align}
Even for a quadratic Hamiltonian, the GFs cannot be computed within
the SPDM approach (unless one defines a two-time SPDM \citep{broers2022detecting},
which is computationally very heavy), but they can be straightforwardly
obtained within DPOA in terms of the dynamical projection matrices
$P$ as
\begin{align}
 & G_{\mathbf{k},n,n^{\prime}}^{R}\left(t,t^{\prime}\right)=-\mathrm{i}\theta\left(t-t^{\prime}\right)\sum_{m}P_{\mathbf{k},n,m}\left(t\right)P_{\mathbf{k},n^{\prime},m}^{\star}\left(t^{\prime}\right),\label{eq:GF_R_P}\\
 & G_{\mathbf{k},n,n^{\prime}}^{<}\left(t,t^{\prime}\right)=\mathrm{i}\sum_{m,m^{\prime}}\left(\delta_{m,m^{\prime}}-\rho_{\mathbf{k},m,m^{\prime}}\left(t_{0}\right)\right)P_{\mathbf{k},n,m}\left(t\right)P_{\mathbf{k},n^{\prime},m^{\prime}}^{\star}\left(t^{\prime}\right),\label{eq:GF_L_P}
\end{align}
where, in the band basis in which the equilibrium Hamiltonian is diagonal,
$\delta_{n,n^{\prime}}-\rho_{\mathbf{k},n,n^{\prime}}\left(t_{0}\right)=\delta_{n,n^{\prime}}f_{+}\left(\varepsilon_{\mathbf{k},n}\right)$.

At equilibrium, the usual way to study the energy bands of the system,
$\varepsilon_{\mathbf{k},n}$, and their corresponding occupations,
is to compute the spectral functions through the imaginary components
of the retarded and of the lesser GFs, respectively. However, out-of-equilibrium,
the spectral functions are not necessarily non-negative quantities
\citep{eckstein2021time} (see App.~\ref{app:SF}). This occurrence
invalidates their physical interpretation of availability and occupation
of the corresponding energies per momentum. Nevertheless, such an
information is of crucial importance to describe and understand the
response of the system to external probes.

Indeed, out of equilibrium, one investigates the TR-ARPES signal \citep{freericks2009theoretical,sentef2013examining,freericks2015gauge,schuler2021theory},
which individuates the occupation of the energy $\omega$ at momentum
$\mathbf{k}$ for a probe pulse centered at time $t_{\mathrm{pr}}$.
The TR-ARPES signal is proportional to

\begin{multline}
I_{\mathbf{k}}^{<}\left(\omega,t_{\mathrm{pr}}\right)=\frac{\tau_{\mathrm{pr}}}{\sqrt{8\pi\ln2}}\int_{-\infty}^{+\infty}dt_{1}\int_{-\infty}^{+\infty}dt_{2}S_{\mathrm{pr}}\left(t_{1}-t_{\mathrm{pr}}\right)\\
S_{\mathrm{pr}}\left(t_{2}-t_{\mathrm{pr}}\right)\Im\left[e^{\mathrm{i}\omega\left(t_{1}-t_{2}\right)}\Tr\left[G_{\mathbf{k}}^{<}\left(t_{1},t_{2}\right)\right]\right],\label{eq:I_L}
\end{multline}
where $S_{\mathrm{pr}}\left(t-t_{\mathrm{pr}}\right)=\frac{2\sqrt{\ln2}}{\sqrt{\pi}\tau_{\mathrm{pr}}}e^{-4\ln2\left(t-t_{\mathrm{pr}}\right)^{2}/\tau_{\mathrm{pr}}^{2}}$
is the probe-pulse envelope which is assumed to be Gaussian with a
FWHM $\tau_{\mathrm{pr}}$. Here we assumed that the TR-ARPES matrix
elements are just constant numerical factors and removed them from
the expression. Moreover, we assumed that the ejected photo-electrons
outside of the sample, originating from orthogonal electronic states
inside of the solid, are described by orthogonal wave functions. This
assumption leads to the presence of the trace ($\Tr$) in Eq.~\ref{eq:I_L}.
At any rate, $\Tr\left[G_{\mathbf{k}}^{<}\left(t_{1},t_{2}\right)\right]$
is invariant with respect to the chosen basis as it is desirable.
Without such assumptions, one would need to carry on a detailed modeling
to get the actual matrix elements \citep{freericks2015gauge,schuler2021theory}.
We have chosen the normalization factor in such a way that $I^{<}\left(\omega,t_{\mathrm{pr}}\right)$
is normalized to the total number of particles at momentum $\mathbf{k}$,
\begin{equation}
\int_{-\infty}^{+\infty}d\omega I_{\mathbf{k}}^{<}\left(\omega,t_{\mathrm{pr}}\right)=\sum_{n}N_{\mathbf{k},n}.\label{eq:I_L_norm}
\end{equation}

$I_{\mathbf{k}}^{<}\left(\omega,t_{\mathrm{pr}}\right)$ gives information
about the occupied states. Instead, to identify the available states
$\left(\omega,\mathbf{k}\right)$, that is the bands out-of-equilibrium
or TR-ARPES \emph{bands}, we use the retarded GF in place of the lesser
one and define
\begin{multline}
I_{\mathbf{k}}^{R}\left(\omega,t_{\mathrm{pr}}\right)=-\frac{\tau_{\mathrm{pr}}}{\sqrt{2\pi\ln2}}\int_{-\infty}^{+\infty}dt_{1}\int_{-\infty}^{+\infty}dt_{2}S_{\mathrm{pr}}\left(t_{1}-t_{\mathrm{pr}}\right)\\
S_{\mathrm{pr}}\left(t_{2}-t_{\mathrm{pr}}\right)\Im\left[e^{\mathrm{i}\omega\left(t_{1}-t_{2}\right)}\Tr\left[G_{\mathbf{k}}^{R}\left(t_{1},t_{2}\right)\right]\right].\label{eq:I_R}
\end{multline}
It is straightforward to show that, in the band basis (see App. ~\ref{app:FD_derivition}),

\begin{align}
I_{\mathbf{k}}^{<}\left(\omega,t_{\mathrm{pr}}\right)= & \sum_{n,n^{\prime}}L_{\mathbf{k},n;n^{\prime}}\left(\omega,t_{\mathrm{pr}}\right)f_{+}\left(\varepsilon_{\mathbf{k},n^{\prime}}\right),\label{eq:FD_1}\\
I_{\mathbf{k}}^{R}\left(\omega,t_{\mathrm{pr}}\right)= & \sum_{n,n^{\prime}}L_{\mathbf{k},n;n^{\prime}}\left(\omega,t_{\mathrm{pr}}\right),\label{eq:FD_2}
\end{align}
where 
\begin{multline}
L_{\mathbf{k},n;n^{\prime}}\left(\omega,t_{\mathrm{pr}}\right)=\\
=\frac{\tau_{\mathrm{pr}}}{2\sqrt{2\pi\ln2}}\left|\int_{-\infty}^{+\infty}dt_{1}S_{\mathrm{pr}}\left(t_{1}-t_{\mathrm{pr}}\right)e^{\mathrm{i}\omega t_{1}}P_{\mathbf{k},nn^{\prime}}\left(t_{1}\right)\right|^{2},\label{eq:L_redux}
\end{multline}
which guarantees that the TR-ARPES signal is always non-negative.

Eqs.~\ref{eq:FD_1} and \ref{eq:FD_2} provide a generalized fluctuation-dissipation
theorem for TR-ARPES signal.

\begin{figure}
\begin{centering}
\begin{tabular}{c}
\includegraphics[width=7.5cm]{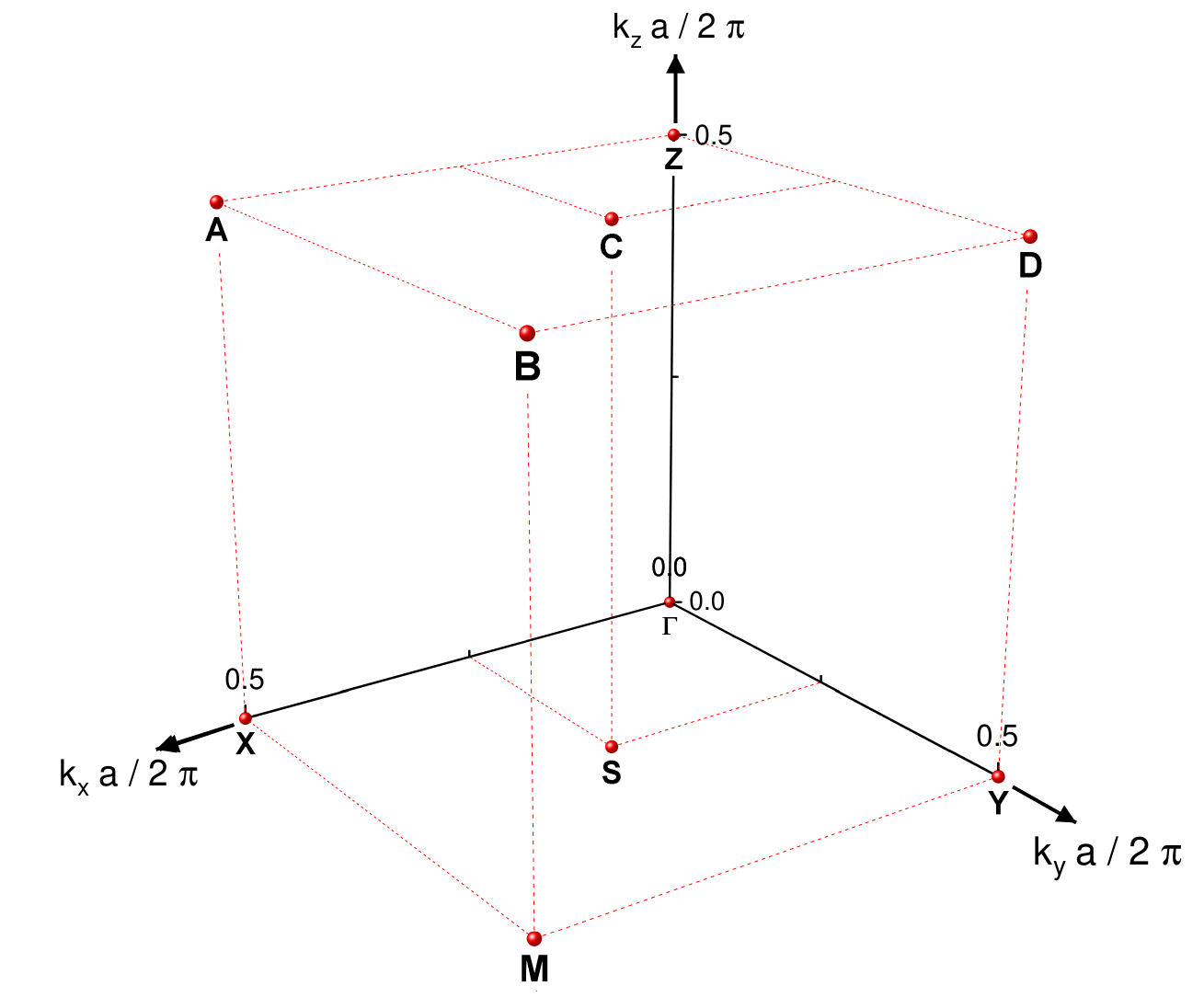}\tabularnewline
\includegraphics[width=7.5cm]{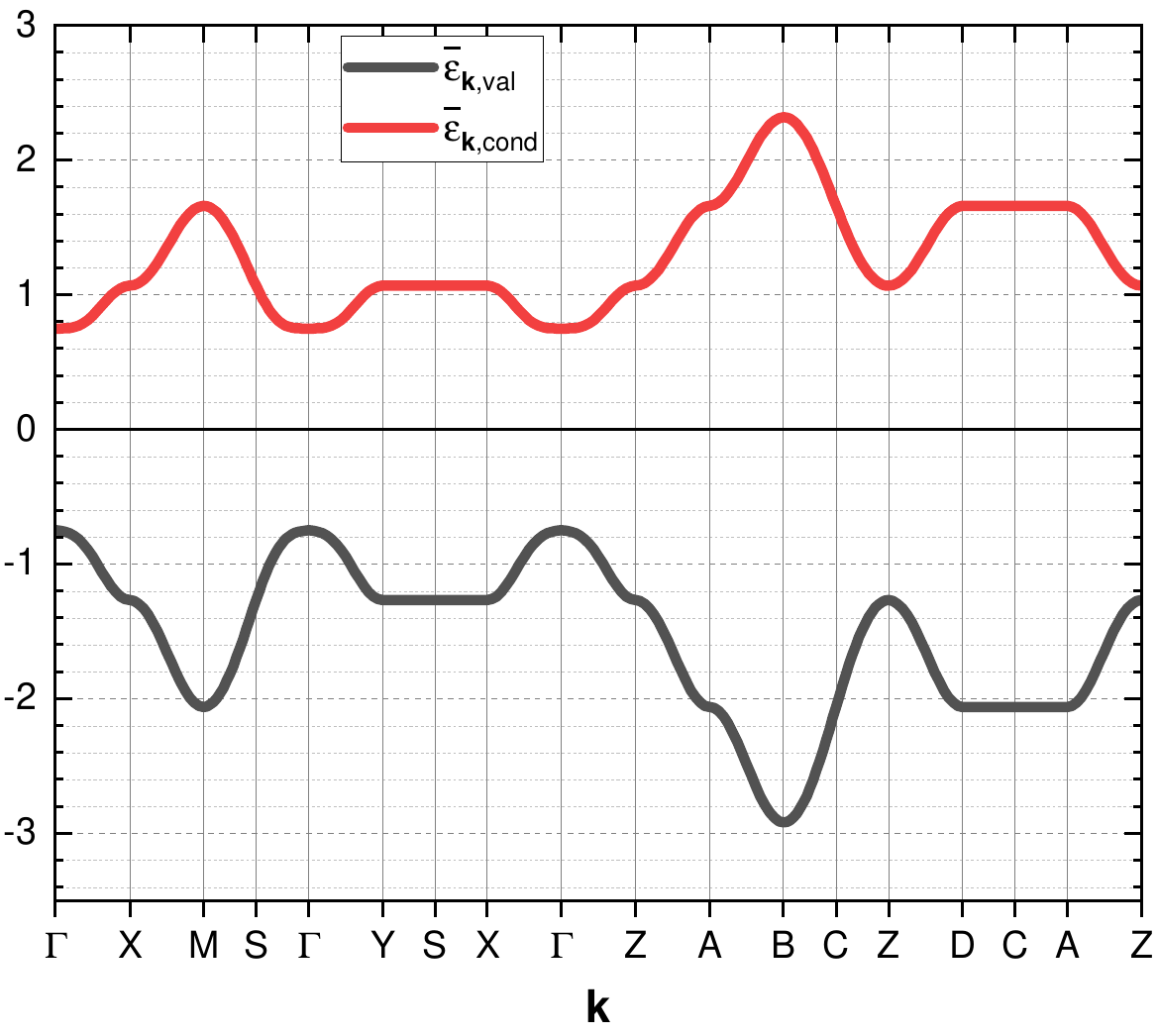}\tabularnewline
\end{tabular}
\par\end{centering}
\caption{(top) High-symmetry points in the first Brillouin zone. (bottom) Equilibrium
energy bands, $\bar{\varepsilon}_{\mathbf{k},\mathrm{val}}$ and $\bar{\varepsilon}_{\mathbf{k},\mathrm{cond}}$,
along the \emph{main} path.\protect\label{fig:kspace_bands}}
\end{figure}

\begin{figure}[t]
\begin{centering}
\begin{tabular}{c}
\includegraphics[width=7.5cm]{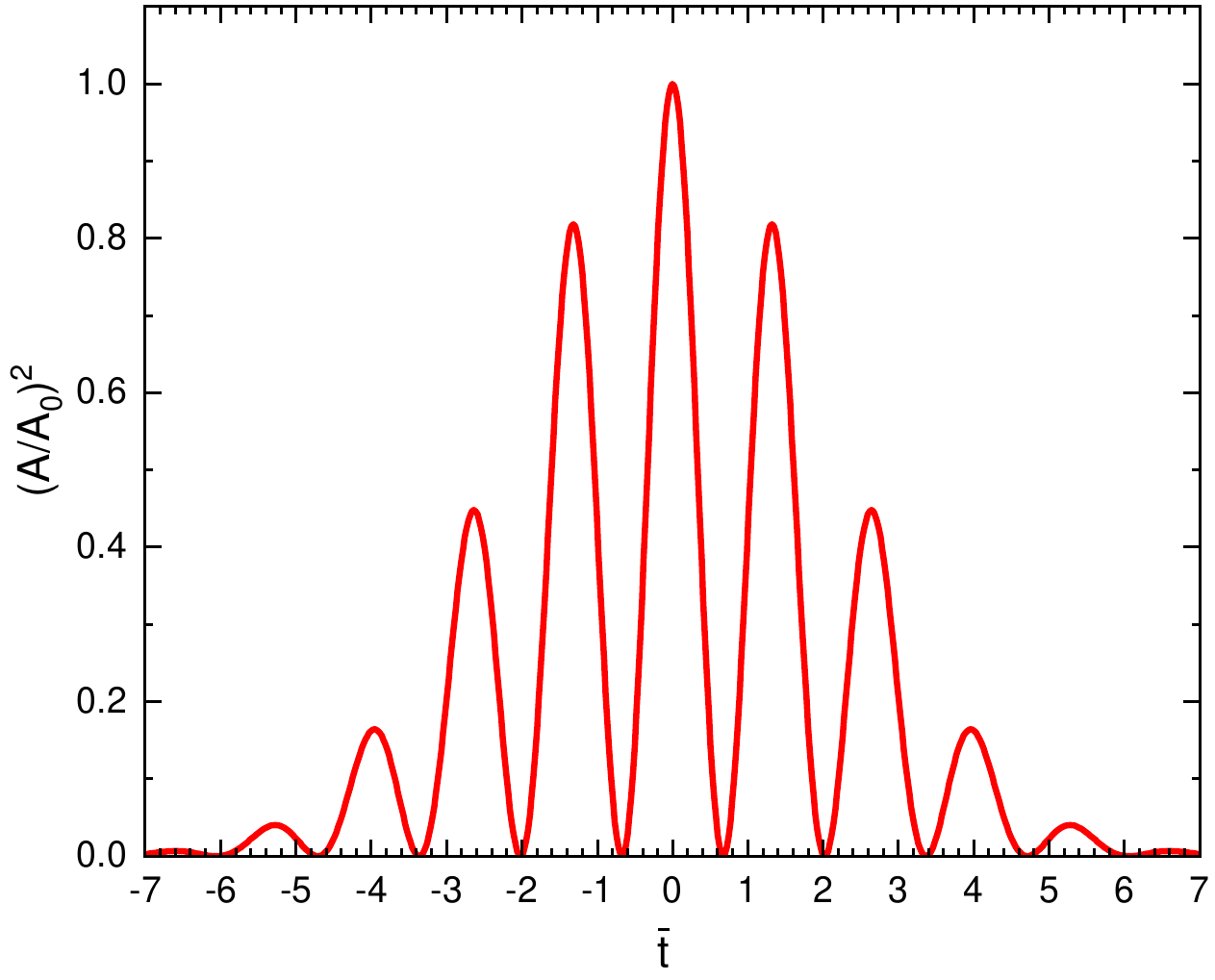}\tabularnewline
\includegraphics[width=7.5cm]{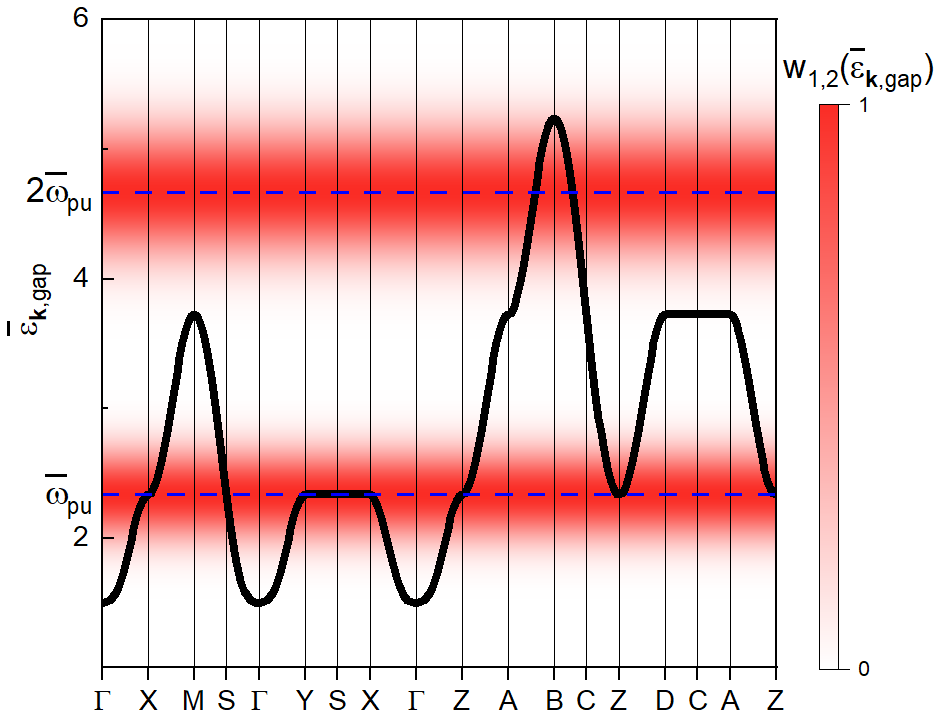}\tabularnewline
\end{tabular}
\par\end{centering}
\caption{(top) The square of the pumping vector potential as a function of
time. (bottom) The energy gaps, $\bar{\varepsilon}_{\mathbf{k},\mathrm{gap}}$,
along the \emph{main} path. The colored map shows $w_{l=1,2}\left(\bar{\varepsilon}_{\mathbf{k},\mathrm{gap}}\right)$
for one- and two-photon resonances.\protect\label{fig:A2_gaps}}
\end{figure}

\begin{figure}[t]
\begin{centering}
\begin{tabular}{c}
\includegraphics[width=7.5cm]{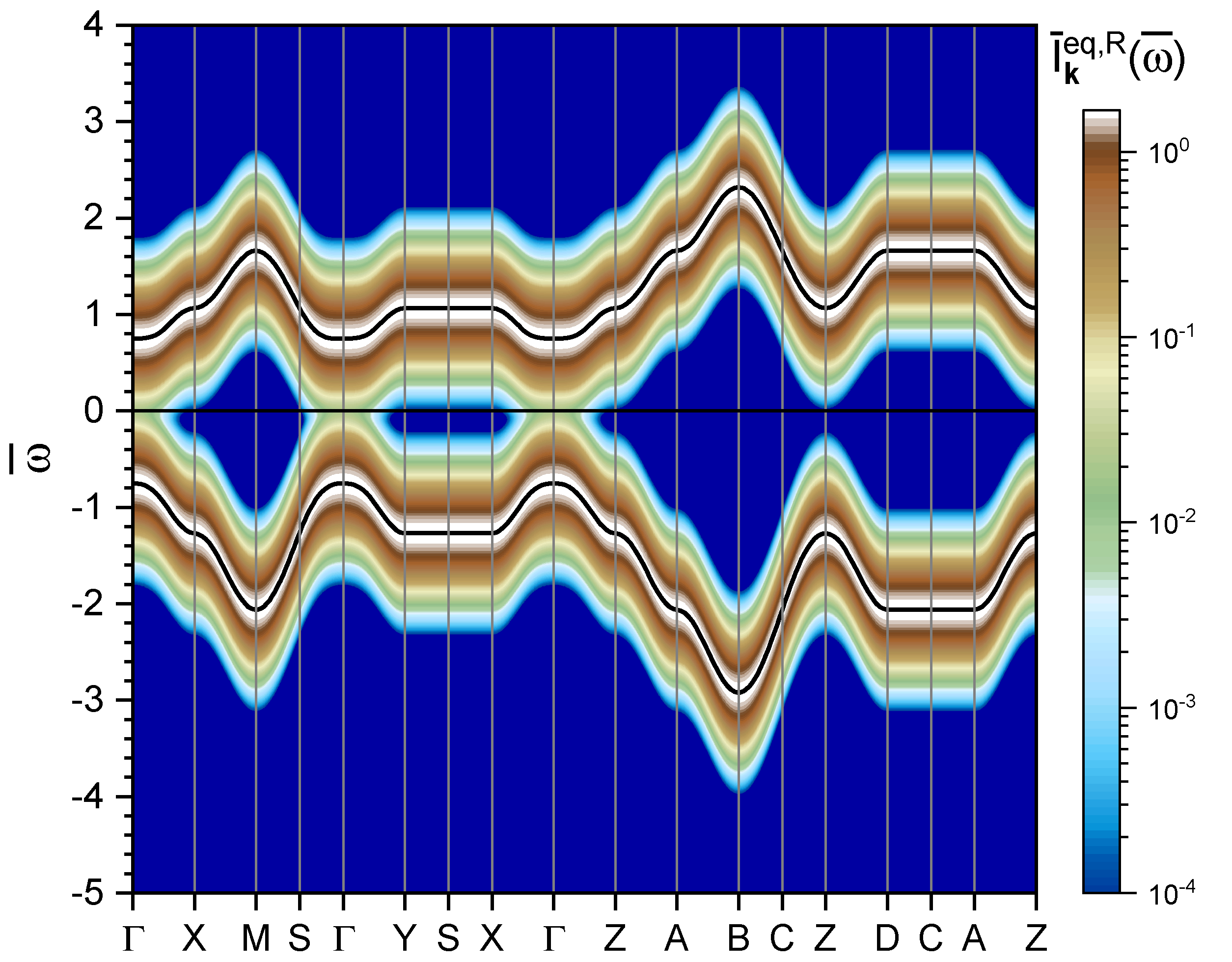}\tabularnewline
\includegraphics[width=7.5cm]{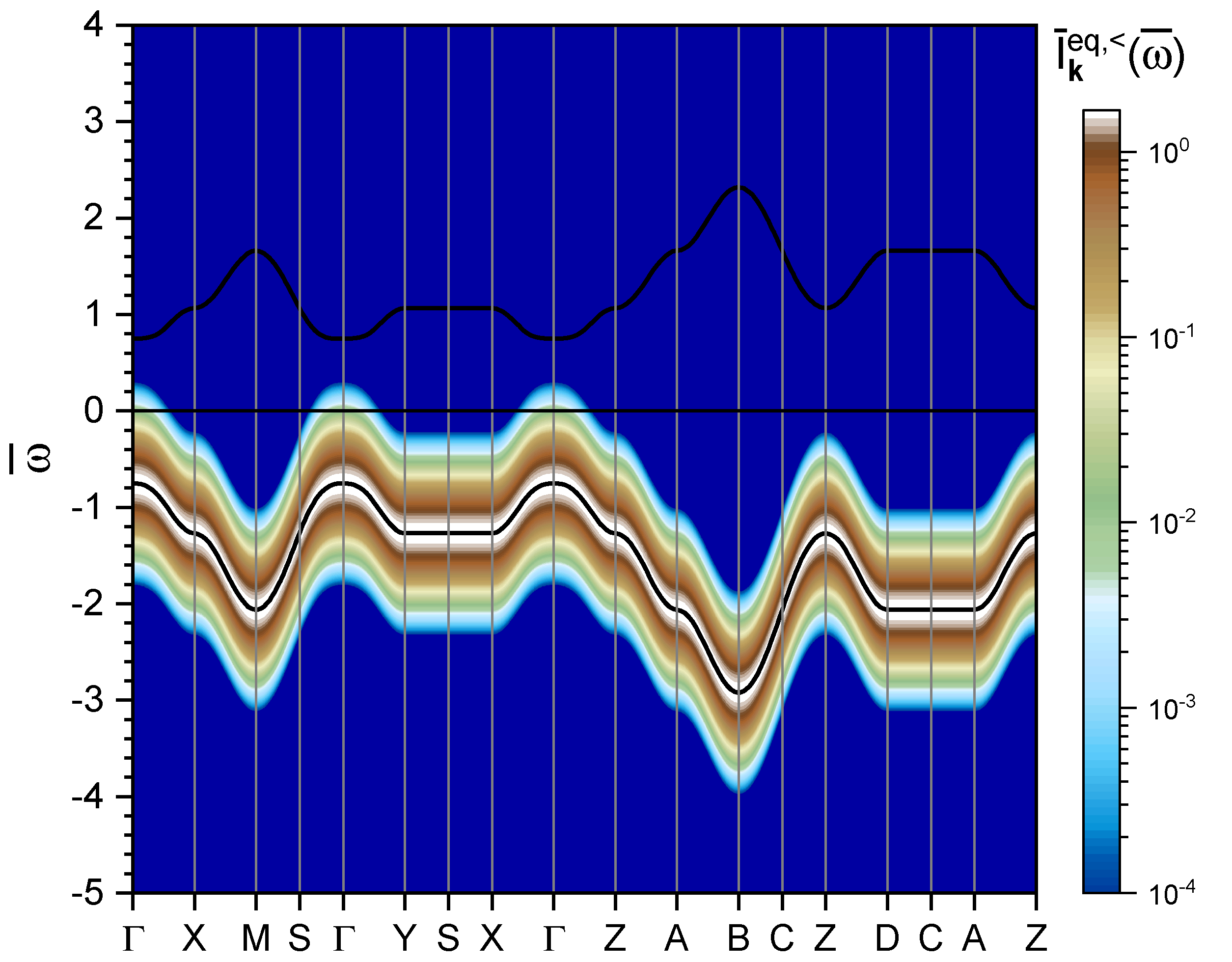}\tabularnewline
\end{tabular}
\par\end{centering}
\caption{TR-ARPES signals at equilibrium, (top) $\bar{I}_{\mathbf{k}}^{\mathrm{eq},R}\left(\bar{\omega}\right)$
and (bottom) $\bar{I}_{\mathbf{k}}^{\mathrm{eq},<}\left(\bar{\omega}\right)$.
The solid black curves show the equilibrium energy bands.\protect\label{fig:I_k-eq}}
\end{figure}

\section{A two-band lattice system: a noteworthy application\protect\label{sec:Two-band_system}}

Very recently, we have proved the capabilities of DPOA in investigating
\emph{real} materials by exploiting it to analyze the actual photo-injection
mechanisms in germanium within an ultrafast (attosecond) pump-probe
setup \citep{inzani2022field}. To discuss the variety of possible
physical phenomena without being limited by the characteristics of
a single particular \emph{real} material, here we choose to study
a toy model. This study will be a guideline for understanding the
complex effects and interplays in realistic setups. We consider a
cubic lattice system, of lattice constant $a$, with two bands corresponding
to the main valence and conduction bands in a semiconductor. We consider
two localized states with the onsite energies $\tilde{T}_{\mathbf{R}=\boldsymbol{0},1,1}=-1.65\Delta$
and $\tilde{T}_{\mathbf{R}=\boldsymbol{0},2,2}=1.35\Delta$, respectively,
diagonal first-neighbor hoppings $\tilde{T}_{\mathbf{R}=\boldsymbol{a},1,1}=0.2\Delta$
and $\tilde{T}_{\mathbf{R}=\boldsymbol{a},2,2}=-0.15\Delta$, and
off-diagonal first-neighbor hoppings $\tilde{T}_{\mathbf{R}=\boldsymbol{a},1,2}=\tilde{T}_{\mathbf{R}=\boldsymbol{a},2,1}=-0.1\Delta$,
where $\tilde{T}_{\boldsymbol{R},\nu,\nu^{\prime}}$ is the hopping
matrix between two sites at distance $\boldsymbol{R}$ and states
$\nu$ and $\nu^{\prime}$, respectively, $\boldsymbol{a}\in\left\{ a\left(\pm1,0,0\right),a\left(0,\pm1,0\right),a\left(0,0,\pm1\right)\right\} $
and $\Delta$ is the unit of energy that can be adjusted to obtain
the desired band gap energy at $\Gamma=\left(0,0,0\right)$. With
our parameters, the band gap at $\Gamma$ is $1.5\Delta$, so that
in order to have a gap of $\unit[0.75]{eV}$ for instance, one should
set $\Delta=\unit[0.5]{eV}$. For the cases that we analyze with a
finite dipole, we consider an on-site (local) and off-diagonal dipole
moment: $\tilde{\mathbf{D}}_{\mathbf{R}=\boldsymbol{0},1,2}=\tilde{\mathbf{D}}_{\mathbf{R}=\boldsymbol{0},2,1}^{*}=\mathrm{i}0.05a\hat{\mathbf{j}}$,
which will lead only to a $0$-th term in its \emph{Peierls expansion}.

In Fig.~\ref{fig:kspace_bands}, top panel, we show the high-symmetry
points of the first Brillouin zone, while in the bottom panel we show
the equilibrium energy bands, $\bar{\varepsilon}_{\mathbf{k},\mathrm{val}}=\varepsilon_{\mathbf{k},1}/\Delta$
and $\bar{\varepsilon}_{\mathbf{k},\mathrm{cond}}=\varepsilon_{\mathbf{k},2}/\Delta$,
for a path which connects these high-symmetry points (the \emph{main}
path hereafter). All energies denoted with a bar on top are divided
by $\Delta$ and hence dimensionless. Having $\Delta$ as the unit
of energy, the unit of time is simply chosen to be $\hbar/\Delta$,
which results in the dimensionless time $\bar{t}=t\Delta/\hbar$ for
each time $t$.

We apply a pump pulse in the form $\boldsymbol{A}\left(\bar{t}\right)=A\left(\bar{t}\right)\hat{\mathbf{j}}$
where $A\left(\bar{t}\right)$ is a wave packet given by

\begin{equation}
A\left(\bar{t}\right)=\frac{2\pi\hbar}{a\mathrm{e}}\bar{A}_{0}e^{-\left(4\ln2\right)\bar{t}^{2}/\bar{\tau}_{\mathrm{pu}}^{2}}\cos\left(\bar{\omega}_{\mathrm{pu}}\bar{t}\right),\label{eq:A_t}
\end{equation}
in which the center of the pump pulse is taken as the origin of the
time axis. The dimensionless frequency of the pump pulse is chosen
to be $\bar{\omega}_{\mathrm{pu}}=\omega_{\mathrm{pu}}\hbar/\Delta=2.33$
and, unless otherwise explicitly stated, the FWHM is chosen to be
$\bar{\tau}_{\mathrm{pu}}=7$ and the dimensionless pump-pulse amplitude
is chosen to be $\bar{A}_{0}=0.2$.

The square of the pumping vector potential as a function of time is
plotted in Fig.~\ref{fig:A2_gaps}, top panel. Fig.~\ref{fig:A2_gaps}
bottom panel shows the energy gaps, $\bar{\varepsilon}_{\mathbf{k},\mathrm{gap}}=\bar{\varepsilon}_{\mathbf{k},\mathrm{cond}}-\bar{\varepsilon}_{\mathbf{k},\mathrm{val}}$,
at the \textbf{k} points along the \emph{main} path, while the colored
map shows $w_{l=1,2}\left(\bar{\varepsilon}_{\mathbf{k},\mathrm{gap}}\right)$,
which indicates the strength of $l$-photon resonance for each energy
gap.

For TR-ARPES signal, we apply a probe pulse with FWHM of $\bar{\tau}_{\mathrm{pr}}=\bar{\tau}_{\mathrm{pu}}=7$,
unless otherwise explicitly stated. We study the dimensionless signals
that are obtained as $\bar{I}_{\mathbf{k}}^{R,<}=\nicefrac{\Delta}{\hbar}I_{\mathbf{k}}^{R,<}$.
In Fig.~\ref{fig:I_k-eq}, we show $\bar{I}_{\mathbf{k}}^{\mathrm{eq},R}\left(\bar{\omega}\right)$
and $\bar{I}_{\mathbf{k}}^{\mathrm{eq},<}\left(\bar{\omega}\right)$
at equilibrium, that is when no pump pulse is applied to the system.
The finite width of the probe pulse results in a broadening of the
levels, which is intrinsic to quantum mechanics and unavoidable. Increasing
the FWHM of the probe pulse, one can decrease this broadening, but
we are not interested in probe pulses much wider than the pump-pulse
envelope. The retarded signal, $\bar{I}_{\mathbf{k}}^{\mathrm{eq},R}\left(\bar{\omega}\right)$,
is peaked around both valence and conduction band energies and shows
the spectrum of the system, while the lesser signal, $\bar{I}_{\mathbf{k}}^{\mathrm{eq},<}\left(\bar{\omega}\right)$,
shows the occupied valence-band levels only, which is the signal measured
in experiments.

\begin{figure*}
\centering{}%
\begin{tabular}{ccc}
\includegraphics[width=6cm]{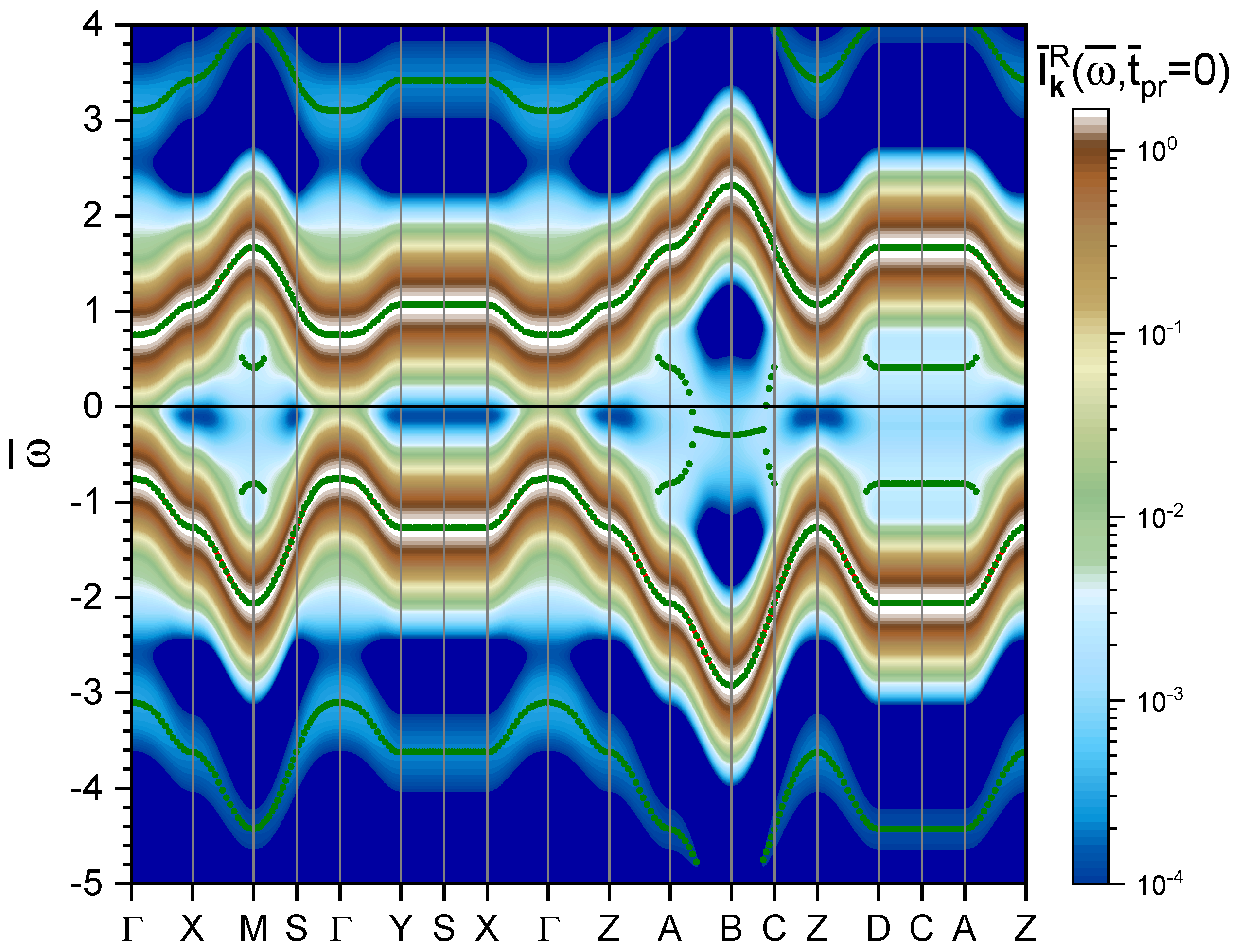} & \includegraphics[width=6cm]{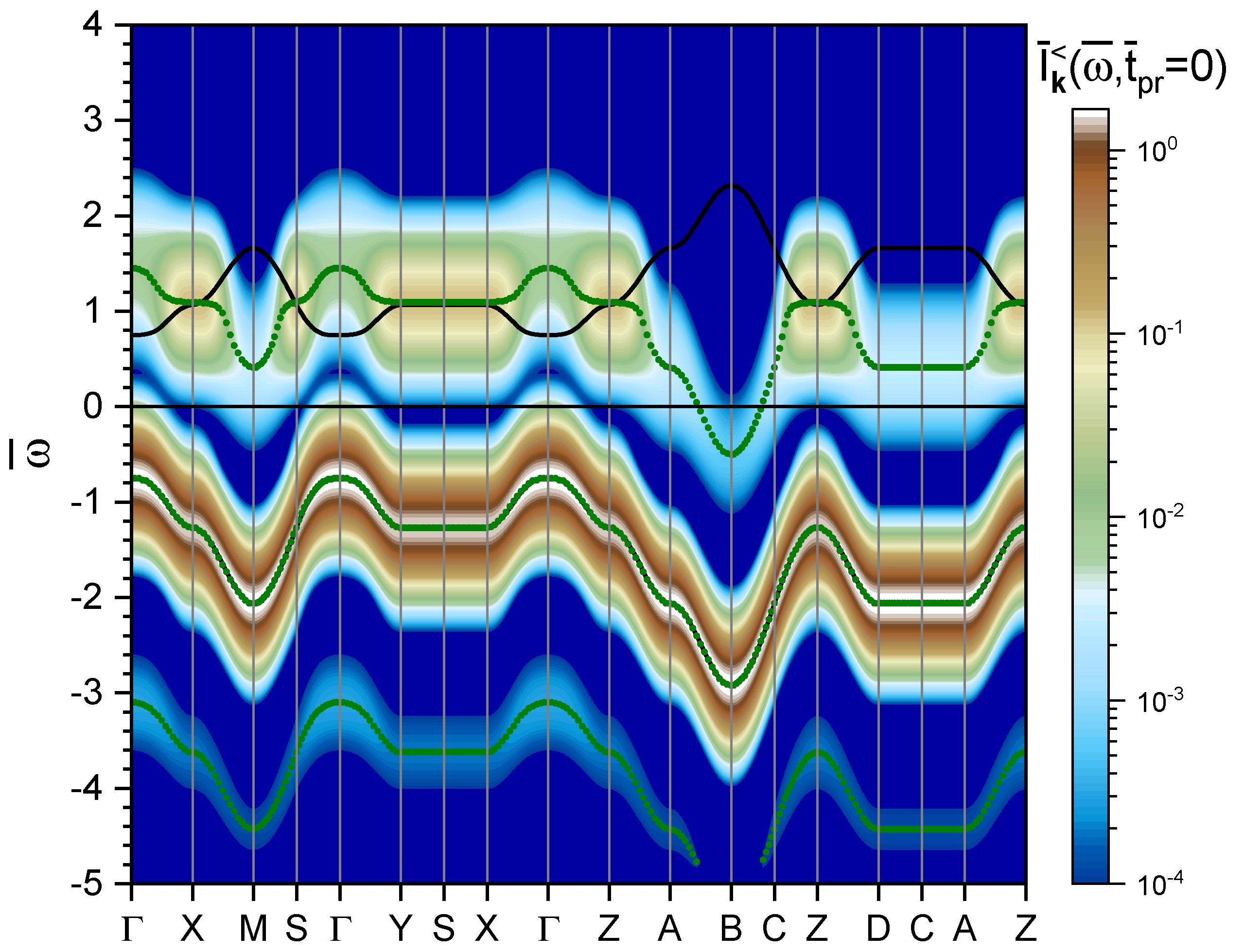} & \includegraphics[width=6cm]{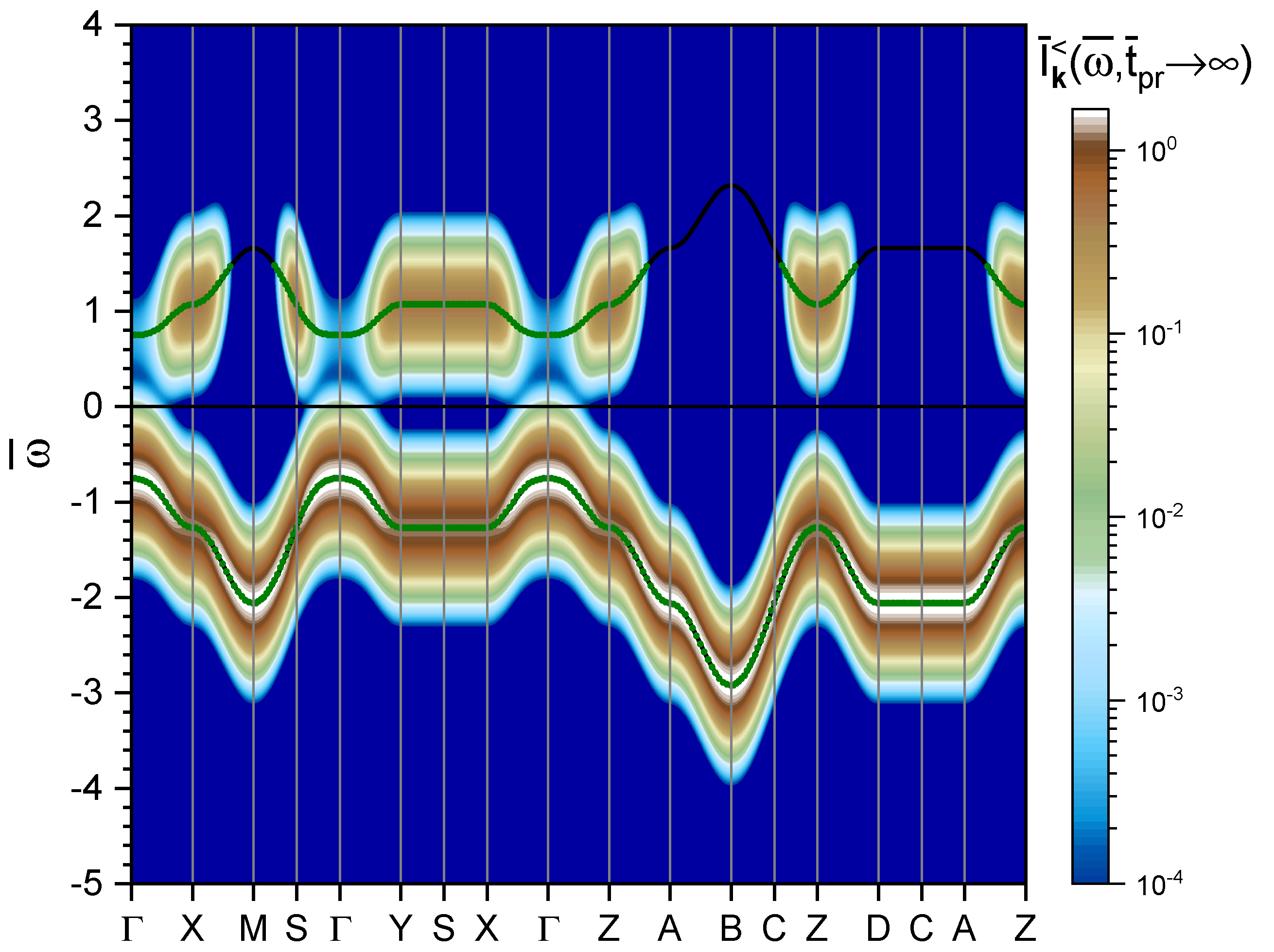}\tabularnewline
\end{tabular}\caption{TR-ARPES signals for the case of local dipole coupling. (left) $\bar{I}_{\mathbf{k}}^{R}\left(\bar{\omega},\bar{t}_{\mathrm{pr}}=0\right)$
and (middle) $\bar{I}_{\mathbf{k}}^{<}\left(\bar{\omega},\bar{t}_{\mathrm{pr}}=0\right)$:
the center of the probe pulse coincides with the center of the pump
pulse. (right) $\bar{I}_{\mathbf{k}}^{<}\left(\bar{\omega},\bar{t}_{\mathrm{pr}}\rightarrow\infty\right)$:
after the pump pulse is turned off. The black solid curves mark the
equilibrium band energies. The green dots indicate the local maxima
in energy of the signals for each $\mathbf{k}$: the (out-of-equilibrium)
bands of TR-ARPES.\protect\label{fig:I-dipole-only}}
\end{figure*}

\begin{figure}[t]
\begin{centering}
\begin{tabular}{c}
\includegraphics[width=7.5cm]{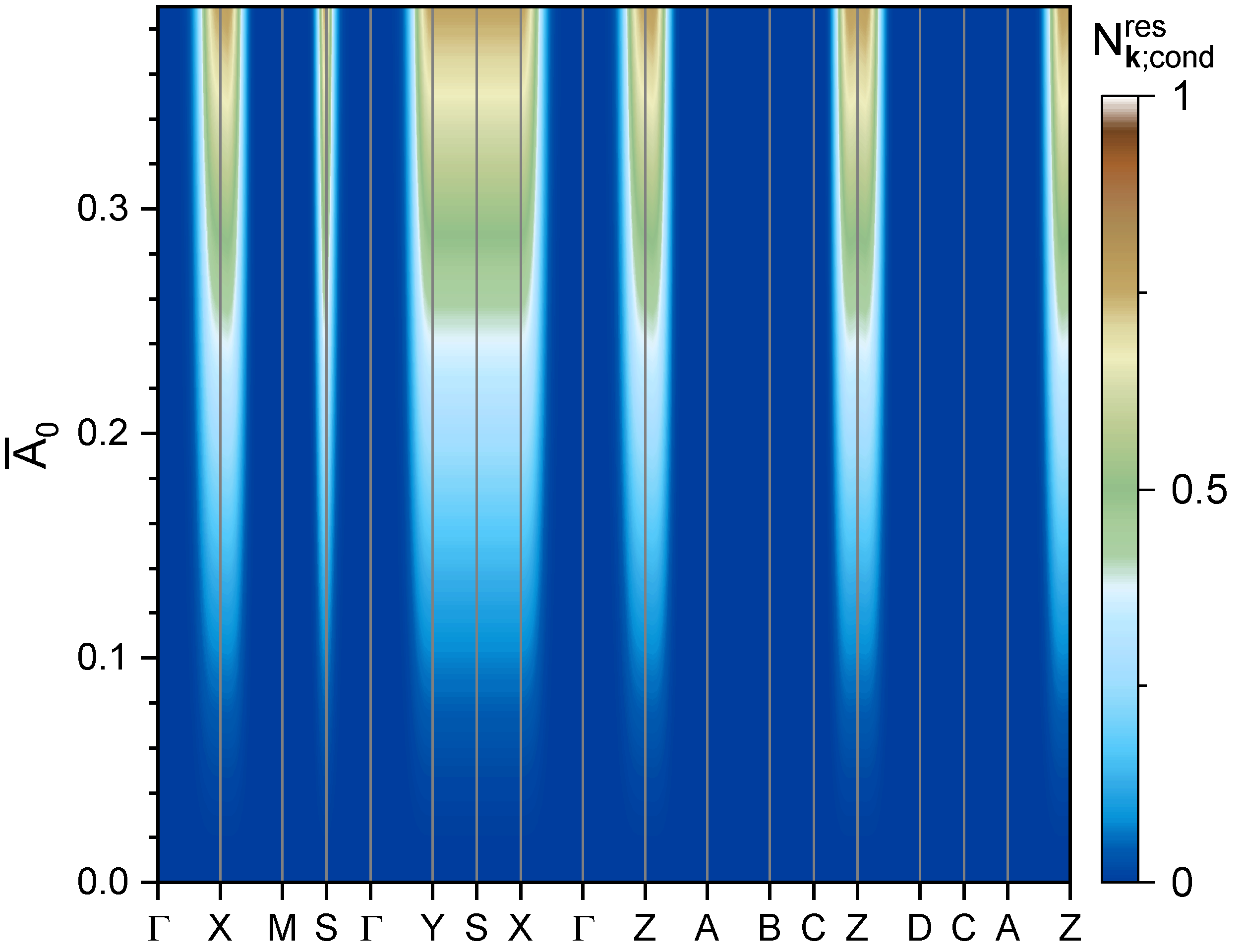}\tabularnewline
\includegraphics[width=7.5cm]{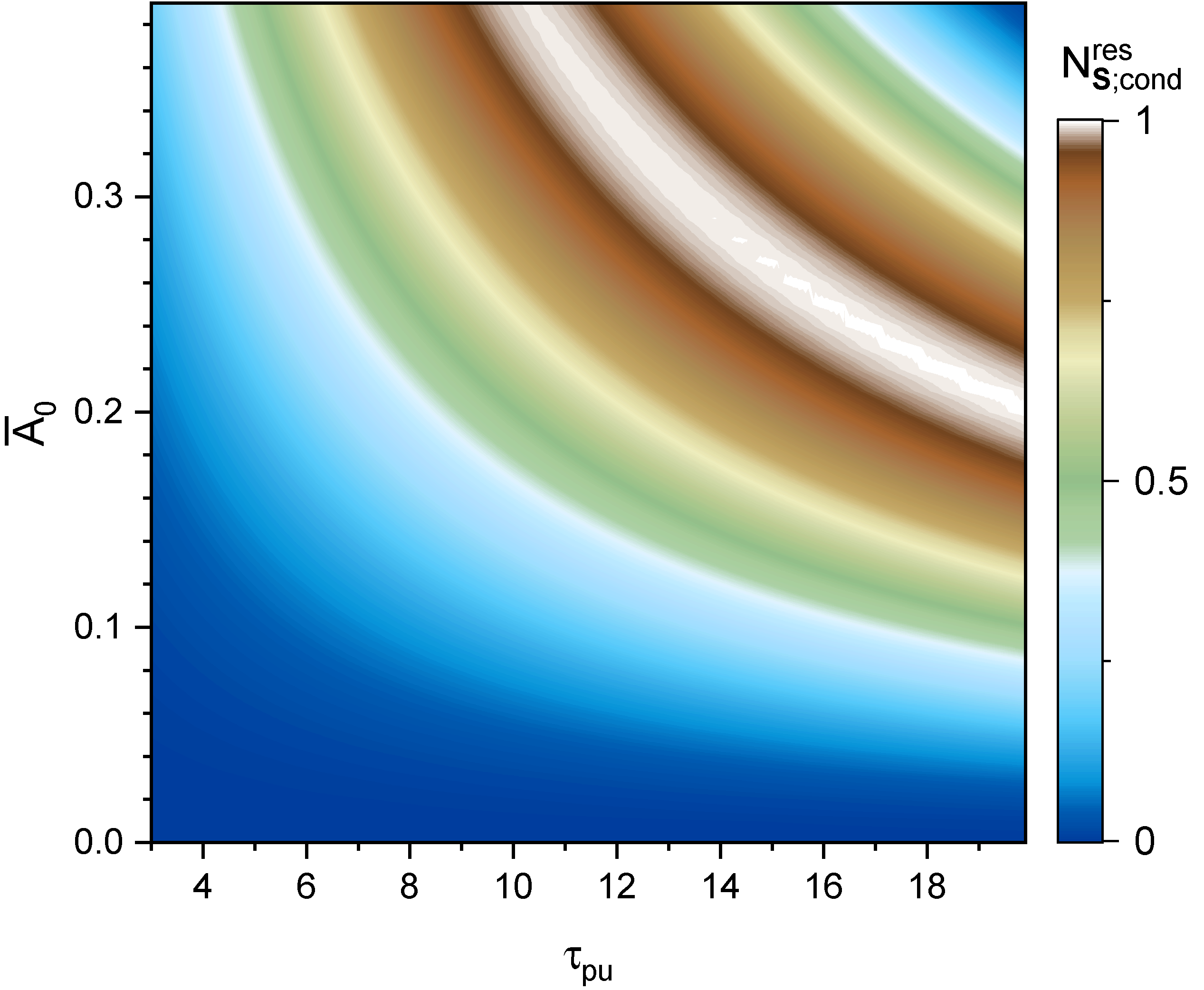}\tabularnewline
\end{tabular}
\par\end{centering}
\caption{Local dipole coupling case: (top) Residual excited electronic population
along the \emph{main} path as a function of the pump-pulse amplitude.
(bottom) Residual excited electronic population at \textbf{S} as a
function of the amplitude and of the FWHM of the pump pulse.\protect\label{fig:resN_dipole_only}}
\end{figure}

\subsection{Local dipole coupling (no Peierls substitution)\protect\label{subsec:Dipole}}

As first case, we consider a Hamiltonian in which the coupling to
the pumping field comes only through a local dipole moment, i.e.,
we neglect the Peierls substitution in the hopping term and in the
dipole one (in Eq.~\ref{eq:ham_dg} we set $\mathbf{k}+\frac{e}{\hbar}\boldsymbol{A}\left(t\right)\rightarrow\mathbf{k}$),
to focus only on the effects of such a coupling on the system and
analyze them in detail. This case is relevant to systems such as quantum
dots and molecules, and low-dimensional systems with transverse pumps.

In Fig.~\ref{fig:I-dipole-only}, we show the maps of TR-ARPES signals
along the \emph{main} path. The left panel shows the retarded signal,
$\bar{I}_{\mathbf{k}}^{R}\left(\bar{\omega},\bar{t}_{\mathrm{pr}}=0\right)$,
for the case where the center of the probe pulse coincides with the
center of the pump pulse. The valence and conduction bands are more
broad than at equilibrium (Fig.~\ref{fig:I_k-eq}) because the electrons
get excited to the conduction band and cannot be assigned to a specific
band anymore, inducing a quantum-mechanical uncertainty in the energy
of the bands themselves.

The photon-side-bands (PSBs) emerge at energies that differ from the
main-band energies of integer multiples of the (dressed) pump-pulse
photon energy. Some PSBs overlap in energy with the conduction and
the valence bands and, therefore, are not distinguishable in the map
of the retarded signal.

On top of the maps, we reported both the equilibrium band energies
(black solid curves) and the local maxima in energy of the signals
at each $\mathbf{k}$ (green dots), that indicate the (out-of-equilibrium)
bands of TR-ARPES. As the retarded signal shows, the equilibrium valence
and conduction bands coincide with TR-ARPES ones: a local dipole,
for realistic intensities, has negligible effects on the TR-ARPES
bands of the system.

Since in equilibrium only the valence band is occupied, the lesser
signal, $\bar{I}_{\mathbf{k}}^{<}\left(\bar{\omega},\bar{t}_{\mathrm{pr}}=0\right)$,
which is reported in the middle panel of Fig.~\ref{fig:I-dipole-only},
shows only the valence band and its corresponding PSBs. Wherever (in
$\mathbf{k}$ space) we have a one-photon resonance, the related resonant
one-photon PSB is definitely stronger than other PSBs as it coincides
with the conduction band in this case. The two-photon PSBs are some
orders of magnitude weaker than the one-photon ones and, in the scale
we have chosen for the maps, it is not possible to see them.

If we probe the system after the pump pulse is turned off, i.e., by
setting a large $\bar{t}_{\mathrm{pr}}\rightarrow+\infty$, but still
much shorter than the time scale of other decoherence and recombination
processes like spontaneous emission or electron-phonon interaction,
the spectrum of the system goes back to equilibrium, so that we have
$\bar{I}_{\mathbf{k}}^{R}\left(\bar{\omega},\bar{t}_{\mathrm{pr}}\rightarrow+\infty\right)=\bar{I}_{\mathbf{k}}^{R}\left(\bar{\omega},\bar{t}_{\mathrm{pr}}\rightarrow-\infty\right)$,
which is already shown in Fig.~\ref{fig:I_k-eq} and we do not repeat
here.

In Fig.~\ref{fig:I-dipole-only}, right panel, we report $\bar{I}_{\mathbf{k}}^{<}\left(\bar{\omega},\bar{t}_{\mathrm{pr}}\rightarrow+\infty\right)$:
contrarily to what happens for the retarded signal, the lesser signal
shows residual effects at the \textbf{k} points for which the pump-pulse
frequency is in one-photon resonance with the equilibrium gap energy.
The multi-photon PSBs do not show any residual signal even though
at $\bar{t}_{\mathrm{pr}}=0$ they are non-vanishing. This is because
we have only a local dipole in the interaction Hamiltonian and such
a term have no $\cos\left(l\bar{\omega}_{\mathrm{pu}}\bar{t}\right)$
term for $l>1$, hence no multi-photon Rabi-like resonances. Moreover,
the local dipole term we considered is completely off-diagonal, and
we have no oscillating diagonal term to result in multi-photon resonances
(see. App.~\ref{app:Oscillations-of-diag}). According to our experience,
this can be overcome having more than two bands in the system (not
shown).

In Fig.~\ref{fig:resN_dipole_only}, top panel, we plot the residual
excited electronic population in the conduction band, $N_{\mathbf{k},\mathrm{cond}}^{\mathrm{res}}$,
for the \textbf{k} points along the \emph{main} path as a function
of the pump-pulse amplitude. Rabi-like oscillations induce residual
excited populations at the \textbf{k} points for which a one-photon
resonance condition is realized. The finite width of the pump pulse
broadens the resonant energies so that, in addition to the exact resonances,
also the \textbf{k} points in the proximity of resonant ones have
some residual excited population (compare with Fig.~\ref{fig:A2_gaps}).

In Fig.~\ref{fig:resN_dipole_only}, bottom panel, we plot the residual
excited population in the conduction band at \textbf{S} as a function
of the amplitude and of the FWHM of the pump pulse. Being (i) the
Rabi frequency, $\omega_{R}$, proportional to the pump-pulse amplitude
and (ii) the overall oscillation time roughly proportional to the
FWHM of the pump pulse, the residual excited population is almost
constant wherever $\bar{A}_{0}\bar{\tau}_{\mathrm{pu}}$ is constant,
that yields the hyperbolic shape of the color contours in the figure.
For the very same reason, on both cuts at fixed $\bar{A}_{0}$ and
at fixed $\bar{\tau}_{\mathrm{pu}}$, one clearly sees the \emph{signature}
of the Rabi-like oscillations. For instance, at fixed $\bar{A}_{0}$,
that is at fixed $\omega_{R}$, the end tail (in time) of the pump-pulse
envelope determines the residual excited population and on changing
$\bar{\tau}_{\mathrm{pu}}$ one can \emph{scan} the Rabi-like oscillating
behavior of the population (roughly $N_{\mathbf{k};\mathrm{cond}}^{\mathrm{res}}\propto\sin^{2}\left(\omega_{R}\tau_{\mathrm{pu}}\right)$).
It is worth reminding that, for small $\omega_{R}\tau_{\mathrm{pu}}$,
which is usually the more relevant case in ultrafast experiments,
one can approximate $\sin\left(\omega_{R}\tau_{\mathrm{pu}}\right)\simeq\omega_{R}\tau_{\mathrm{pu}}$,
which results in $N_{\mathbf{k};\mathrm{cond}}^{\mathrm{res}}\propto\bar{A}_{0}^{2}\bar{\tau}_{\mathrm{pu}}^{2}$.

\begin{figure*}
\centering{}%
\begin{tabular}{cc}
\includegraphics[width=7.5cm]{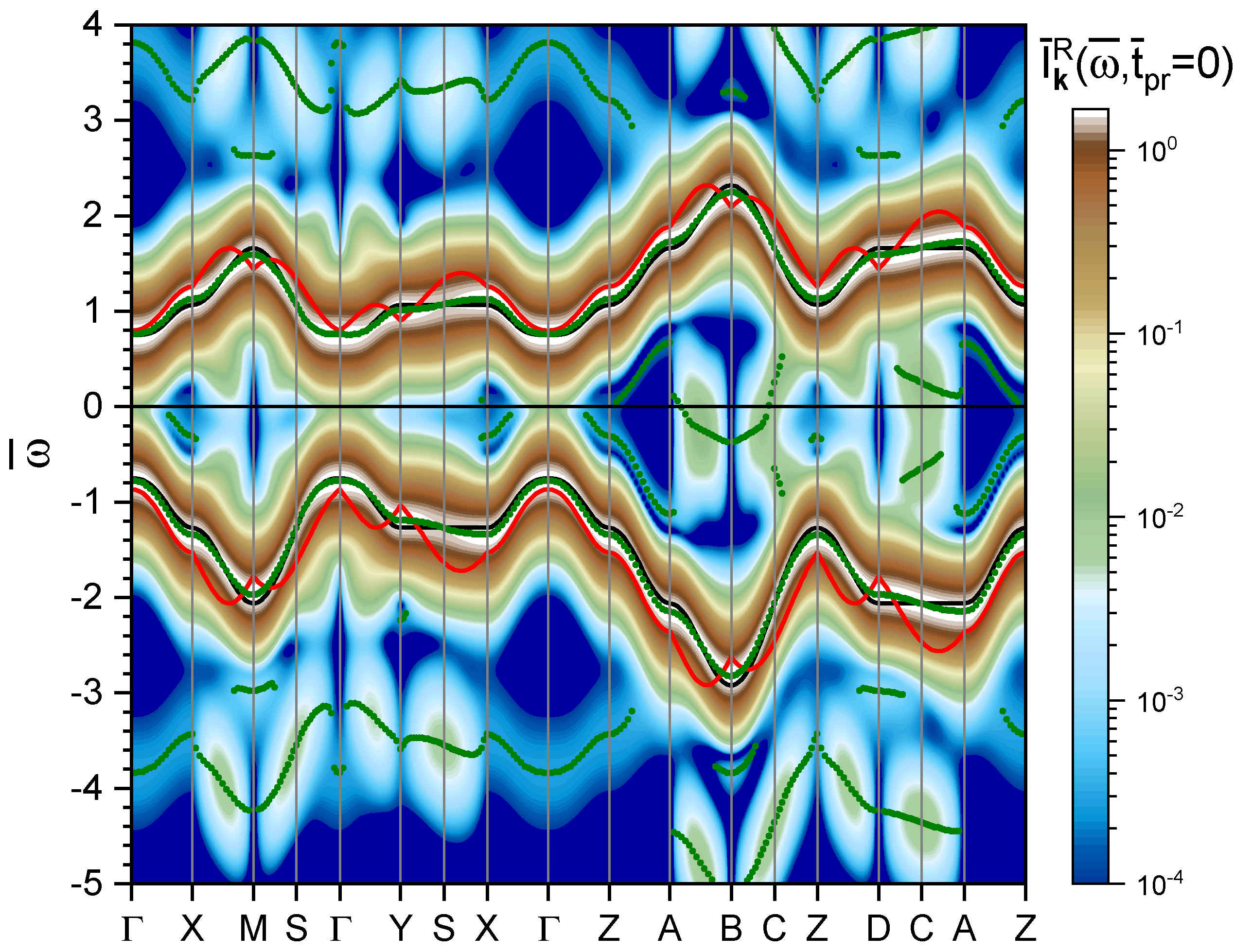} & \includegraphics[width=7.5cm]{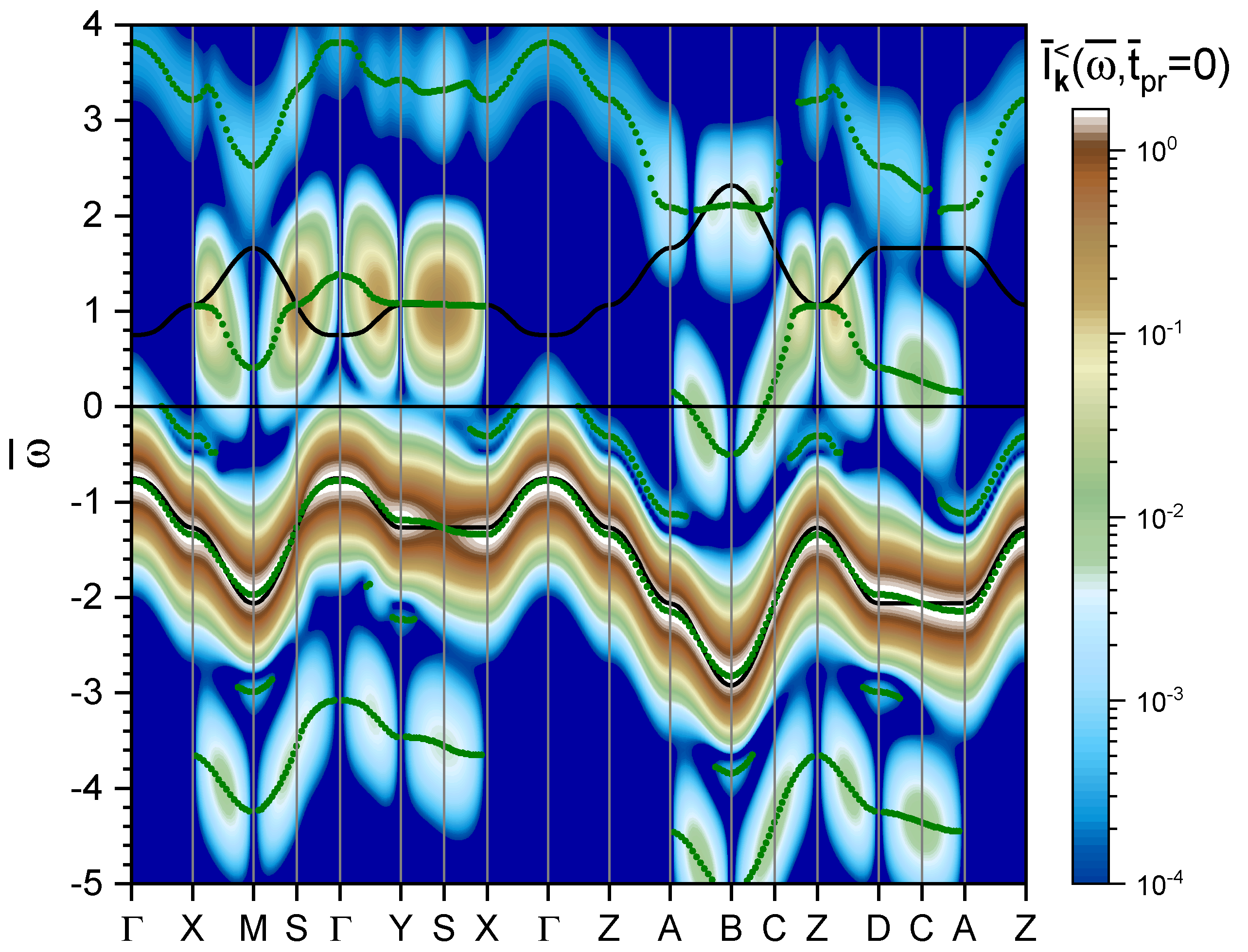}\tabularnewline
\includegraphics[width=7.5cm]{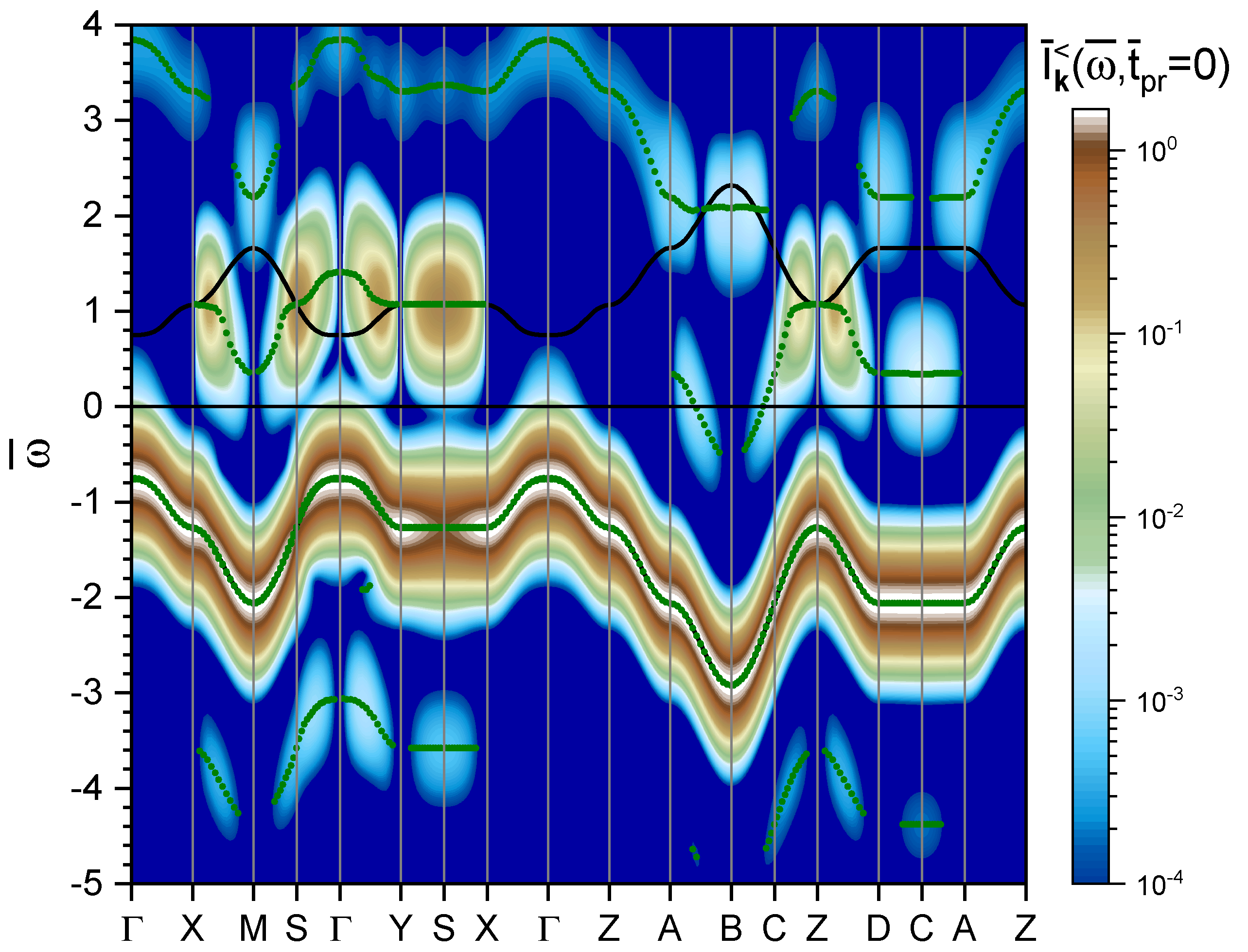} & \includegraphics[width=7.5cm]{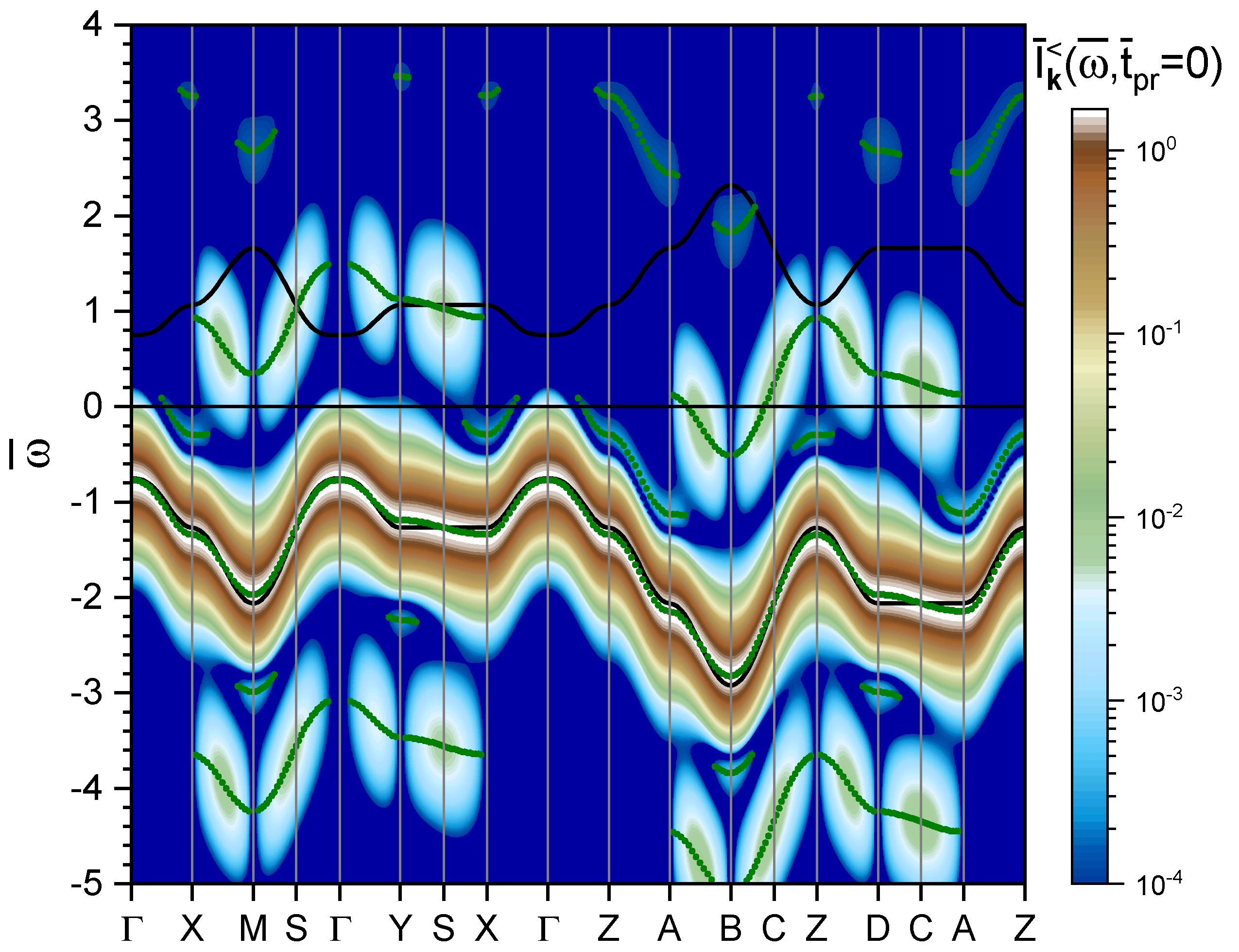}\tabularnewline
\end{tabular}\caption{TR-ARPES signals along the \emph{main} path for Peierls substitution
in the hopping term and no dipole, with the center of probe pulse
coinciding with the one of pump pulse ($\bar{t}_{\mathrm{pr}}=0$
). (top-left) $\bar{I}_{\mathbf{k}}^{R}\left(\bar{\omega},\bar{t}_{\mathrm{pr}}=0\right)$
and (top-right) $\bar{I}_{\mathbf{k}}^{<}\left(\bar{\omega},\bar{t}_{\mathrm{pr}}=0\right)$
for the full dynamics, (bottom-left) $\bar{I}_{\mathbf{k}}^{<}\left(\bar{\omega},\bar{t}_{\mathrm{pr}}=0\right)$
for the dynamics with inter-band-only transitions, and (bottom-right)
$\bar{I}_{\mathbf{k}}^{<}\left(\bar{\omega},\bar{t}_{\mathrm{pr}}=0\right)$
for the dynamics with intra-band-only transitions. The black-solid
curves show the equilibrium band energies while the green dots indicate
the local maxima of the signals for each \textbf{k} point, which are
the (out-of-equilibrium) bands of TR-ARPES. The red-solid curves show
the instantaneous eigenenergies at time zero.\protect\label{fig:I-t0-NoD}}
\end{figure*}

\begin{figure}
\begin{centering}
\begin{tabular}{c}
\includegraphics[width=7.5cm]{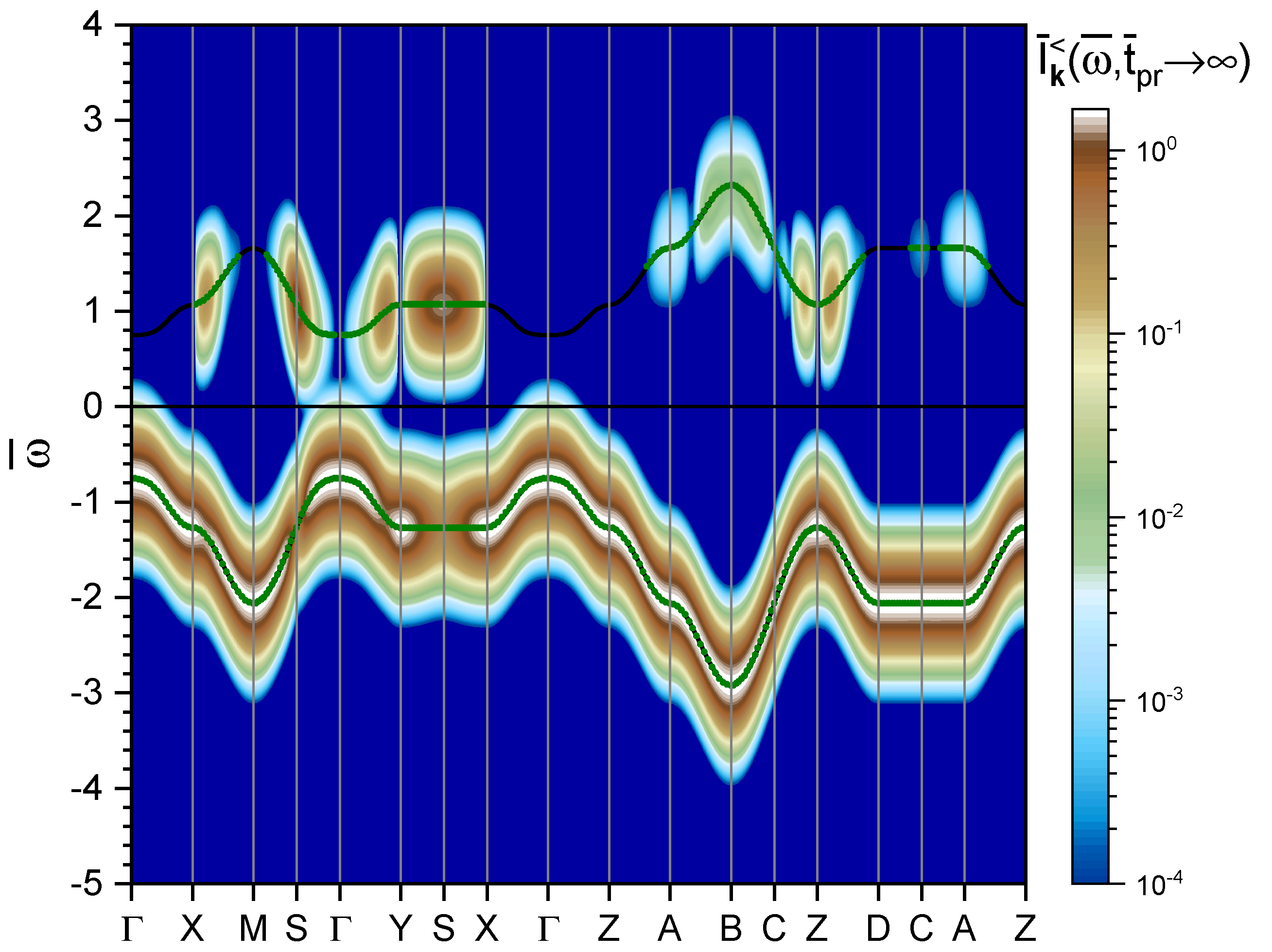}\tabularnewline
\includegraphics[width=7.5cm]{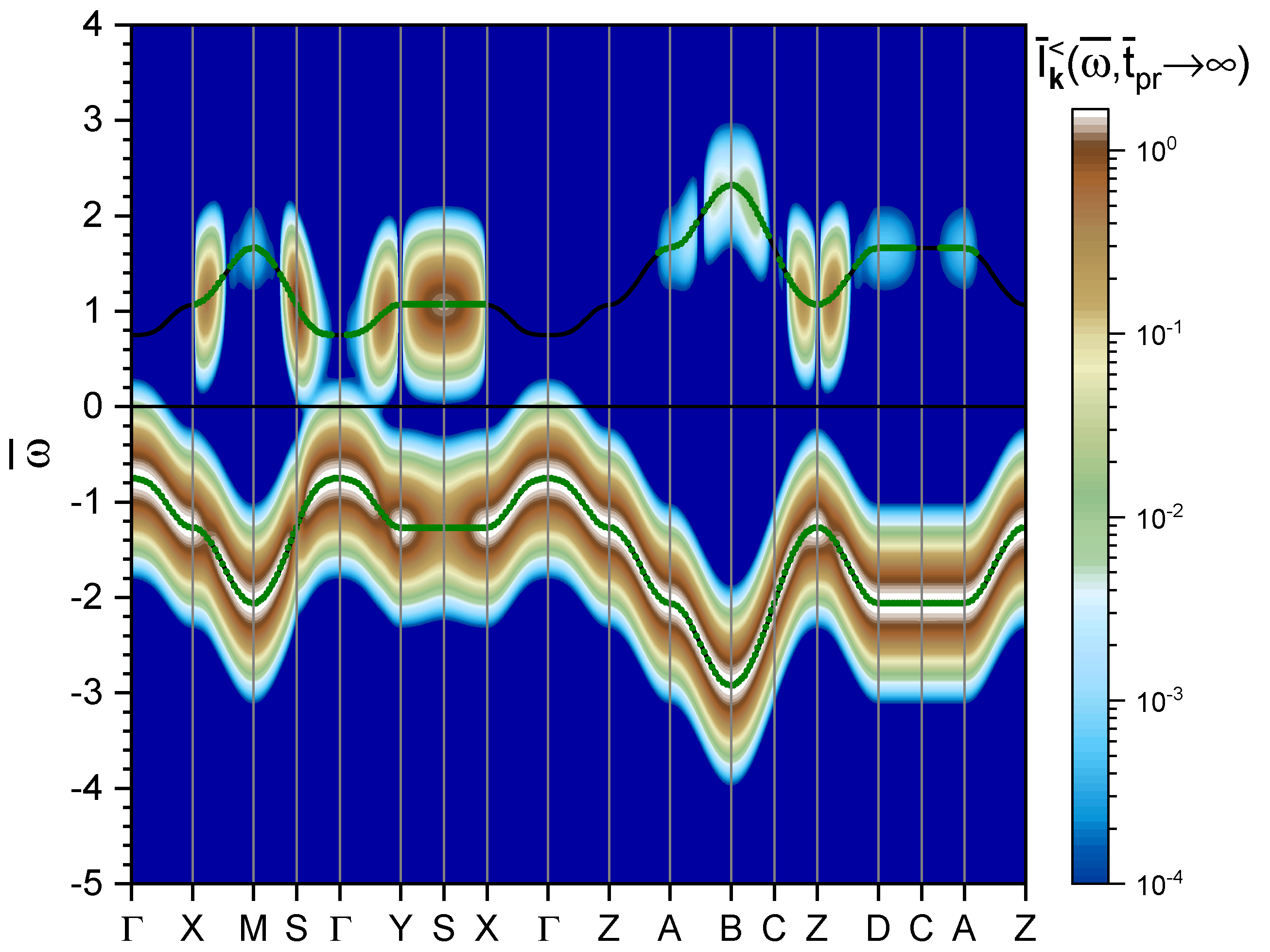}\tabularnewline
\end{tabular}
\par\end{centering}
\caption{Residual lesser TR-ARPES signal along the \emph{main} path for only
Peierls substitution with zero dipole, considering (top) full dynamics,
and (bottom) inter-band-only dynamics. The black-solid curves show
the equilibrium band energies while the green dots show the local
maximum of the signals for a fixed \textbf{k} point.\protect\label{fig:I-res-NoD}}
\end{figure}

\begin{figure}
\begin{centering}
\begin{tabular}{c}
\includegraphics[width=7.5cm]{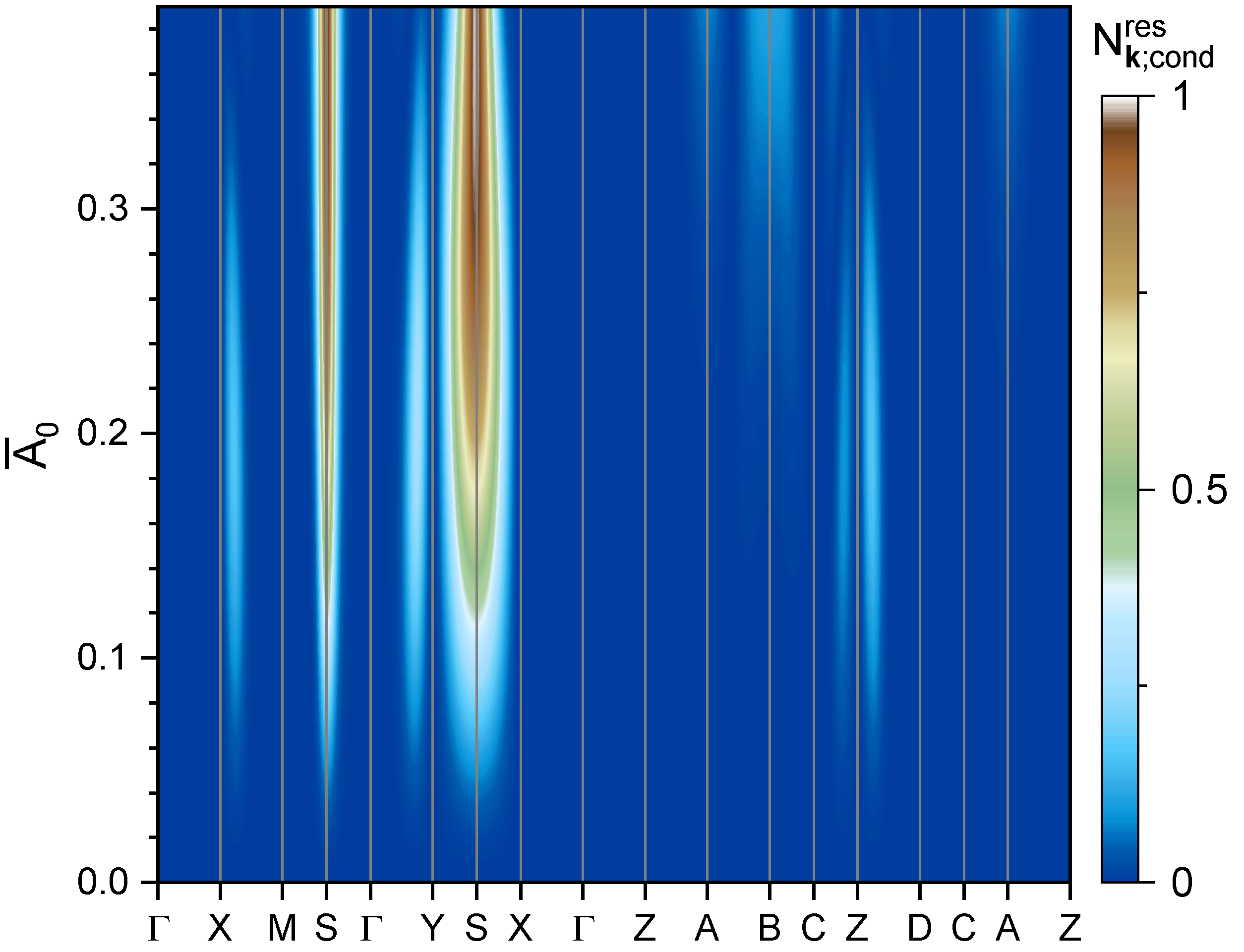}\tabularnewline
\includegraphics[width=7.5cm]{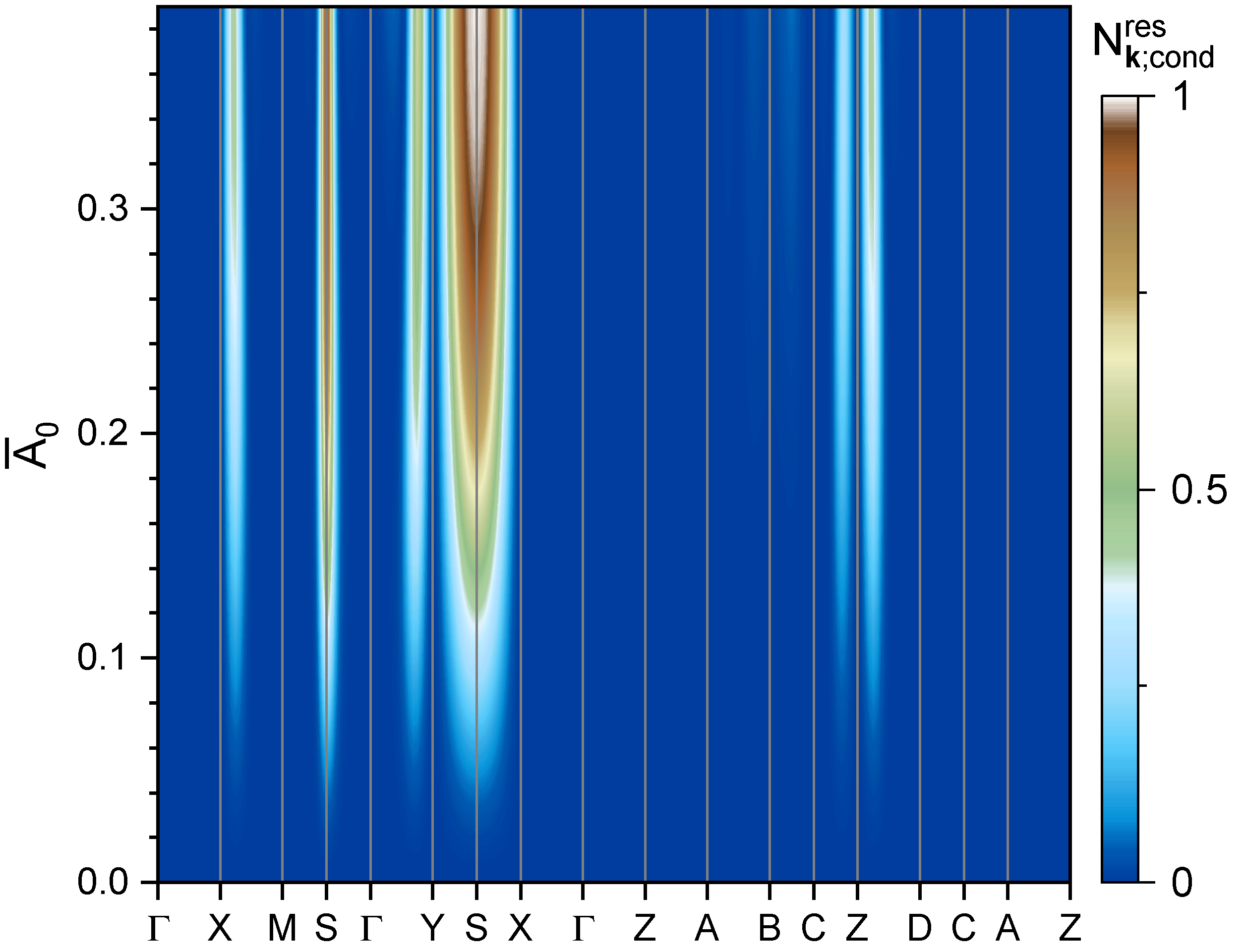}\tabularnewline
\end{tabular}
\par\end{centering}
\caption{Residual excitations along the \emph{main} path as a function of the
pump-pulse amplitude for the case of Peierls substitution in the hopping
term and zero dipole. (top) The full dynamics. (bottom) The dynamics
considering inter-band transitions only.\protect\label{fig:resN_NoD}}
\end{figure}

\begin{figure}
\begin{centering}
\begin{tabular}{c}
\includegraphics[width=7.5cm]{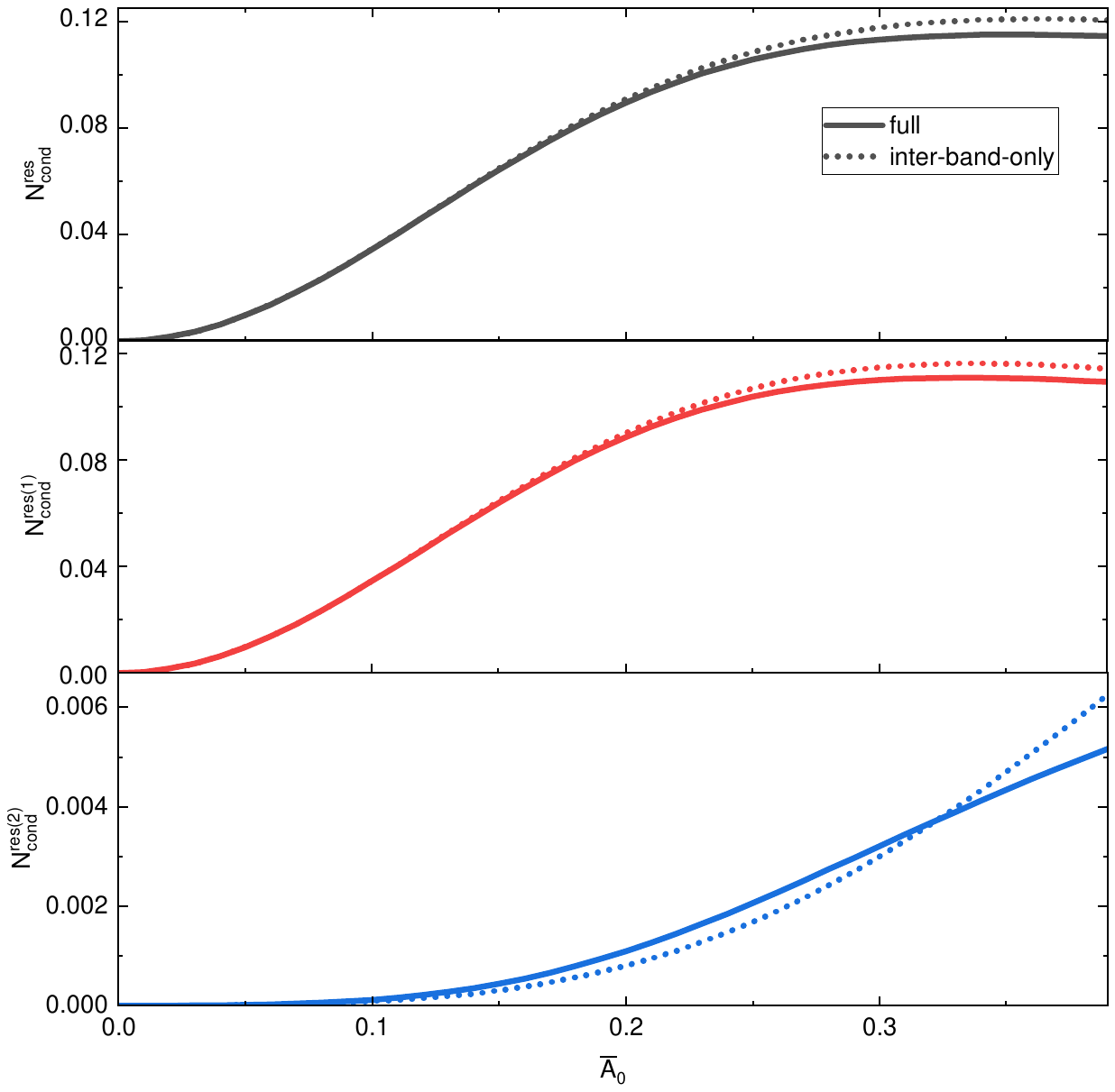}\tabularnewline
\end{tabular}
\par\end{centering}
\caption{Residual excited population per unit cell as a function of the pump-pulse
amplitude for an $8\times8\times8$ \textbf{k} grid over the first
Brillouin zone, considering Peierls substitution in the hopping term
and no dipole. (top) all (any-photon) resonant contributions, (middle)
only one-photon resonant contributions and (bottom) only two-photon
resonant contributions. In all panels, we report the two cases of
the full original Hamiltonian (solid) and inter-band-only Hamiltonians
(dots).\protect\label{fig:resN_cell_NoD}}
\end{figure}

\subsection{Peierls substitution in hopping (no dipole)\protect\label{subsec:Peierls}}

In this case, we consider an interaction with the pump pulse via the
Peierls substitution in the hopping term and set the dipole to zero.
This is very relevant as the dipole term is often negligible in many
realistic cases. Moreover, neglecting the dipole we can focus on the
effects of band symmetries on TR-ARPES signal and electronic excitations
and analyze them in detail.

\begin{figure}
\begin{centering}
\begin{tabular}{c}
\includegraphics[width=7.5cm]{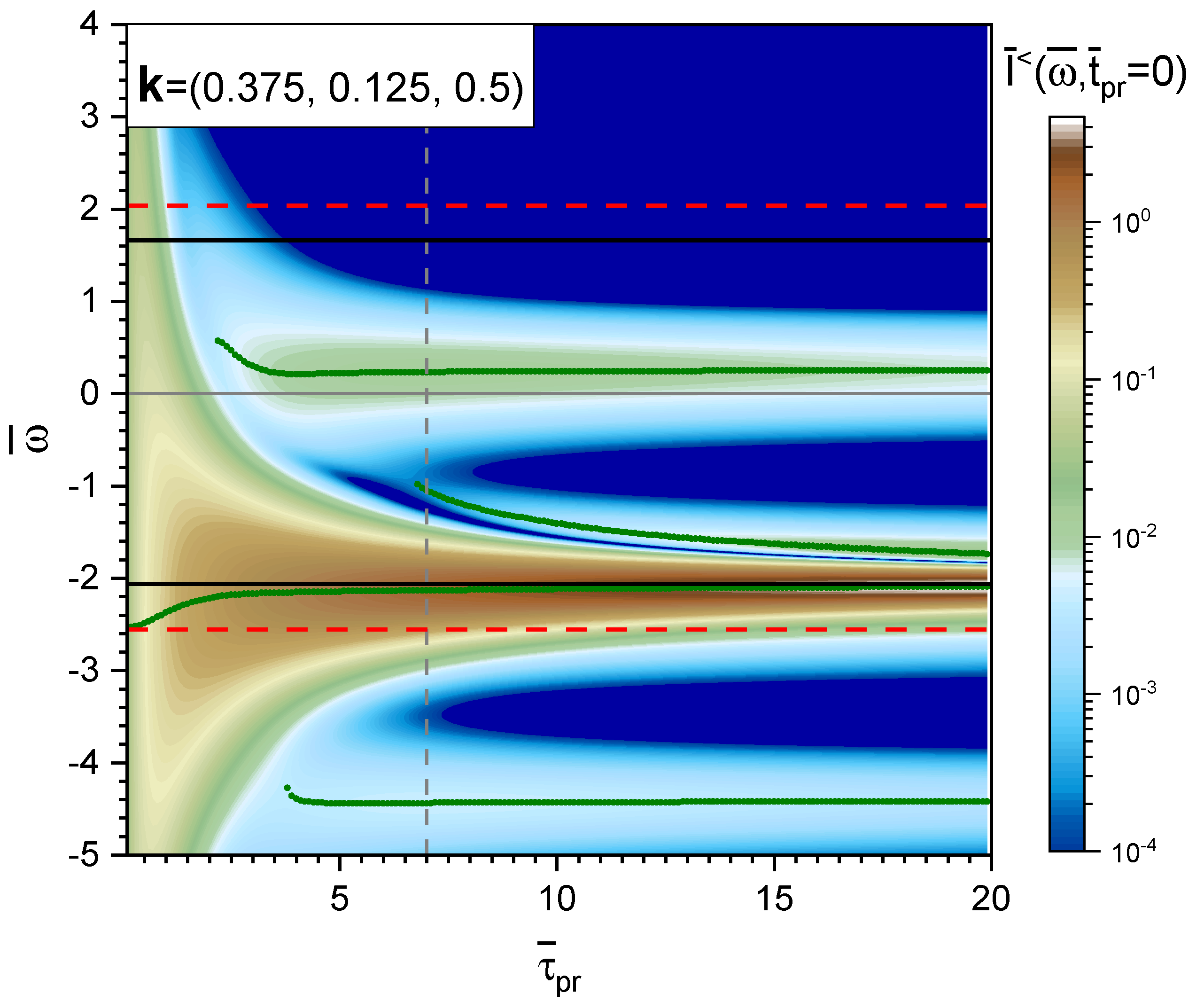}\tabularnewline
\includegraphics[width=7.5cm]{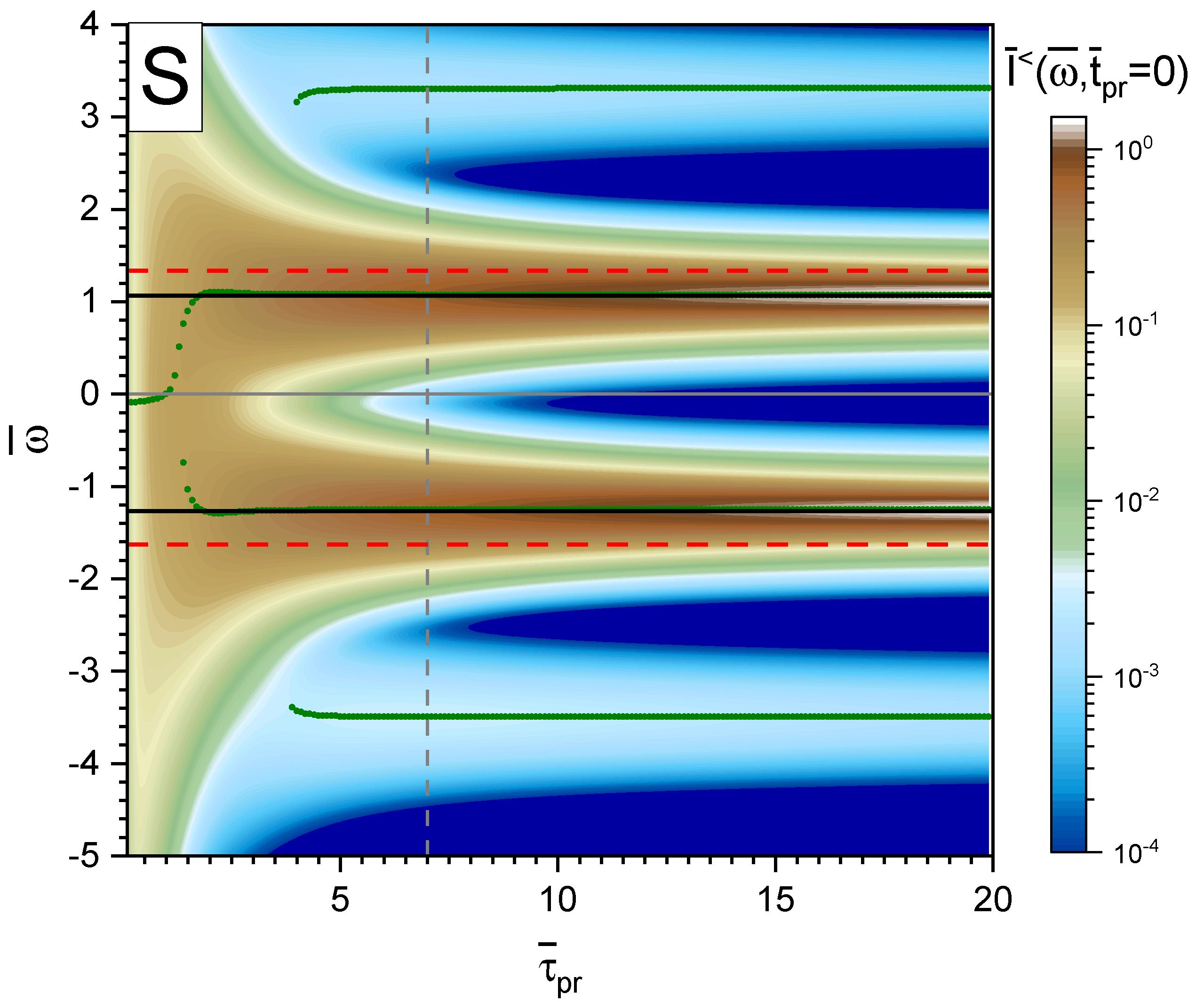}\tabularnewline
\end{tabular}
\par\end{centering}
\caption{The lesser TR-ARPES signals vs the FWHM of probe-pulse envelope, $\bar{\tau}_{\mathrm{pr}}$,
while the center of probe pulse is kept at $\bar{t}_{\mathrm{pr}}=0$,
for (top) the \textbf{k} point $\mathbf{k}=\left(0.375,0.125,0.5\right)$
and (bottom) the high-symmetry point \textbf{S}. The vertical dashed
gray lines determine $\bar{\tau}_{\mathrm{pu}}$. The black-solid
lines show the equilibrium band energies while the green dots show
the local maximum of the signals for a fixed \textbf{k} point. The
red-solid lines show the instantaneous eigenenergies at time zero.
\protect\label{fig:IL-vs-taupr_PD}}
\end{figure}

In Fig.~\ref{fig:I-t0-NoD} top-left (top-right), we show the map
of $\bar{I}_{\mathbf{k}}^{R}\left(\bar{\omega},\bar{t}_{\mathrm{pr}}=0\right)$
($\bar{I}_{\mathbf{k}}^{<}\left(\bar{\omega},\bar{t}_{\mathrm{pr}}=0\right)$).
The higher local maxima of the TR-ARPES signal, that one can consider
the main TR-ARPES bands, are slightly shifted with respect to the
equilibrium valence and conduction bands and show almost no correspondence
to the instantaneous eigenenergies at time zero. This is expected
since TR-ARPES measures the system over a time period and not at a
specific instant of time. We will shed more light on this issue later
on (Fig.~\ref{fig:IL-vs-taupr_PD} and related discussion). For the
retarded signal, which shows the full TR-ARPES spectrum of the system,
we can see both valence and conduction bands and all of their sidebands.
Because of the finite broadening of the bands, they overlap and distinguishing
them in the case of retarded signal can be very difficult. Obviously,
in the lesser signal, we see only the valence band and its sidebands.

The one-photon PSBs originate from the velocity term in the \emph{Peierls
expansion}, which is proportional to $\sin\left(ak_{y}\right)$ and,
therefore, identically vanishes on the planes $\boldsymbol{\Gamma}$-\textbf{X}-\textbf{A}-\textbf{Z}
and \textbf{Y}-\textbf{M}-\textbf{B}-\textbf{D}, yielding no one-photon
PSB there. Instead, on \textbf{S}, \textbf{C}, middle points of the
lines \textbf{X}-\textbf{M}, \textbf{A}-\textbf{B}, \textbf{Z}-\textbf{D}
and $\boldsymbol{\Gamma}$-\textbf{Y}, the second order (inverse-mass)
term -- as well as all other even terms -- of the \emph{Peierls
expansion} vanishes as it is proportional to $\cos\left(ak_{y}\right)$.
Recall that the polarization of the pump-pulse has been chosen along
the $y$ direction.

However, even though at some of these points the two-photon PSBs are
very weak, at some others (such as \textbf{S}), where we have a strong
one-photon PSB, the two-photon PSB is also strong, which shows that
the second order signal is assisted by multiple actions of the first
order terms of the Hamiltonian (as in the case of Floquet staircase).
At the \textbf{k} points where the band gap is at either one- or two-photon
resonance with the pump pulse, the corresponding PSB is much stronger
than non resonant ones, provided that it is not zero by symmetry.

At the \textbf{k} points where the inverse-mass vanishes, we have
practically no shift of the TR-ARPES bands with respect to the equilibrium
ones. The shift in the bands is mainly due to the non-oscillating
components that appear in the even order terms of \emph{Peierls expansion}
($\sum_{m=1}^{\infty}\Theta_{0,m}\left(t\right)$ in Eq.~\ref{eq:Peierls_exp_L}),
which identically vanish when the inverse-mass term vanishes by symmetry.
Moreover, the higher order effects of the same term result in some
weak side-bands near the main bands as it is more clear in the map
of $\bar{I}_{\mathbf{k}}^{<}\left(\bar{\omega},\bar{t}_{\mathrm{pr}}=0\right)$
(top-right panel). It is worth noting that if we had an infinitely
oscillating pump pulse without an envelope (that is, a pump-pulse
FWHM extremely longer than the probe-pulse FWHM), there would have
been no higher order effects and the non-oscillating components would
have just resulted in rigid shifts of the bands. Therefore, we dub
these new side-bands as envelope-Peierls side-bands (EPSBs): they
are due to both the envelope and the even terms of \emph{Peierls expansion}.\textbf{
}However, according to our experience, EPSBs can also result from
the odd terms of the \emph{Peierls expansion} in multi-band systems.

In Fig.~\ref{fig:I-t0-NoD}, bottom-left (bottom-right) panel, we
show the map of $\bar{I}_{\mathbf{k}}^{<}\left(\bar{\omega},\bar{t}_{\mathrm{pr}}=0\right)$
for the dynamics with inter-band-only (intra-band-only) transitions.
Interestingly, the main TR-ARPES bands are practically on top of the
equilibrium ones in the case of inter-band-only dynamics. On the other
hand, for the case of intra-band-only dynamics, we see the same shift
as for the full dynamics. This is consistent with the inter-band transitions
governing the electronic transitions between the bands and not altering
the bands noticeably, while the intra-band transitions change the
band energies dynamically. As we already mentioned above, the shift
in the main bands have the same origin as the EPSBs and since we do
not have band shifts for inter-band-only transitions, the EPSBs disappear
as well.

PSBs have different behaviors depending on being one-photon or two-photon,
and in resonance or off resonance. The resonant PSBs are much stronger
in the inter-band-only case (bottom-left panel) than in the intra-band-only
case (bottom-right panel), because in order to differentiate between
in resonance and off resonance, one needs the inter-band transitions.
On the contrary, the off-resonant one-photon PSBs are stronger in
the intra-band-only case than in the inter-band-only one, which shows
that for the system parameters that we have chosen, out of resonance,
the inter-band transitions have very negligible effects on the system,
while intra-band transitions obviously still induce one-photon PSBs.
In fact, in the intra-band-only case, our system is equivalent to
a single-band (the valence band) Floquet one as the conduction band
is obviously empty and not coupled to the valence band. However, in
our system, in the inter-band-only case (bottom-left panel), two-photon
off-resonance PSBs can be noticeable in comparison to the case of
full dynamics (top-right panel).

The resonant one-photon (two-photon) PSBs are stronger (weaker) in
the case of inter-band-only dynamics than in the case of the full
dynamics. This can be understood by noticing that removing intra-band
transitions pins down the electrons at one-photon resonant \textbf{k}
points and helps them to get more and more excited, while lack of
the assistance provided by the intra-band motions to the inter-band
transitions, reduces the two-photon resonant PSBs. In App.~\ref{app:Oscillations-of-diag},
it is shown that oscillating diagonal terms in a Hamiltonian, which
are the origin of intra-band transitions, yield another mechanism
for multi-photon resonant transitions.

Another important property to be studied is the residual signal of
TR-ARPES. As we already mentioned, after the action of pump pulse,
the spectrum which is given by the retarded signal is exactly the
one of equilibrium (Fig.~\ref{fig:I_k-eq}), while the lesser signal
is different. As shown in Fig.~\ref{fig:I-res-NoD} top panel, where
we plot $\bar{I}_{\mathbf{k}}^{<}\left(\bar{\omega},\bar{t}_{\mathrm{pr}}\rightarrow\infty\right)$,
at the one- or two-photon resonant \textbf{k} points we have the corresponding
residual signals at PSBs, unless the PSB is prohibited by symmetry.
For instance, this condition realizes for one-photon PSBs at \textbf{X},
\textbf{Y}, and \textbf{Z}, where we have exact one-photon resonances,
and for two-photon PSBs at the middle of \textbf{A}-\textbf{B} and
at \textbf{C}, where we have non-exact two-photon resonances.

Fig.~\ref{fig:I-res-NoD} bottom panel, shows the residual lesser
signal for the dynamics given by inter-band-only transitions. The
one-photon (two-photon) residual PSBs are stronger (weaker) for the
inter-band-only dynamics than for the full one according to the very
same reasoning reported above. It is noteworthy that even though the
intra-band-only transitions induce PSBs within the pump-pulse envelope
(see Fig.~\ref{fig:I-t0-NoD}, bottom-right panel), they yield no
residual in the TR-ARPES signal, which return to equilibrium after
the pump pulse is turned off (Fig.~\ref{fig:I_k-eq}).

In Fig.~\ref{fig:resN_NoD}, top panel, we plot the residual excited
populations along the \emph{main} path as a function of the pump-pulse
amplitude. One- and two-photon resonances have residual excited populations
that show Rabi-like oscillations with respect to changing the pump-pulse
amplitude. Moreover, unlike in the former case (only local dipole),
\textbf{k} points with the same gap energies (for example those along
the path \textbf{Y}-\textbf{S}-\textbf{X}) have different behaviors
as their velocities and inverse-masses are different, which yields
different couplings to the pumping field. In particular, at the $\mathbf{k}$
points where we have one-/two-photon resonances, but the velocity/inverse
mass vanishes (see the related discussion above on the TR-ARPES signal
regarding the relevant regions of first Brillouin zone), there are
no residual excitations. However, the points at their immediate proximities
with non-exact resonant gaps, but non-zero velocities/inverse masses,
host some residual excited populations.

In Fig.~\ref{fig:resN_NoD}, bottom panel, we see that for the case
of a dynamics with just the inter-band transitions, the residual excited
population coming from one-photon resonances gets larger (on the contrary,
if one keeps only the intra-band transitions there would be no excitations
at all). In this case, one-photon resonances get stronger because
we have removed the intra-band transitions that drive them transiently
out of resonance and leads to a smaller residual excited population.
However, there are some weak one-photon resonances that benefit from
intra-band transitions since they are far from the exact resonant
points and by the intra-band transitions they can get transiently
closer to resonance. The net effect for them is to gain some residual
excited population so that the related resonant region in $\mathbf{k}$-space
appears wider compared to the case of inter-band-only transitions.
An example can be the proximities of the \textbf{S} point on the path
\textbf{M}-\textbf{S}-$\boldsymbol{\Gamma}$. On the other hand, for
the two-photon resonances considering only the inter-band transitions
results in smaller residual excited populations. The two-photon resonances
are assisted by the intra-band transitions, similar to TR-ARPES signal
(see above).

In Fig.~\ref{fig:resN_cell_NoD}, top panel, we plot the total residual
excited population per unit cell, Eq.~\ref{eq:N_}, as a function
of the pump-pulse amplitude. We have considered an $8\times8\times8$
k-grid to sample the first Brillouin zone, even though we checked
the robustness of the results with respect to the size of the grid
by using also $16\times16\times16$ and $32\times32\times32$ k-grids
for a larger step in the pump-pulse amplitude (not shown). We compare
the two cases: the full Hamiltonian and the one with inter-band transitions
only. For all values of the pump-pulse amplitude, the total residual
excited population with only inter-band transitions is larger. The
middle (bottom) panel of Fig.~\ref{fig:resN_cell_NoD}, shows the
contribution from one-photon (two-photon) resonances, i.e., Eq.~\ref{eq:N_l}
with $l=1$ ($l=2$). In our system, the largest contribution comes
from the one-photon resonances (middle panel). Computing the relative
multi-photon resonance strengths (see Eq.~\ref{eq:eff_res_l}) in
our grid, we find out that the relative strength of one-photon ($l=1$)
resonances is 64\%, while for two-photon ($l=2$) resonances is 36\%
($W_{l}/\left(W_{1}+W_{2}\right)$, see Eq.~\ref{eq:eff_res_l}).
Clearly, these numbers do not take into account the actual strength
of the system-pump couplings at these resonant $\mathbf{k}$ points
and that the second order transitions are generally weaker than the
first order ones.

\begin{figure*}
\centering{}%
\begin{tabular}{ccc}
\includegraphics[width=6cm]{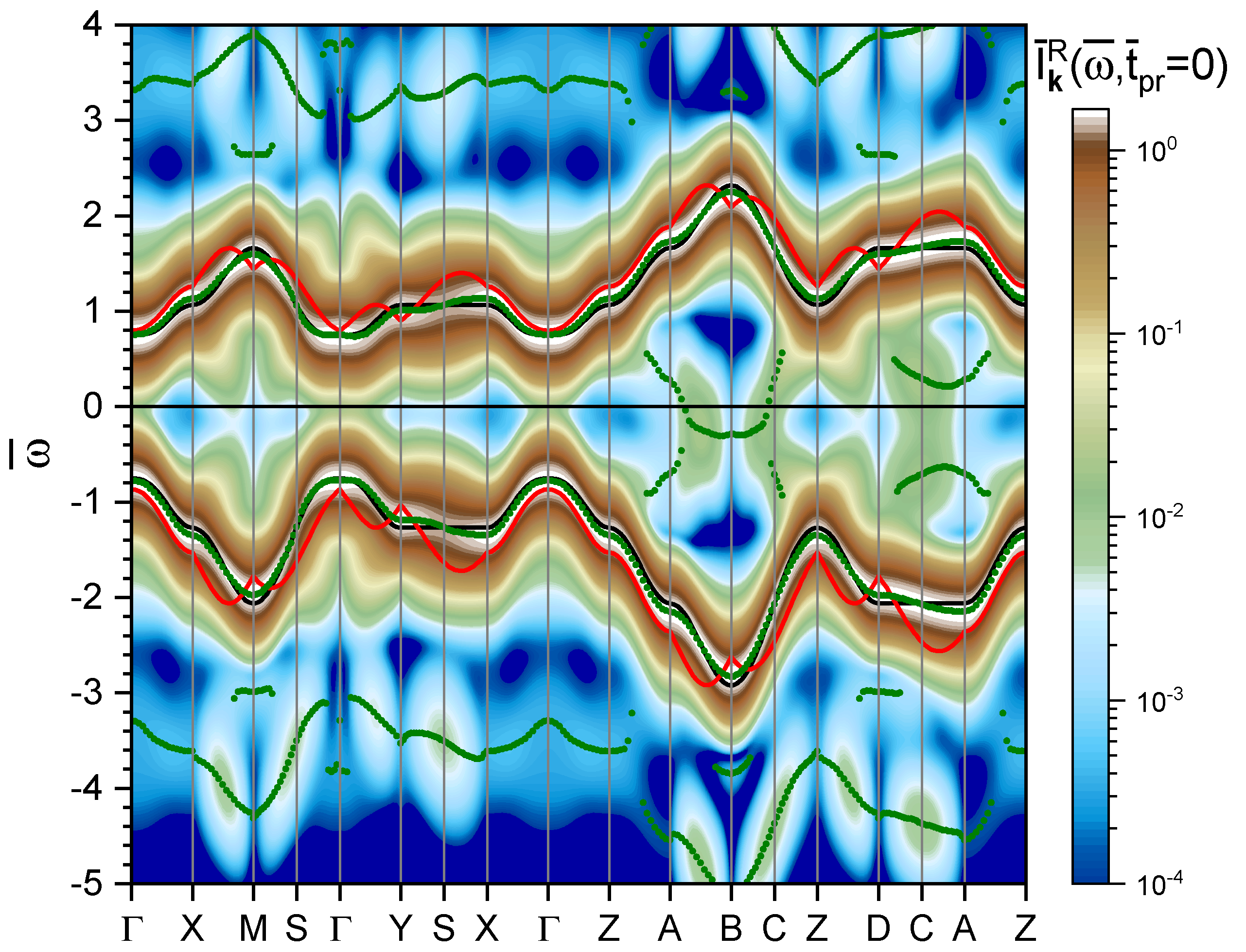} & \includegraphics[width=6cm]{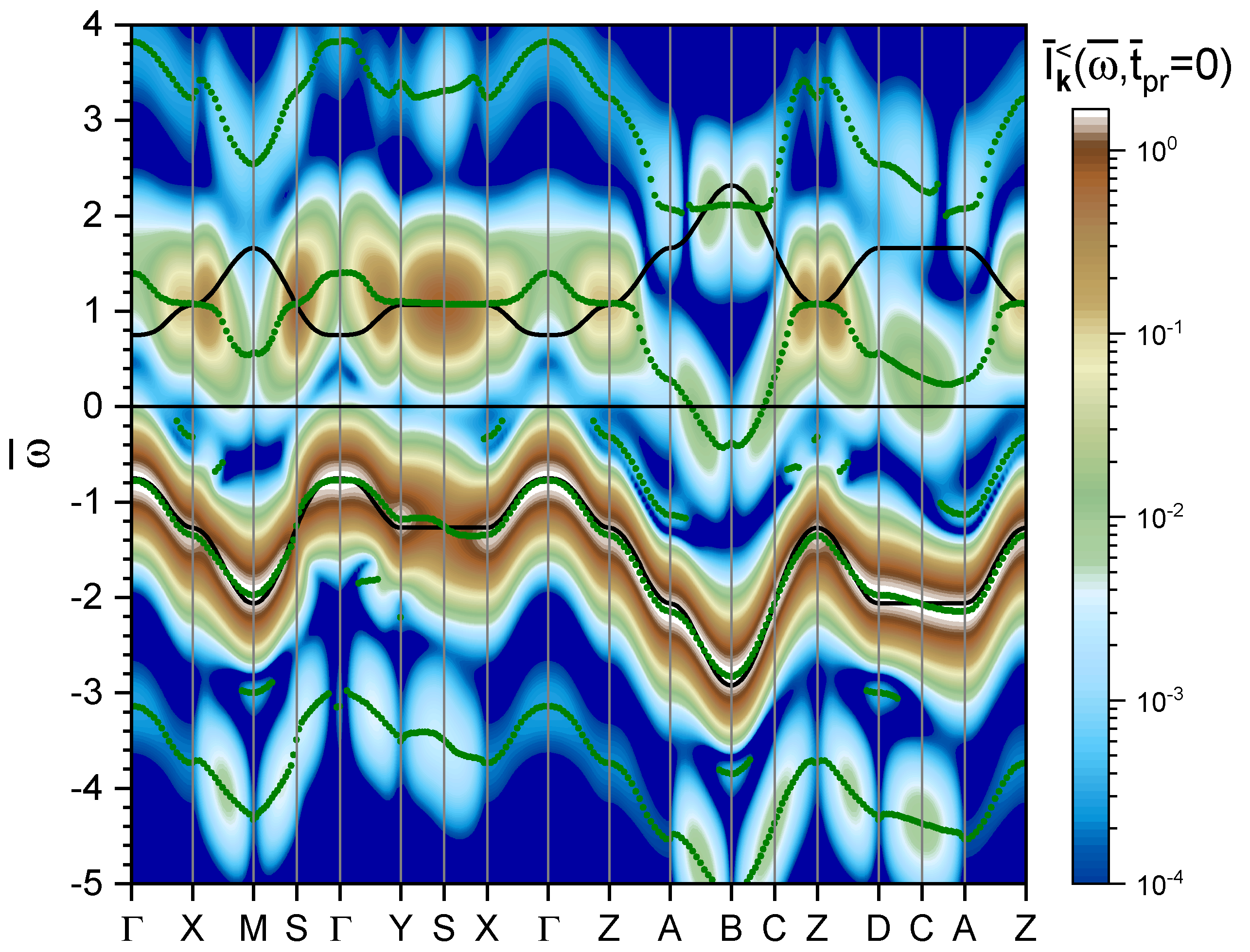} & \includegraphics[width=6cm]{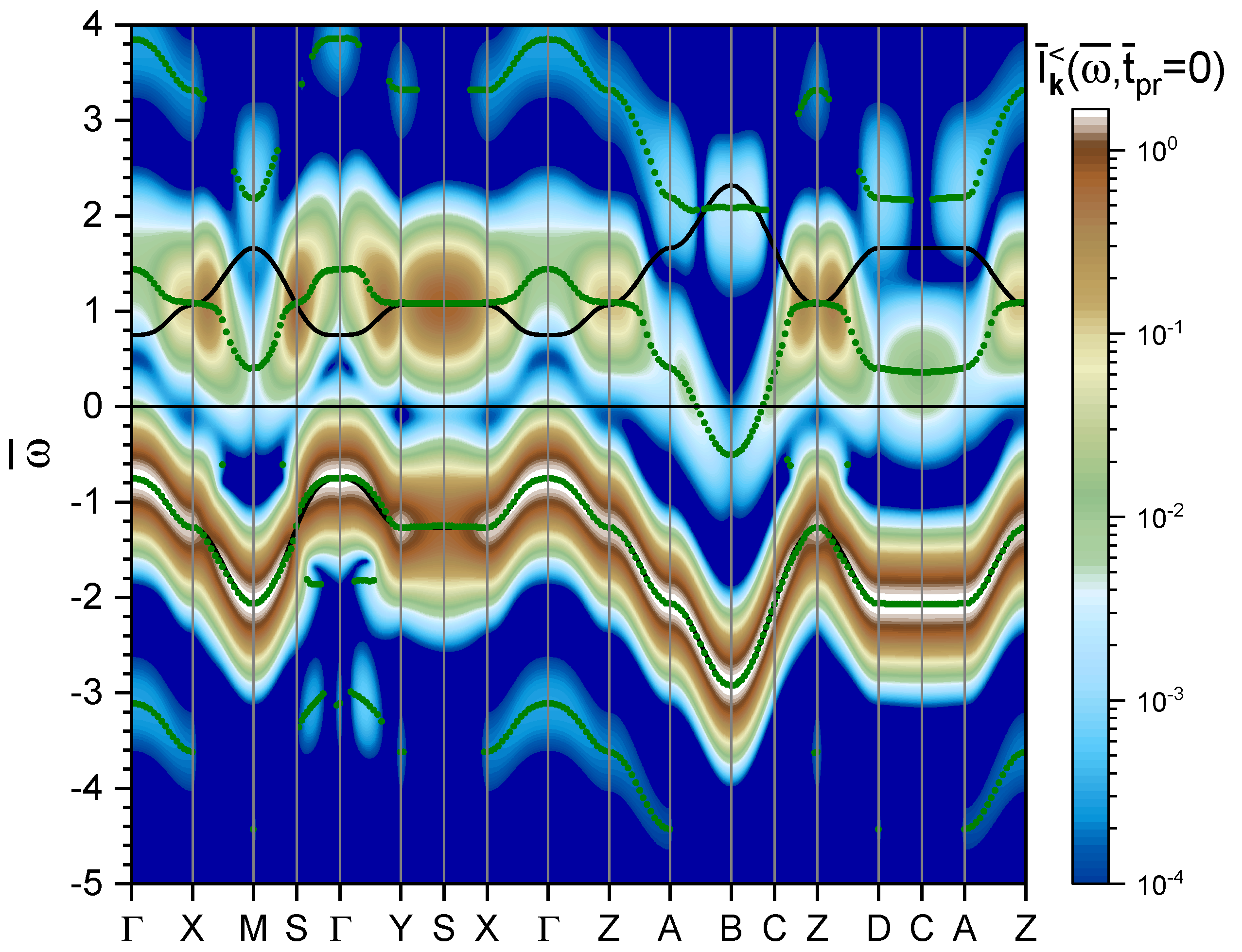}\tabularnewline
\end{tabular}\caption{The TR-ARPES signal considering both Peierls substitution and local
dipole in the Hamiltonian, with the probe and pump pulses having the
same center, i.e., $\bar{t}_{\mathrm{pr}}=0$. (right) The retarded
signal, (middle) the lesser signal, and (left) the lesser signal for
inter-band-only transitions. The black-solid curves show the equilibrium
band energies while the green dots show the local maxima of the signals
for a fixed \textbf{k} point. The red-solid curves show the instantaneous
eigenenergies at time zero. \protect\label{fig:I-t0-PD}}
\end{figure*}

As we already explained above in detail, for the one-photon resonances
(middle panel), the removal of intra-band transitions increases the
residual excited populations, while for the two-photon resonances
(bottom panel), the residual excited populations get reduced by removal
of the intra-band transitions. In the latter case, increasing the
pump-pulse amplitude to high values, the behavior changes and the
results of inter-band-only Hamiltonian overcome the full Hamiltonian
ones. This can be understood by noting that, upon removing intra-band
transitions, the Rabi-like oscillations become in average slower over
all of the two-photon resonant \textbf{k} points and of the sin-like
shape we see only the monotonously increasing behavior that eventually
manages to overcome the usual bending-over sin-like behavior in the
case of the full Hamiltonian. This results in a higher residual excitation
for very high amplitudes of the pump pulse in the case of inter-band-only
and two-photon resonances. It is noteworthy that such very high pump-pulse
intensities are not affordable in realistic setups as they would damage
the sample.

\subsection{Both Peierls substitution and local dipole\protect\label{subsec:Dipole_Peierls}}

In this section, we consider both local dipole and Peierls substitution
with the related Hamiltonian parameter values given in the former
two cases. In this case, even though some of the effects can be explained
by simply considering the mere addition of the effects yielded by
the individual coupling terms, we see that the interplay between the
two interaction terms is also relevant.

The retarded TR-ARPES signal along the \emph{main} path at $\bar{t}_{\mathrm{pr}}=0$
is shown in Fig.~\ref{fig:I-t0-PD}, left panel, while the lesser
TR-ARPES signal is shown in the middle panel. The local dipole strengthens
both one-photon and two-photon PSBs. In this case, the \textbf{k}
points with zero velocity do have one-photon PSBs, because of the
local dipole which does not follow the symmetry of the bands. The
TR-ARPES bands are definitely closer to the equilibrium bands rather
than to the instantaneous eigenenergies. The presence of both coupling
terms augments the broadening of the signals as it increases the excited
population overall and, in particular, at the main resonant \textbf{k}
point, \textbf{S}.

Looking at the inter-band-only lesser TR-ARPES signal, which is shown
in the right panel, we see similar behaviors to the case of zero dipole,
except for one main difference: the reduction in the two-photon resonant
signal is much stronger. As explained above (see also App.~\ref{app:Oscillations-of-diag}),
the intra-band transition term assists the two-photon inter-band resonances.
The addition of the local dipole term to the Peierls substitution
strengthens the inter-band transition (stemming from off-diagonal
terms in the Hamiltonian) and, in turn, makes even more effective
the assistance mechanism.

It is worth noticing that we considered the local dipole to be just
of inter-band form, therefore, the intra-band-only results are exactly
the same as the case of zero dipole, which were presented in Fig.~\ref{fig:I-t0-NoD},
bottom-right panel.

\begin{figure}
\begin{centering}
\begin{tabular}{c}
\includegraphics[width=7.5cm]{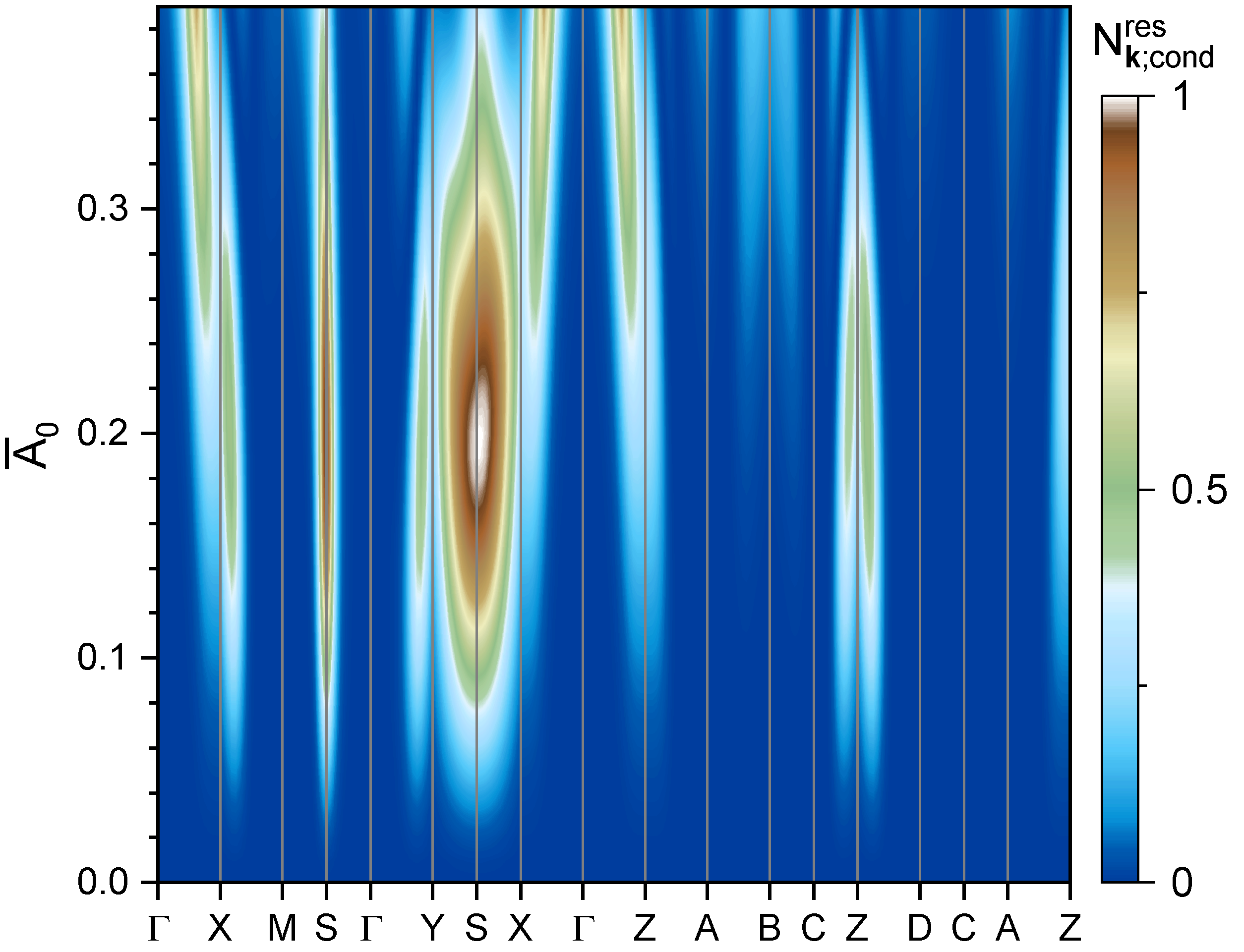}\tabularnewline
\includegraphics[width=7.5cm]{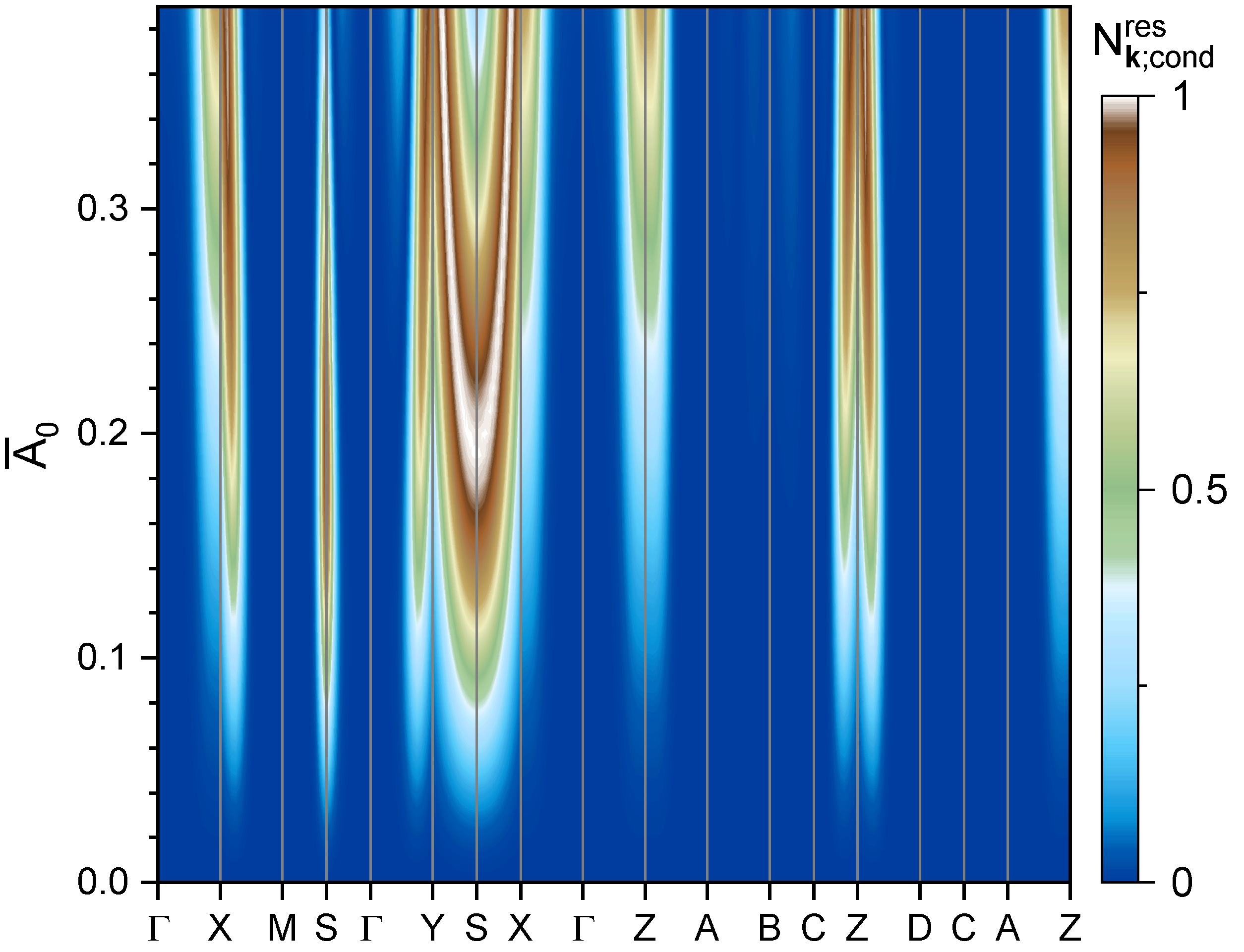}\tabularnewline
\end{tabular}
\par\end{centering}
\caption{Residual excitation along the \emph{main} path as a function of the
pump-pulse amplitude for the case of both Peierls substitution and
local dipole. (top) Full dynamics, (bottom) dynamics with inter-band
transitions only.\protect\label{fig:resN_PD}}
\end{figure}

\begin{figure}
\begin{centering}
\begin{tabular}{c}
\includegraphics[width=7.5cm]{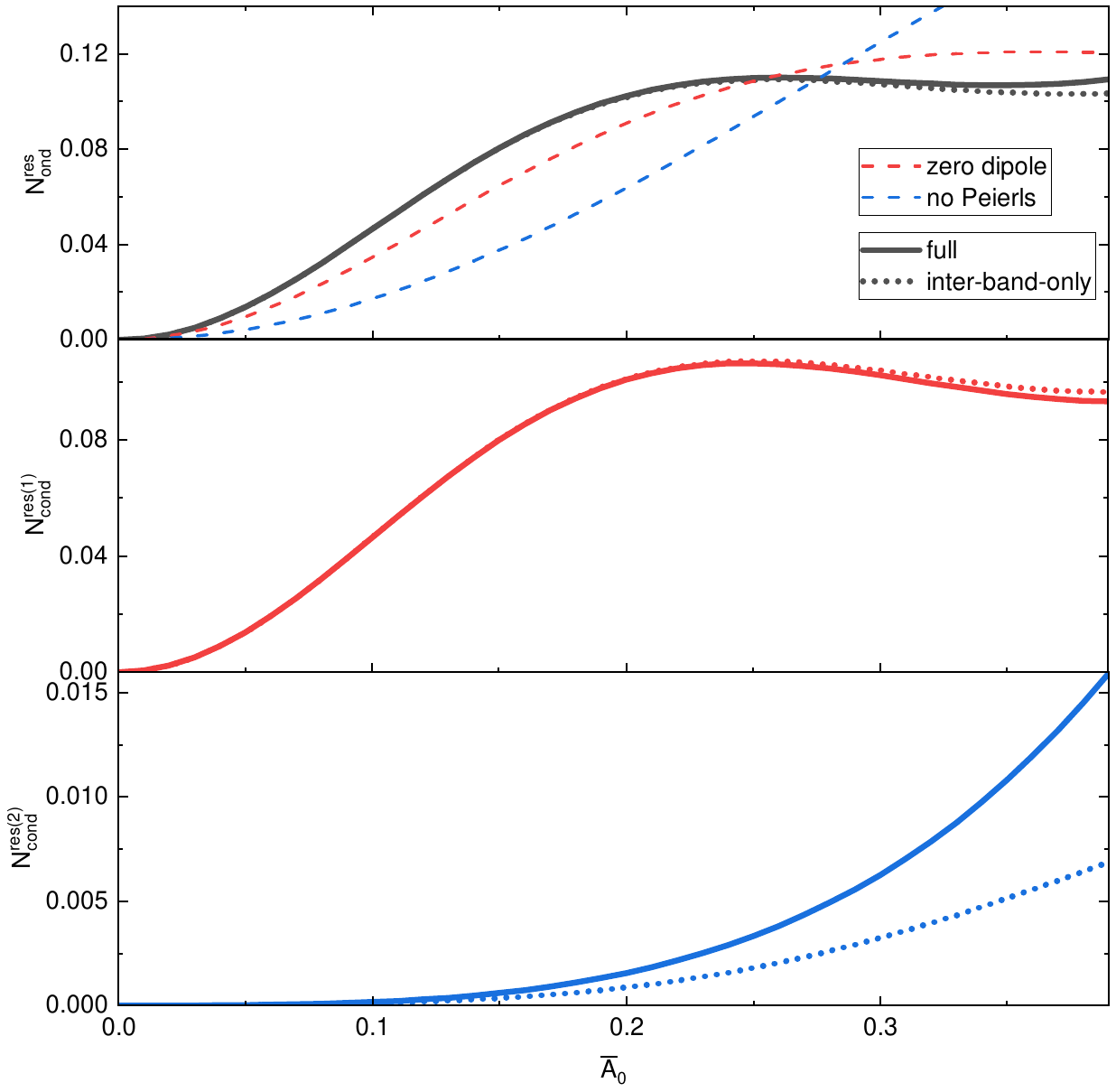}\tabularnewline
\end{tabular}
\par\end{centering}
\caption{Residual excitation per unit cell as a function of the pump-pulse
amplitude for an $8\times8\times8$ \textbf{k} grid over the first
Brillouin zone, considering both Peierls substitution and local dipole.
(top) All (any-photon) resonant contributions, (middle) one-photon
and (bottom) two-photon resonance contributions. In all panels, we
report the two cases of the full original Hamiltonian (solid) and
inter-band-only Hamiltonians (dots). In the top panel, the results
of zero dipole (red dashed) and Peierls-substitution-only (blue dashed)
cases are also reported.\protect\label{fig:resN_cell_PD}}
\end{figure}

In Fig.~\ref{fig:resN_PD} top panel, we plot the residual excited
population along the \emph{main} path vs the pump-pulse amplitude.
The first important change with respect to the zero dipole case is
that the one-photon resonant \textbf{k} points with zero velocity
(\textbf{X}, \textbf{Y}, and \textbf{Z}) do have residual excited
population now: the symmetry protection is lost in the presence of
the dipole term (as discussed for the TR-ARPES signal). Moreover,
having both local dipole and Peierls substitution increases the Rabi
frequency on the line \textbf{X}-\textbf{S}-\textbf{Y} which yields
the residual excited population at \textbf{S} to have a maximum at
around the pump-pulse peak amplitude of $\bar{A}_{0}\simeq0.19$,
showing more clearly the Rabi-like behavior.

\begin{figure*}
\centering{}%
\begin{tabular}{cc}
\includegraphics[width=7.5cm]{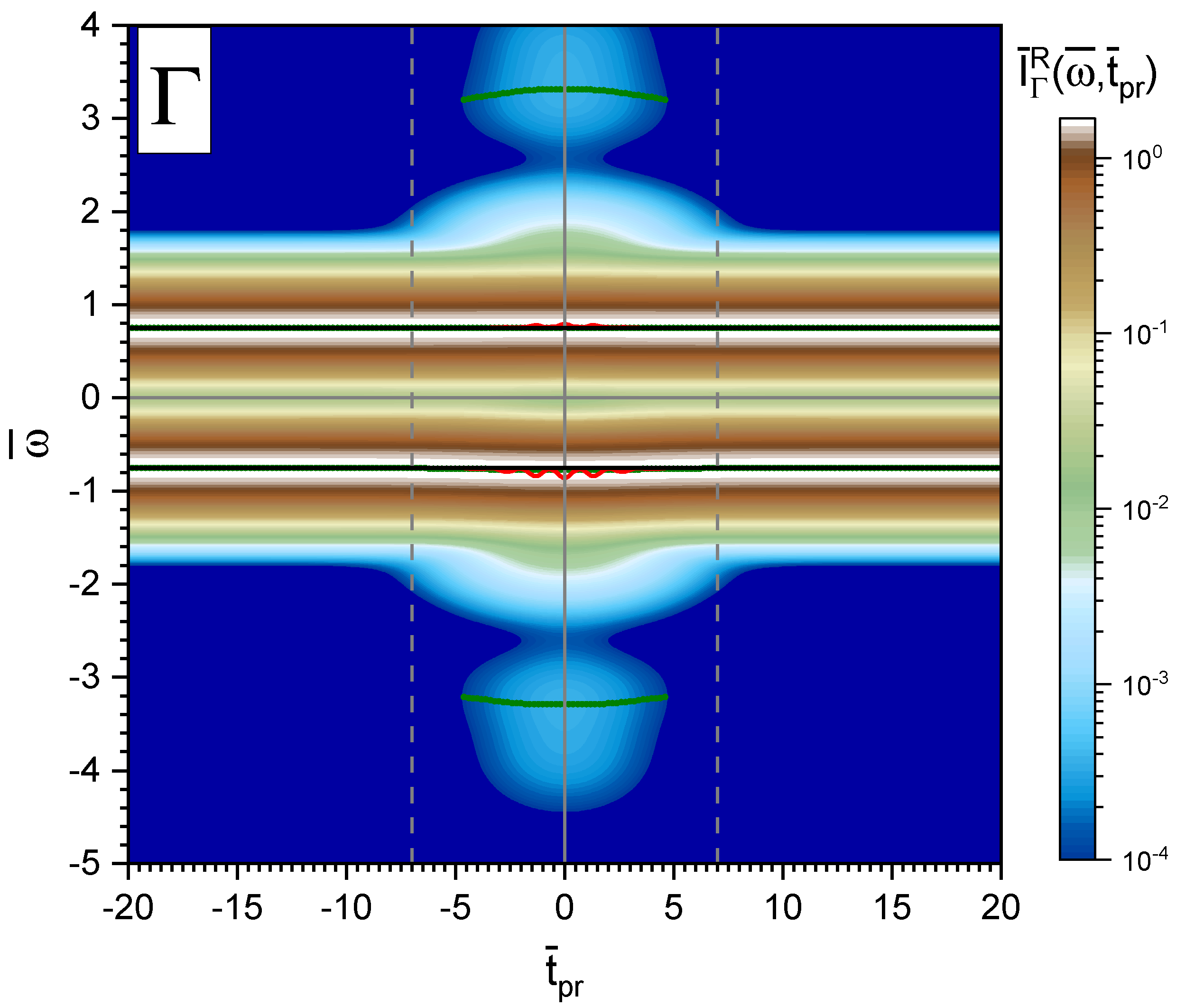} & \includegraphics[width=7.5cm]{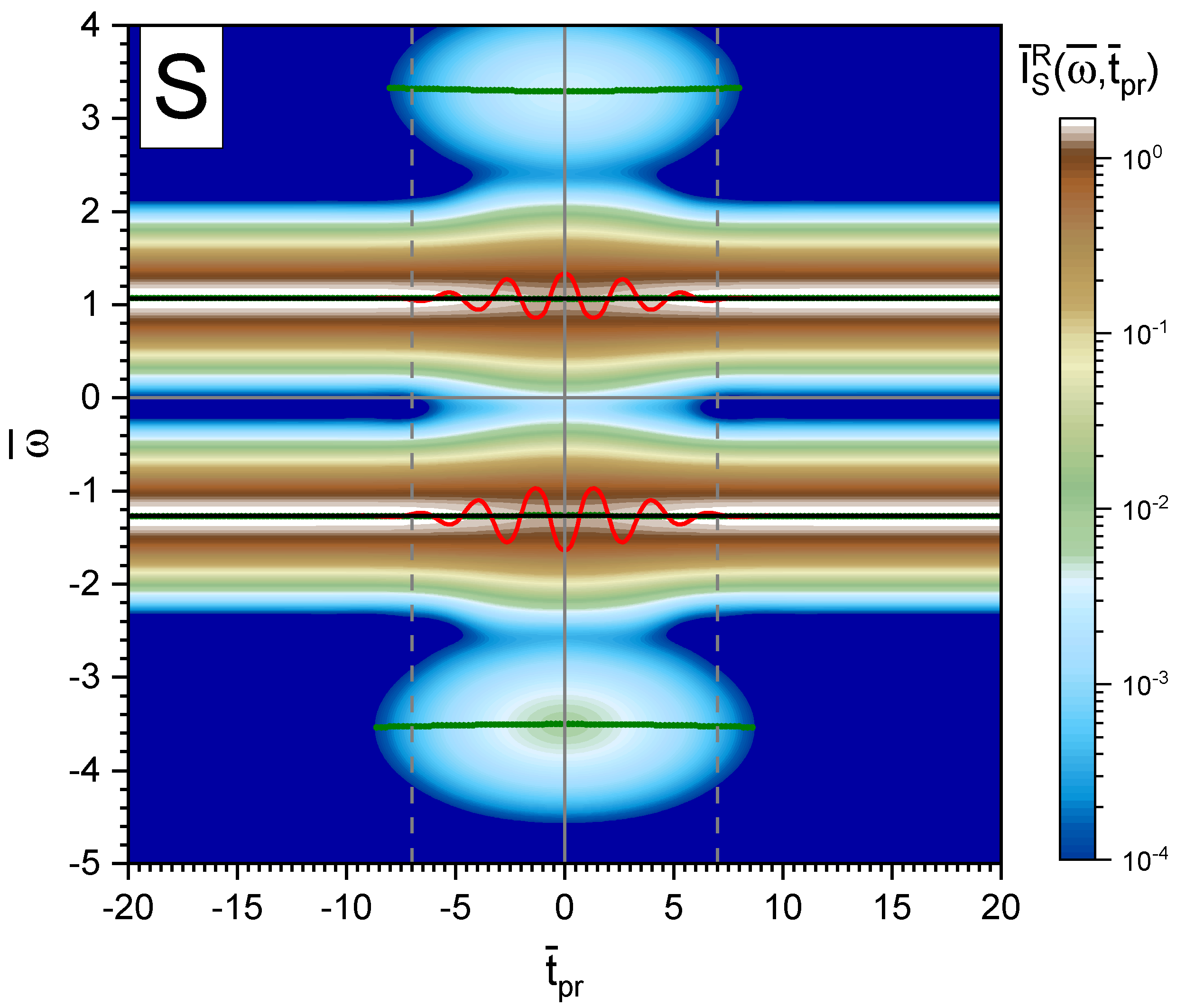}\tabularnewline
\includegraphics[width=7.5cm]{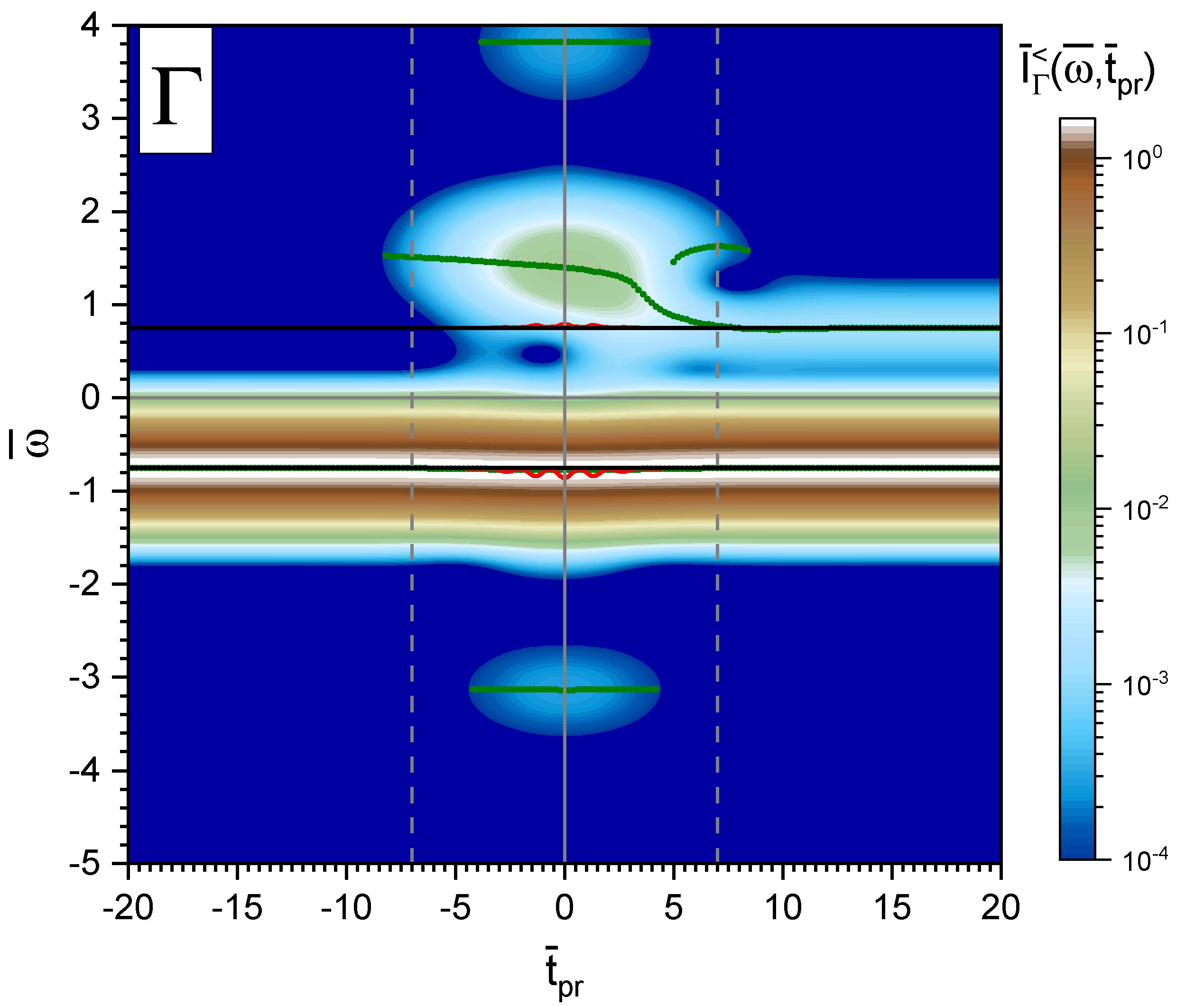} & \includegraphics[width=7.5cm]{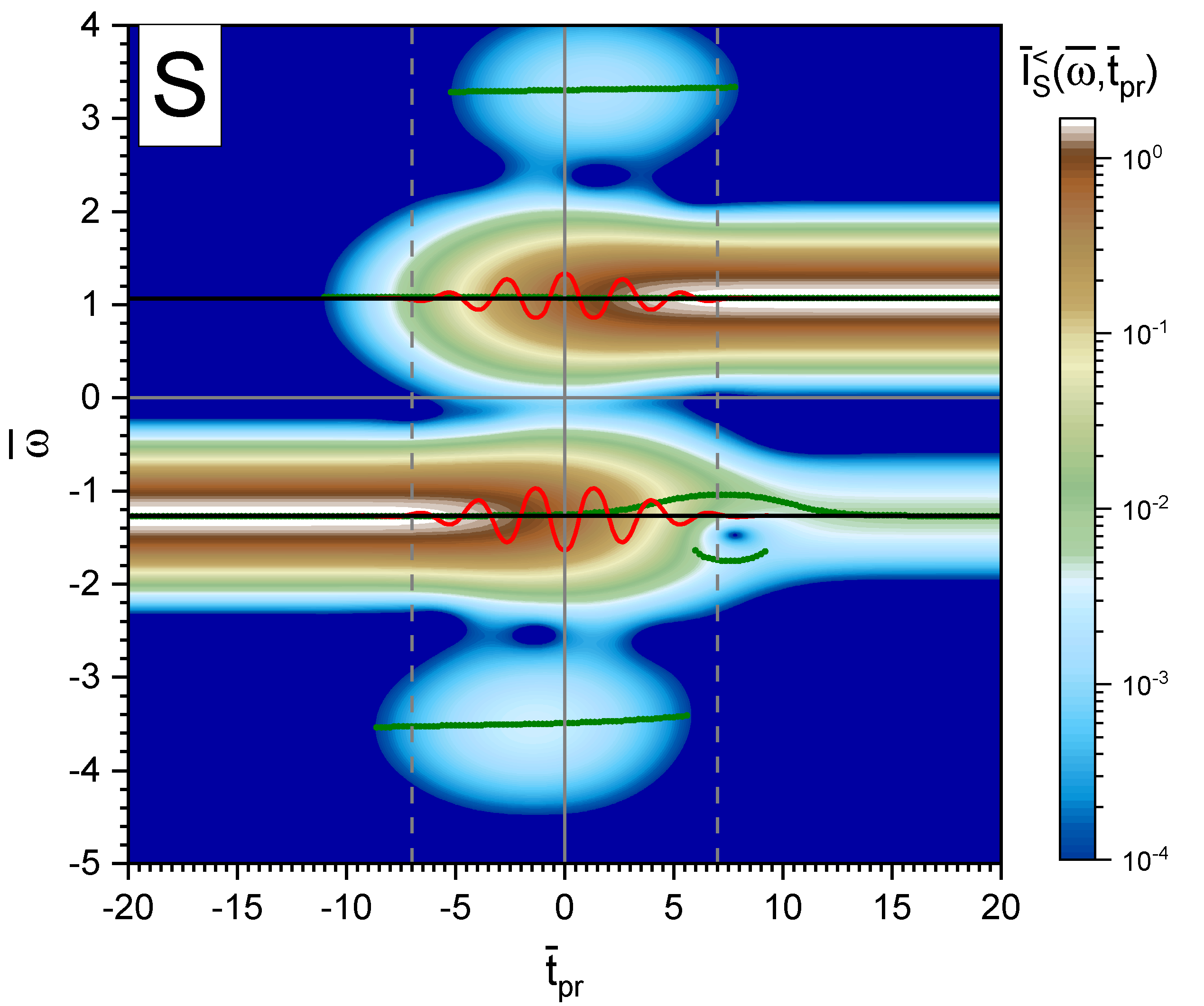}\tabularnewline
\end{tabular}\caption{(top) The retarded and (bottom) the lesser TR-ARPES signals vs the
center of probe-pulse envelope, $\bar{t}_{\mathrm{pr}}$, for the
high-symmetry points (left) $\boldsymbol{\Gamma}$ and (right) \textbf{S}.
The vertical dashed gray lines determine $\pm\bar{\tau}_{\mathrm{pu}}$.
The black-solid lines show the equilibrium band energies while the
green dots show the local maxima of the signals for a fixed \textbf{k}
point. The red-solid curves show the instantaneous eigenenergies as
functions of $\bar{t}_{\mathrm{pr}}$. \protect\label{fig:I-vs-t0-GS}}
\end{figure*}

In Fig.~\ref{fig:resN_PD} bottom panel, we plot the residual excited
population keeping only inter-band transitions in the dynamics. Removing
the intra-band transitions, noticeably increases the residual excited
population at the resonance points near \textbf{X} and \textbf{Y},
so that they can also reach the maximum of full population inversion.
Moreover, for the two-photon resonant \textbf{k} points, the difference
between full and inter-band-only dynamics is much larger than in the
case of zero dipole (as discussed for TR-ARPES signal).

After investigating the residual excitation on the \emph{main} path,
we discuss the excitation per unit cell, which is obtained using an
$8\times8\times8$ \textbf{k} grid to sample the first Brillouin zone
and plotted in the top panel of Fig.~\ref{fig:resN_cell_PD}. Comparison
with the two former cases of considering no Peierls substitution and
having zero dipole (both shown in the same panel), we see the maximum
occurs at a smaller pump-pulse amplitude, as the local dipole adds
up to the Peierls substitution which increases the Rabi frequencies.
Another relevant feature is related to only-inter-band transitions
in the dynamics that seem to reduce the residual excited population,
which is apparently in contradiction with the result of Fig.~\ref{fig:resN_cell_NoD}.
However, the behavior of one-photon and two-photon resonance contributions,
as plotted in the middle and bottom panels of Fig.~\ref{fig:resN_cell_PD},
reveals that similar to the case of zero dipole, the inter-band-only
dynamics gives more (less) residual excited population for the one-photon
(two-photon) resonance contributions, but the difference between the
inter-band-only and full dynamics of the two-photon resonances are
much larger in this case, as we explained in the discussion of Fig.~\ref{fig:I-t0-PD}.

\subsection{More on the characteristics of the TR-ARPES signal\protect\label{subsec:More_TR-ARPES}}

In this section, we get more insights about the behavior of the system
out of equilibrium by changing the probe-pulse parameters. For the
coupling Hamiltonian, we consider both local dipole and Peierls substitution,
but the general conclusions we will draw are independent of this choice.

First, we study how the TR-ARPES signal changes on varying the center
of probe pulse, $\bar{t}_{\mathrm{pr}}$, from before to after the
pump-pulse envelope. The retarded and lesser TR-ARPES signals for
two high-symmetry \textbf{k} points, $\boldsymbol{\Gamma}$ and \textbf{S},
are reported in Fig.~\ref{fig:I-vs-t0-GS}, top panels. For both
\textbf{k} points, the PSBs are detected as soon as the probe-pulse
center enters the pump-pulse envelope, that is when the instantaneous
eigenenergies become different from the equilibrium band energies.

At equilibrium, as expected, the lesser signal (bottom panels) shows
that the electrons reside in the valence band, while they get excited
into the conduction band during the pump-pulse application. At \textbf{S},
which is exactly one-photon resonant, we register an almost complete
population inversion.

After the pump pulse is turned off, the residual signal at $\boldsymbol{\Gamma}$
is very weak ($\boldsymbol{\Gamma}$ is not in resonance), while at
\textbf{S} we have a very strong residual signal because of the resonance
condition. Having both local dipole and Peierls substitution, yields
slightly more residual signal compared to the cases of removing one
of the two coupling terms.

Moreover, at\textbf{ S }that is a resonant point, we have the splitting
of the valence band. Such a splitting is not visible at time $\bar{t}_{\mathrm{pr}}=0$,
because the two wide split bands overlap with each other. Increasing
either the local dipole or the pump-pulse amplitude one can see the
splitting even at time $\bar{t}_{\mathrm{pr}}=0$ (not shown).

So far, the FWHM of the probe pulse was kept constant and equal to
the one of the pump pulse, $\bar{\tau}_{\mathrm{pu}}=\bar{\tau}_{\mathrm{pr}}=7$.
Now, we study the effect of varying $\bar{\tau}_{\mathrm{pu}}$ while
keeping $\bar{t}_{\mathrm{pr}}=0$. In Fig.~\ref{fig:IL-vs-taupr_PD},
the lesser signal is reported at an off-resonant \textbf{k} point
$\left(0.375,0.125,0.5\right)$ (top panel) and at the resonant \textbf{k}
point \textbf{S} (bottom panel). The former point is chosen in order
to have an energy difference between the instantaneous eigenenergies
at time zero and the equilibrium-band energies quite noticeable to
better illustrate the phenomenology we are going to discuss.

First, we analyze the behavior at the off-resonant \textbf{k} point
(top panel). For very narrow probes, that is small $\bar{\tau}_{\mathrm{pr}}$
with respect to $\bar{\tau}_{\mathrm{pu}}$, by decreasing\textbf{
$\bar{\tau}_{\mathrm{pr}}$, }the signal gets wider and its peaks
tend to the instantaneous eigenenergies. This indicates that the system
is practically in the lower eigenstate, which is predominantly valence-band-like
as there is no excitation to the higher eigenstate. At any rate, the
peaks will never exactly coincide with the instantaneous eigenenergies
(even though they are very close to them) as the process is not adiabatic.
We expect this also in \emph{real} semiconductors and insulators as
there the off-diagonal terms of the coupling Hamiltonian are usually
much smaller than the energy gaps determined by the total Hamiltonian.

Instead, increasing the width of the probe-pulse envelope, $\bar{\tau}_{\mathrm{pr}}$,
corresponds to measuring the system over a finite time interval and,
practically, to performing a time average over such an interval. This
\emph{averaging} process results in the emergence of side-bands while
the main peaks tend to the equilibrium bands. The shifts of the bands
are due to the high non-linearity of the processes and to the non-zero
average of the oscillating pump pulse. The PSBs remain at almost fixed
energies after they emerge because they are related to the oscillating
component of the pumping field and, if the probe pulse is wide enough
to see the oscillations, it does not matter how much wider it becomes.
On the other hand, EPSBs change their energies by changing the width
of the probe-pulse envelope, because they are driven by the non-oscillating
component of the pumping field.

By increasing the width of the probe pulse to very high values, the
resolution in energy increases and the peaks become very sharp. However,
having such a large probe-pulse FWHM corresponds to (i) reducing more
and more the time resolution of the measurement and (ii) including
more and more equilibrium behavior (before pump-pulse envelope) and
residual effects (after pump-pulse envelope) into the measurement.
Therefore, we cannot obtain enough information about the transient
out-of-equilibrium dynamics of the system. On the other hand, on decreasing
the width of the probe pulse, the signals become very wide in energy.
This requires more and more experimental resolution in energy to determine
the position of the peaks and understand the physics. Consequently,
one needs to choose some intermediate value in order to cope with
the unavoidable intrinsic time-energy uncertainty relationship of
the underlying quantum mechanical system.

The situation at resonance is quite different. As it is shown in the
bottom panel of Fig.~\ref{fig:IL-vs-taupr_PD}, even for the smallest
values of $\bar{\tau}_{\mathrm{pu}}$, the peak of lesser TR-ARPES
signal does not coincide with the lower eigenenergy as the resonant
dynamics forces the electrons to evolve in a superposition of valence
and conduction band states. The superposition of two eigenstates results
in the overlap of the TR-ARPES signals and, consequently, gives a
peak somewhere in the middle of the two eigenenergies. Increasing
the width of probe pulse, again the PSBs emerge and the one-photon
PSB is highly populated. It is noteworthy that the inverse mass at
\textbf{S} is zero and this is why we do not have any shifting of
the bands and no EPSB emerges.

\section{Summary and perspectives\protect\label{sec:Summary}}

In this manuscript, we have reported on a novel model-Hamiltonian
approach that we have recently devised and developed to study out-of-equilibrium
\emph{real} materials \citep{inzani2022field}, the dynamical projective
operatorial approach (DPOA). Its internals have been illustrated in
detail and the theory is extended to the calculations of TR-ARPES
signal in pumped lattice systems. As a noteworthy prototypical application,
a pumped two-band (valence-conduction) system, is discussed extensively.
DPOA relies on many-body second-quantization formalism and composite
operators in order to be capable of handling both weakly and strongly
correlated systems. DPOA exploits the tight-binding approach and the
wannierization of DFT band structures in order to cope with the complexity
and the very many degrees of freedom of \emph{real} materials \citep{inzani2022field}.
We have devised an ad hoc \emph{Peierls expansion} in order to increase
the efficiency of numerical calculations. This expansion makes clear
how multi-photon resonances, rigid shifts, band dressings and different
types of sidebands naturally emerge and allows to understand deeply
the related phenomenologies. We also reported another mechanism to
explain the emergence of multi-photon resonances relaying on the intra-band
motion (oscillating diagonal terms in the Hamiltonian).

We have defined a protocol for evaluating the strength of single/multi-photon
resonances and for assigning the residual excited electronic population
at each \textbf{k} point and band to a specific single/multi-photon
process. Comparing DPOA to the single-particle density-matrix approach
and the Houston method, which we have generalized to second-quantization
formalism and rephrased in the DPOA framework to compute exactly its
dynamics, we have shown that DPOA goes much beyond both of them in
terms of computing capabilities (multi-particle multi-time correlators)
and complexity handling. To study the injection processes and the
out-of-equilibrium electronic dynamics, we have expressed the relevant
out-of-equilibrium Green's functions and the (lesser) TR-ARPES signal
within the DPOA framework. Then, defining a retarded TR-ARPES signal,
which allows to analyze the behavior of the dynamical bands independently
from their occupation, we have shown that it is possible to obtain
an out-of-equilibrium version of the fluctuation-dissipation theorem.
Another relevant aspect that we have thoroughly considered resides
in the possibility to analyze intra- and inter-band transitions in
the TR-ARPES signal and in the residual electronic excited population
by selectively inhibiting them in the model Hamiltonian.

We have studied the three most relevant cases of light-matter coupling
within the dipole gauge, which has been derived in the second-quantization
formalism: only a local dipole (relevant to systems such as quantum
dots and molecules, and low-dimensional systems with transverse pumps),
only the Peierls substitution in the hopping term (relevant to many
\emph{real} materials), and both terms at once. Within the framework
of a pumped two-band lattice system, we have analyzed in detail the
TR-ARPES signal and the residual electronic excited population with
respect to the band energies and their symmetries as well as their
dependence on the pump/probe-pulse characteristics. We have studied:
(i) how the first-order (in the pump-pulse amplitude) terms of the
two types of light-matter couplings assist the higher-order ones;
(ii) how their decomposition in terms of intra- and inter-band components
can allow to understand the actual photo-injection process; (iii)
how the symmetries of the system rule the actual behavior of the lesser
and the retarded TR-ARPES signals as well as of the residual excited
populations; (iv) how the (dynamical) bands broaden out-of-equilibrium
and shift with respect to the equilibrium ones; (v) how different
kinds of photon (resonant, non resonant) and envelope-Peierls sidebands
emerge and vanish in relation to band symmetries and how dipole term
breaks this symmetry \emph{protection}; (vi) how residual electronic
excited population accumulate in the conduction band induced by Rabi-like
oscillations at the multi-photon resonant \emph{non-symmetry-protected}
\textbf{k} points and the characteristics of such oscillations in
terms of the pump-pulse features; (vi) how the width and the delay
of the probe pulse affect the TR-ARPES signal.

Recently, we applied DPOA to unveil the different charge-injection
mechanisms in ultrafast (attosecond) pumped germanium proving its
efficiency and relevance to \emph{real} experimental setups \citep{inzani2022field}.
Moreover, we have obtained, within DPOA, the expressions for the time-dependent
optical response (transient reflectivity and absorption) in pump-probe
setups \citep{eskandari2024generalized}. This kind of analyses is
fundamental to advance the physical understanding of complex materials
and the capability to eventually turn this knowledge into actual industrial
and commercial applications, such as the recently proposed novel types
of electronics.
\begin{acknowledgments}
The authors thank Claudio Giannetti, Matteo Lucchini, Stefano Pittalis,
and Carlo Andrea Rozzi for the insightful discussions. The authors
acknowledge support by MIUR under Project No. PRIN 2017RKWTMY.
\end{acknowledgments}

\appendix

\section{Velocity and dipole gauges:\protect \\
Hamiltonian, density and current operators\protect\label{app:gauges}}

\subsection{System\protect\label{subapp:System}}

Let us start from the single-particle Hamiltonian operator in first
quantization, $\hat{H}_{0}$, for an electron of charge $-e$ and
mass $m$ in the periodic potential $V\left(\mathbf{r}+\mathbf{R}_{\mathbf{i}}\right)=V\left(\mathbf{r}\right)$
generated by the Bravais lattice $\left\{ \mathbf{R}_{\mathbf{i}}\right\} $
of ions of a solid state system {[}$\mathbf{R}_{\mathbf{i}}=\sum_{\lambda=1}^{3}i_{\lambda}\mathbf{a}_{\lambda}$
where $\mathbf{i}=\left(i_{1},i_{2},i_{3}\right)$, $i_{\lambda}\in\mathbb{Z}$
and $\mathbf{a}_{\lambda}$ are the lattice vectors{]}:
\begin{equation}
\hat{H}_{0}=\frac{\hat{p}^{2}}{2m}+V\left(\hat{\mathbf{r}}\right),\label{eq:H_vg}
\end{equation}
where $\hat{\mathbf{p}}$ and $\hat{\mathbf{r}}$ are the momentum
and the position operators of the electron, respectively, that satisfy
the canonical commutation relation $\left[\hat{r}_{\eta},\hat{p}_{\eta^{\prime}}\right]=\mathrm{i}\hbar\delta_{\eta\eta^{\prime}}$,
where $\eta,\eta^{\prime}\in\left\{ x,y,z\right\} $. In this appendix,
we denote the operators in first-quantization formulation by the hat
($\mathcircumflex$) over-script. The Bloch theorem states that we
can find a solution $\k{\phi_{\mathbf{k},n}}=\mathrm{e}^{\mathrm{i}\mathbf{k}\bullet\hat{\mathbf{r}}}\k{u_{\mathbf{k},n}}$,
parametrized by the band index $n$ and the momentum $\mathbf{k}$,
of the Schrödinger equation, $\hat{H}_{0}\k{\phi_{\mathbf{k},n}}=\varepsilon_{\mathbf{k},n}\k{\phi_{\mathbf{k},n}}$,
where $u_{\mathbf{k},n}\left(\mathbf{r}\right)=\bk{\mathbf{r}}{u_{\mathbf{k},n}}$
has the periodicity of the Bravais lattice and $\varepsilon_{\mathbf{k},n}$
is the $n$-th band-energy dispersion. We also have $\hat{H}_{0,\mathbf{k}}\k{u_{\mathbf{k},n}}=\varepsilon_{\mathbf{k},n}\k{u_{\mathbf{k},n}}$,
where $\hat{H}_{0,\mathbf{k}}=\mathrm{e}^{-\mathrm{i}\mathbf{k}\bullet\hat{\mathbf{r}}}\hat{H}_{0}\mathrm{e}^{\mathrm{i}\mathbf{k}\bullet\hat{\mathbf{r}}}$,
and $\phi_{\mathbf{k},n}\left(\mathbf{r}\right)=\bk{\mathbf{r}}{\phi_{\mathbf{k},n}}$.

\subsection{Velocity gauge\protect\label{subapp:Velocity}}

In the dipole approximation (i.e., for wavelengths much larger than
the unit cell extent in the direction of propagation), an electromagnetic
wave interacting with the system (the electrons) can be described
by a homogeneous vector potential $\mathbf{A}\left(t\right)$. Then,
according to the minimal coupling protocol $\hat{\mathbf{p}}\rightarrow\hat{\boldsymbol{\pi}}=\hat{\mathbf{p}}+e\mathbf{A}\left(t\right)$,
the Hamiltonian operator reads as
\begin{equation}
\hat{H}=\frac{\hat{\pi}^{2}}{2m}+V\left(\hat{\mathbf{r}}\right)=\hat{H}_{0}+\frac{e}{m}\mathbf{A}\left(t\right)\bullet\hat{\mathbf{p}}+\frac{e^{2}}{2m}A^{2}\left(t\right),\label{eq:H_t_vg}
\end{equation}
where $\bullet$ is the scalar product in direct space. This scenario
is known as \emph{velocity gauge} after the electron-field interaction
term in the Hamiltonian: $e\mathbf{A}\left(\hat{\mathbf{r}},t\right)\bullet\frac{\hat{\mathbf{p}}}{m}$.
Let us suppose that $\k{\psi}$ is the solution of the time-dependent
Schrödinger equation, $\mathrm{i}\hbar\frac{\partial}{\partial t}\k{\psi}=\hat{H}\k{\psi}$.
Then, the dynamics of the charge density operator $\hat{\rho}=-e\p{\mathbf{r}}$
and, in particular, of its average $\e{\hat{\rho}}=\ef{\hat{\rho}}{\psi}=-e\ns{\psi\left(\mathbf{r},t\right)}$
(recall that $\psi\left(\mathbf{r},t\right)=\bk{\mathbf{r}}{\psi}$)
is given by
\begin{equation}
\frac{\partial}{\partial t}\e{\hat{\rho}}=-\frac{\mathrm{i}}{\hbar}\e{\left[\hat{\rho},\hat{H}\right]}=-\frac{1}{2m}\sum_{\eta=x,y,z}\nabla_{\eta}\e{\left[\hat{\rho}\hat{\pi}_{\eta}+\hat{\pi}_{\eta}\hat{\rho}\right]}.\label{eq:EOM_rho_vg}
\end{equation}
Next, the continuity equation, $\frac{\partial}{\partial t}\e{\hat{\rho}}+\boldsymbol{\nabla}\bullet\e{\hat{\mathbf{J}}}=0$,
calls for the following definition for the current operator
\begin{equation}
\hat{\mathbf{J}}=\frac{1}{2m}\left(\hat{\rho}\hat{\boldsymbol{\pi}}+\hat{\boldsymbol{\pi}}\hat{\rho}\right)=\frac{1}{2m}\left(\hat{\rho}\hat{\mathbf{p}}+\hat{\mathbf{p}}\hat{\rho}\right)+\frac{e}{m}\mathbf{A}\left(t\right)\hat{\rho},\label{eq:J_vg}
\end{equation}
where we can distinguish the paramagnetic (first) and the diamagnetic
(second) terms. It is worth noticing that the continuity equation
can be equivalently written as follows: 
\begin{equation}
\frac{\partial}{\partial t}\e{\hat{\rho}}=\frac{\mathrm{i}}{\hbar}\sum_{\eta=x,y,z}\e{\left[\hat{\pi}_{\eta},\hat{J}_{\eta}\right]}.\label{eq:cont_2}
\end{equation}

In order to move to second quantization in the Bloch basis, we need
\begin{multline}
\mathbf{v}_{n,n^{\prime}}\left(\mathbf{k}\right)=\frac{1}{m}\me{\phi_{\mathbf{k},n}}{\hat{\mathbf{p}}}{\phi_{\mathbf{k},n^{\prime}}}\\
=\frac{1}{\hbar}\me{u_{\mathbf{k},n}}{\boldsymbol{\nabla}_{\mathbf{k}}\hat{H}_{0,\mathbf{k}}}{u_{\mathbf{k},n^{\prime}}}\\
=\delta_{nn^{\prime}}\frac{1}{\hbar}\boldsymbol{\nabla}_{\mathbf{k}}\varepsilon_{\mathbf{k},n}-\frac{\mathrm{i}}{\hbar}\left(\varepsilon_{\mathbf{k},n^{\prime}}-\varepsilon_{\mathbf{k},n}\right)\mathbf{B}_{n,n^{\prime}}\left(\mathbf{k}\right),\label{eq:v_vg}
\end{multline}
where we have used the relation $\hat{\mathbf{p}}=-\mathrm{i}\frac{m}{\hbar}\left[\hat{\mathbf{r}},\hat{H}_{0}\right]$
and $\mathbf{B}_{n,n^{\prime}}\left(\mathbf{k}\right)=\bk{u_{\mathbf{k},n}}{\boldsymbol{\nabla}_{\mathbf{k}}u_{\mathbf{k},n^{\prime}}}$
is the Berry connection. It is worth noticing that the last expression
requires that the Bloch basis used in the actual numerical calculations
is complete. Then, we have
\begin{multline}
\mathcal{H}=\sum_{\mathbf{k},n,n^{\prime}}\me{\phi_{\mathbf{k},n}}{\hat{H}}{\phi_{\mathbf{k},n^{\prime}}}c_{\mathbf{k},n}^{\dagger}c_{\mathbf{k},n^{\prime}}=\sum_{\mathbf{k},n}\varepsilon_{\mathbf{k},n}c_{\mathbf{k},n}^{\dagger}c_{\mathbf{k},n}\\
+\sum_{\mathbf{k},n,n^{\prime}}\left(e\mathbf{A}\left(t\right)\bullet\mathbf{v}_{n,n^{\prime}}\left(\mathbf{k}\right)+\delta_{nn^{\prime}}\frac{e^{2}}{2m}A^{2}\left(t\right)\right)c_{\mathbf{k},n}^{\dagger}c_{\mathbf{k},n^{\prime}},\label{eq:ham_2q_vg}
\end{multline}
where $c_{\mathbf{k},n}$ is the annihilation operator related to
the single-particle state $\k{\phi_{\mathbf{k},n}}$. We also have,
in second quantization,
\begin{align}
 & \mathcal{\rho}\left(\mathbf{r}\right)=-e\sum_{\mathbf{k},n}\ns{\phi_{\mathbf{k},n}\left(\mathbf{r}\right)}c_{\mathbf{k},n}^{\dagger}c_{\mathbf{k},n},\label{eq:rho_2q_vg}\\
 & \mathbf{J}\left(\mathbf{r},t\right)=\frac{1}{2}\sum_{\mathbf{k},n,n^{\prime}}\ns{\phi_{\mathbf{k},n}\left(\mathbf{r}\right)}\mathbf{v}_{n,n^{\prime}}\left(\mathbf{k}\right)c_{\mathbf{k},n}^{\dagger}c_{\mathbf{k},n^{\prime}}\nonumber \\
 & +\frac{1}{2}\sum_{\mathbf{k},n,n^{\prime}}\ns{\phi_{\mathbf{k},n^{\prime}}\left(\mathbf{r}\right)}\mathbf{v}_{n,n^{\prime}}\left(\mathbf{k}\right)c_{\mathbf{k},n}^{\dagger}c_{\mathbf{k},n^{\prime}}\nonumber \\
 & +\frac{e}{m}\mathbf{A}\left(t\right)\mathcal{\rho}\left(\mathbf{r}\right).\label{eq:J_2q_vg}
\end{align}
It is worth noting that, in principle, for any \emph{real} material,
$\mathbf{v}_{n,n^{\prime}}\left(\mathbf{k}\right)$ can be obtained
through the outputs of the majority of the available DFT codes.

\subsection{Dipole gauge\protect\label{subapp:Dipole}}

Now, we can move from computing the average of the velocity operator
$\frac{1}{m}\hat{\mathbf{p}}$ and, consequently, the Berry connection,
to computing the average of the operator $\hat{\mathbf{r}}$ and,
therefore, the dipole operator $\hat{\mathbf{D}}$. In order to do
that, one can apply the following unitary transformation:
\begin{equation}
\hat{U}=\mathrm{e}^{-\mathrm{i}\hat{S}}\quad\textrm{where}\quad\hat{S}=-\frac{e}{\hbar}\mathbf{A}\left(t\right)\bullet\hat{\mathbf{r}}.\label{eq:U_vg2dg}
\end{equation}
Let us recall the following general relations:
\begin{align}
 & \hat{\underline{O}}=\hat{U}\hat{O}\hat{U}^{\dagger}\label{eq:O_U}\\
 & =\sum_{n=0}^{\infty}\frac{\left(-\mathrm{i}\right)^{n}}{n!}\left[\hat{S},\left[\hat{S},\ldots,\left[\hat{S},\left[\hat{O}\right]_{0}\right]_{1}\ldots\right]_{n-1}\right]_{n},\\
 & \k{\underline{\psi}}=\hat{U}\k{\psi},\quad\mathrm{i}\hbar\frac{\partial}{\partial t}\k{\underline{\psi}}=\overline{\hat{H}}\k{\underline{\psi}},\label{eq:psi_U}\\
 & \overline{\hat{H}}=\hat{\underline{H}}+\left(\mathrm{i}\hbar\frac{\partial}{\partial t}\hat{U}\right)\hat{U}^{\dagger}.\label{eq:ham_U}
\end{align}
Where any first-quantization operator $\hat{O}$ transformed by $\hat{U}$
is denoted by $\hat{\underline{O}}$, and $\k{\underline{\psi}}$
is the transformation of the wave function by $\hat{U}$. The time-evolution
of $\k{\underline{\psi}}$ is given by the effective Hamiltonian $\overline{\hat{H}}$
, as given in Eq.~\ref{eq:ham_U}, rather than $\hat{\underline{H}}$.
It is straightforward to show that,
\begin{align}
 & \hat{\underline{\mathbf{r}}}=\hat{\mathbf{r}},\\
 & \hat{\underline{\mathbf{p}}}=\hat{\mathbf{p}}-e\mathbf{A}\left(t\right),\\
 & \hat{\underline{\boldsymbol{\pi}}}=\hat{\boldsymbol{\pi}}-e\mathbf{A}\left(t\right)=\hat{\mathbf{p}},\\
 & \hat{\underline{H}}=\hat{H}_{0},\label{eq:H1}\\
 & \frac{\partial}{\partial t}\hat{U}=-\mathrm{i}\frac{e}{\hbar}\left(\mathbf{E}\left(t\right)\bullet\hat{\mathbf{r}}\right)\hat{U},\\
 & \overline{\hat{H}}=\hat{H}_{0}+e\mathbf{E}\left(t\right)\bullet\hat{\mathbf{r}},\\
 & \hat{\underline{\rho}}=\hat{\rho},\\
 & \hat{\underline{\mathbf{J}}}=\hat{\mathbf{J}}-\frac{e}{m}\mathbf{A}\left(t\right)\hat{\rho}=\frac{1}{2m}\left(\hat{\rho}\hat{\mathbf{p}}+\hat{\mathbf{p}}\hat{\rho}\right),\\
 & \frac{\partial}{\partial t}\e{\hat{\underline{\rho}}}=-\frac{\mathrm{i}}{\hbar}\e{\left[\hat{\underline{\rho}},\overline{\hat{H}}\right]}=-\boldsymbol{\nabla}\bullet\e{\hat{\underline{\mathbf{J}}}}\\
 & =\frac{\mathrm{i}}{\hbar}\sum_{\eta=x,y,z}\e{\left[\hat{\underline{\pi}}_{\eta},\hat{\underline{J}}_{\eta}\right]},
\end{align}
where $\mathbf{E}\left(t\right)=-\frac{\partial}{\partial t}\mathbf{A}\left(t\right)$
is the electric field applied to the system. It is just Eq.~\ref{eq:H1},
as sought outcome, that inspired the transformation. This scenario
is known as \emph{dipole gauge} after the electron-field interaction
term in the Hamiltonian: $e\mathbf{E}\left(t\right)\bullet\hat{\mathbf{r}}$.

Moving from the Bloch states to the localized Wannier ones,
\begin{equation}
\k{\varphi_{\mathbf{i},\nu}}=\frac{1}{\sqrt{N}}\sum_{\mathbf{k},n}a_{\mathbf{k},\nu,n}\mathrm{e}^{-\mathrm{i}\mathbf{k}\bullet\mathbf{R}_{\mathbf{i}}}\k{\phi_{\mathbf{k},n}},
\end{equation}
where $N$ is generically the number of lattice sites and $a_{\mathbf{k},\nu,n}$
can be chosen to refer to a localized state (for instance, a \emph{maximally}
localized Wannier functions), $\varphi_{\mathbf{i},\nu}\left(\mathbf{r}-\mathbf{R}_{\mathbf{i}}\right)=\bk{\mathbf{r}}{\varphi_{\mathbf{i},\nu}}$,
around the specific Bravais lattice site $\mathbf{R}_{\mathbf{i}}$.
Hereafter, this is the choice that has been made in order to properly
compute the dipole term of the Hamiltonian as we will see in the following.
It is easy to demonstrate that the $\hat{\mathbf{r}}$ operator has
an ill-defined average on a Bravais lattice,
\begin{equation}
\me{\varphi_{\mathbf{i},\nu}}{\hat{\mathbf{r}}}{\varphi_{\mathbf{i},\nu}}=\int d\mathbf{r}\mathbf{r}\left|\varphi_{\mathbf{i},\nu}\left(\mathbf{r}-\mathbf{R}_{\mathbf{i}}\right)\right|^{2}\underset{\left|\mathbf{R}_{\mathbf{i}}\right|\rightarrow\infty}{\rightarrow}\infty,
\end{equation}
implying that the Hamiltonian is also ill defined. $\me{\varphi_{\mathbf{i},\nu}}{\hat{\mathbf{r}}-\hat{\mathbf{R}}}{\varphi_{\mathbf{i},\nu}}$
can be instead always finite if the states $\k{\varphi_{\mathbf{i},\nu}}$
are localized enough; actually, if it is not so, the following procedure
cannot be adopted. Here, we have defined the operator $\hat{\mathbf{R}}$
as follows: $\hat{\mathbf{R}}\k{\varphi_{\mathbf{i},\nu}}=\mathbf{R}_{\mathbf{i}}\k{\varphi_{\mathbf{i},\nu}}$.
Accordingly, the elements of the localized basis, $\k{\varphi_{\mathbf{i},\nu}}$,
are its eigenfunctions and the Bravais lattice sites, $\mathbf{R}_{\mathbf{i}}$,
are the corresponding eigenvalues.

This problem calls for the application of one more unitary transformation,
\begin{equation}
\hat{U}^{\prime}=\mathrm{e}^{-\mathrm{i}\hat{S}^{\prime}}\quad\textrm{where}\quad\hat{S}^{\prime}=+\frac{e}{\hbar}\mathbf{A}\left(t\right)\bullet\hat{\mathbf{R}}.
\end{equation}
We can exploit the following relation to apply this transformation
to the relevant operators and write them in the dipole gauge\textbf{:}
\begin{equation}
\underline{\underline{\hat{O}}}_{\mathbf{i},\nu;\mathbf{j},\nu^{\prime}}=\me{\varphi_{\mathbf{i},\nu}}{\underline{\underline{\hat{O}}}}{\varphi_{\mathbf{j},\nu^{\prime}}}=\mathrm{e}^{-\mathrm{i}\frac{e}{\hbar}\mathbf{A}\left(t\right)\bullet\mathbf{R}_{\mathbf{ij}}}\underline{\hat{O}}_{\mathbf{i},\nu;\mathbf{j},\nu^{\prime}},
\end{equation}
and get their first-quantization form in the dipole gauge as follow,
\begin{align}
 & \underline{\underline{\mathbf{r}}}_{\mathbf{i},\nu;\mathbf{j},\nu^{\prime}}=\mathrm{e}^{-\mathrm{i}\frac{e}{\hbar}\mathbf{A}\left(t\right)\bullet\mathbf{R}_{\mathbf{ij}}}\mathbf{r}_{\mathbf{i},\nu;\mathbf{j},\nu^{\prime}},\\
 & \underline{\underline{\mathbf{p}}}_{\mathbf{i},\nu;\mathbf{j},\nu^{\prime}}=\mathrm{e}^{-\mathrm{i}\frac{e}{\hbar}\mathbf{A}\left(t\right)\bullet\mathbf{R}_{\mathbf{ij}}}\left[\mathbf{p}_{\mathbf{i},\nu;\mathbf{j},\nu^{\prime}}-e\mathbf{A}\left(t\right)\delta_{\mathbf{i},\nu;\mathbf{j},\nu^{\prime}}\right],\\
 & \underline{\underline{\boldsymbol{\pi}}}_{\mathbf{i},\nu;\mathbf{j},\nu^{\prime}}=\mathrm{e}^{-\mathrm{i}\frac{e}{\hbar}\mathbf{A}\left(t\right)\bullet\mathbf{R}_{\mathbf{ij}}}\mathbf{p}_{\mathbf{i},\nu;\mathbf{j},\nu^{\prime}},\\
 & \underline{\overline{\hat{H}}}_{\mathbf{i},\nu;\mathbf{j},\nu^{\prime}}=\mathrm{e}^{-\mathrm{i}\frac{e}{\hbar}\mathbf{A}\left(t\right)\bullet\mathbf{R}_{\mathbf{ij}}}\left[\hat{H}_{0}+e\mathbf{E}\left(t\right)\bullet\hat{\mathbf{r}}\right]_{\mathbf{i},\nu;\mathbf{j},\nu^{\prime}},\\
 & \frac{\partial}{\partial t}\hat{U}^{\prime}=+\mathrm{i}\frac{e}{\hbar}\left(\mathbf{E}\left(t\right)\bullet\hat{\mathbf{R}}\right)\hat{U}^{\prime},\\
 & \overline{\overline{\hat{H}}}=\underline{\overline{\hat{H}}}+\mathrm{i}\frac{e}{\hbar}\mathbf{E}\left(t\right)\bullet\hat{\mathbf{R}},\label{eq:ham_U_2_1}\\
 & \overline{\overline{\hat{H}}}_{\mathbf{i},\nu;\mathbf{j},\nu^{\prime}}=\mathrm{e}^{-\mathrm{i}\frac{e}{\hbar}\mathbf{A}\left(t\right)\bullet\mathbf{R}_{\mathbf{ij}}}\left[\hat{H}_{0}+e\mathbf{E}\left(t\right)\bullet\hat{\mathbf{D}}\right]_{\mathbf{i},\nu;\mathbf{j},\nu^{\prime}},\label{eq:ham_U_2_2}\\
 & \underline{\underline{\rho}}_{\mathbf{i},\nu;\mathbf{j},\nu^{\prime}}=\mathrm{e}^{-\mathrm{i}\frac{e}{\hbar}\mathbf{A}\left(t\right)\bullet\mathbf{R}_{\mathbf{ij}}}\rho_{\mathbf{i},\nu;\mathbf{j},\nu^{\prime}},\\
 & \underline{\underline{\mathbf{J}}}_{\mathbf{i},\nu;\mathbf{j},\nu^{\prime}}=\mathrm{e}^{-\mathrm{i}\frac{e}{\hbar}\mathbf{A}\left(t\right)\bullet\mathbf{R}_{\mathbf{ij}}}\underline{\mathbf{J}}_{\mathbf{i},\nu;\mathbf{j},\nu^{\prime}},\\
 & \frac{\partial}{\partial t}\e{\underline{\underline{\hat{\rho}}}}=-\frac{\mathrm{i}}{\hbar}\e{\left[\underline{\underline{\hat{\rho}}},\overline{\overline{\hat{H}}}\right]}=-\boldsymbol{\nabla}\bullet\e{\underline{\underline{\mathbf{J}}}}\\
 & =\frac{\mathrm{i}}{\hbar}\e{\sum_{\eta=x,y,z}\left[\underline{\underline{\pi_{\eta}}},\underline{\underline{J_{\eta}}}\right]},
\end{align}
where $\mathbf{R}_{\mathbf{ij}}=\mathbf{R}_{\mathbf{i}}-\mathbf{R}_{\mathbf{j}}$
and $\hat{\mathbf{D}}=\hat{\mathbf{r}}-\hat{\mathbf{R}}$, and any
first-quantization operator $\underline{\hat{O}}$ transformed by
$\hat{U}^{\prime}$ is denoted by $\underline{\underline{\hat{O}}}$.
The transformation of the Hamiltonian $\overline{\hat{H}}$ is denoted
by $\underline{\overline{\hat{H}}}$. The true effective Hamiltonian
is $\overline{\overline{\hat{H}}}$, as given in Eqs.~\ref{eq:ham_U_2_1}
and~\ref{eq:ham_U_2_2}. The consecutive action of $\hat{U}$ and
$\hat{U}^{\prime}$ gives the operators in the dipole gauge. Moving
to the second quantization, the effective Hamiltonian in the dipole
gauge is written as
\begin{align}
\mathcal{H} & =\frac{1}{M}\sum_{\mathbf{i},\nu;\mathbf{j},\nu^{\prime}}\overline{\overline{\hat{H}}}_{\mathbf{i},\nu;\mathbf{j},\nu^{\prime}}\tilde{c}_{\mathbf{i},\nu}^{\dagger}\tilde{c}_{\mathbf{j},\nu^{\prime}}.
\end{align}
Substituting Eq.~\ref{eq:ham_U_2_2}, one obtains 
\begin{align}
 & \mathcal{H}=\frac{1}{M}\sum_{\mathbf{i},\nu;\mathbf{j},\nu^{\prime}}\mathrm{e}^{-\mathrm{i}\frac{e}{\hbar}\mathbf{A}\left(t\right)\bullet\mathbf{R}_{\mathbf{ij}}}\tilde{T}_{\mathbf{R}_{\mathbf{ij}},\nu,\nu^{\prime}}\tilde{c}_{\mathbf{i},\nu}^{\dagger}\tilde{c}_{\mathbf{j},\nu^{\prime}}\nonumber \\
 & +\frac{1}{M}\sum_{\mathbf{i},\nu;\mathbf{j},\nu^{\prime}}\mathrm{e}^{-\mathrm{i}\frac{e}{\hbar}\mathbf{A}\left(t\right)\bullet\mathbf{R}_{\mathbf{ij}}}e\mathbf{E}\left(t\right)\bullet\tilde{\mathbf{D}}_{\mathbf{R}_{\mathbf{ij}},\nu,\nu^{\prime}}\tilde{c}_{\mathbf{i},\nu}^{\dagger}\tilde{c}_{\mathbf{j},\nu^{\prime}}.\label{eq:Ham_DG_W}
\end{align}
And similarly, for the density and current operators in the dipole
gauge, it is straightforward to show,
\begin{align}
 & \hat{\mathcal{\rho}}\left(\mathbf{q}=\mathbf{0}\right)=-e\frac{1}{M}\sum_{\mathbf{i},\nu}c_{\mathbf{i},\nu}^{\dagger}c_{\mathbf{i},\nu}=-e\frac{N}{M},\label{eq:rho_DG_W}\\
 & \hat{\mathbf{J}}\left(\mathbf{q}=\mathbf{0},t\right)=\nonumber \\
 & =\mathrm{i}\frac{e}{\hbar}\frac{1}{M}\sum_{\mathbf{i},\nu;\mathbf{j},\nu^{\prime}}\mathrm{e}^{-\mathrm{i}\frac{e}{\hbar}\mathbf{A}\left(t\right)\bullet\mathbf{R}_{\mathbf{ij}}}\mathbf{R}_{\mathbf{ij}}\tilde{T}_{\mathbf{R}_{\mathbf{ij}},\nu,\nu^{\prime}}\tilde{c}_{\mathbf{i},\nu}^{\dagger}\tilde{c}_{\mathbf{j},\nu^{\prime}}\nonumber \\
 & +\mathrm{i}\frac{e}{\hbar}\frac{1}{M}\sum_{\mathbf{i},\nu;\mathbf{j},\nu^{\prime}}\mathrm{e}^{-\mathrm{i}\frac{e}{\hbar}\mathbf{A}\left(t\right)\bullet\mathbf{R}_{\mathbf{ij}}}\left[\hat{\mathbf{D}},\hat{H}_{0}\right]_{\mathbf{i},\nu;\mathbf{j},\nu^{\prime}}\tilde{c}_{\mathbf{i},\nu}^{\dagger}\tilde{c}_{\mathbf{j},\nu^{\prime}},\label{eq:J_DG_W}
\end{align}
 where
\begin{align}
 & \left[\hat{\mathbf{D}},\hat{H}_{0}\right]_{\mathbf{i},\nu;\mathbf{j},\nu^{\prime}}=\sum_{\mathbf{i^{\prime}},\nu^{\prime\prime}}\tilde{\mathbf{D}}_{\mathbf{R}_{\mathbf{ii^{\prime}}},\nu,\nu^{\prime\prime}}\tilde{T}_{\mathbf{R}_{\mathbf{i^{\prime}j}},\nu^{\prime\prime},\nu^{\prime}}\nonumber \\
 & -\sum_{\mathbf{i^{\prime}},\nu^{\prime\prime}}\tilde{T}_{\mathbf{R}_{\mathbf{ii^{\prime}}},\nu,\nu^{\prime\prime}}\tilde{\mathbf{D}}_{\mathbf{R}_{\mathbf{i^{\prime}j}},\nu^{\prime\prime},\nu^{\prime}},
\end{align}
$\tilde{T}_{\mathbf{R}_{\mathbf{ij}},\nu,\nu^{\prime}}=\left(\hat{H}_{0}\right)_{\mathbf{i},\nu;\mathbf{j},\nu^{\prime}}$
is known as the hopping matrix, and $\tilde{\mathbf{D}}_{\mathbf{R}_{\mathbf{ij}},\nu,\nu^{\prime}}=\left(\hat{\mathbf{D}}\right)_{\mathbf{i},\nu;\mathbf{j},\nu^{\prime}}$
is the dipole matrix. We have considered a homogeneous lattice so
that both the hoping and the dipole matrices depend on the difference
$\mathbf{R}_{\mathbf{ij}}$. $\tilde{c}_{\mathbf{i},\nu}$ is the
annihilation operator related to the single-particle state $\k{\varphi_{\mathbf{i},\nu}}$,
$M$ is the number of lattice sites, $N$ is the total number of electrons
in the system, and we have used the relations $\hat{\mathbf{p}}=-\mathrm{i}\frac{m}{\hbar}\left[\hat{\mathbf{r}},\hat{H}_{0}\right]$
and $\me{\varphi_{\mathbf{i},\nu}}{\left[\hat{\mathbf{R}},\hat{O}\right]}{\varphi_{\mathbf{j},\nu^{\prime}}}=\mathbf{R}_{\mathbf{ij}}\hat{O}_{\mathbf{i},\nu;\mathbf{j},\nu^{\prime}}$.

Then, we move to the momentum space using Fourier transformation,
\begin{equation}
\tilde{c}_{\mathbf{k},\nu}=\frac{1}{\sqrt{M}}\sum_{\mathbf{i}}\mathrm{e}^{\mathrm{i}\mathbf{k}\bullet\mathbf{R}_{\mathbf{i}}}\tilde{c}_{\mathbf{i},\nu},
\end{equation}
 and obtain
\begin{align}
 & \mathcal{H}=\sum_{\mathbf{k},\nu,\nu^{\prime}}\tilde{T}_{\mathbf{k}+\frac{e}{\hbar}\mathbf{A}\left(t\right),\nu,\nu^{\prime}}\tilde{c}_{\mathbf{k},\nu}^{\dagger}\tilde{c}_{\mathbf{k},\nu^{\prime}}\nonumber \\
 & +e\mathbf{E}\left(t\right)\bullet\sum_{\mathbf{k},\nu,\nu^{\prime}}\tilde{\mathbf{D}}_{\mathbf{k}+\frac{e}{\hbar}\mathbf{A}\left(t\right),\nu,\nu^{\prime}}\tilde{c}_{\mathbf{k},\nu}^{\dagger}\tilde{c}_{\mathbf{k},\nu^{\prime}},\\
 & \rho\left(\mathbf{q}=\mathbf{0}\right)=-e\frac{1}{M}\sum_{\mathbf{k},\nu}\tilde{c}_{\mathbf{k},\nu}^{\dagger}\tilde{c}_{\mathbf{k},\nu^{\prime}}=-e\frac{N}{M},\\
 & \mathbf{J}\left(\mathbf{q}=\mathbf{0},t\right)=-\frac{e}{\hbar}\sum_{\mathbf{k},\nu^{\prime},\nu}\left[\boldsymbol{\nabla}_{\mathbf{k}}\tilde{T}_{\mathbf{k}+\frac{e}{\hbar}\mathbf{A}\left(t\right),\nu,\nu^{\prime}}\right]\tilde{c}_{\mathbf{k},\nu}^{\dagger}\tilde{c}_{\mathbf{k},\nu^{\prime}}\nonumber \\
 & +\mathrm{i}\frac{e}{\hbar}\sum_{\mathbf{k},\nu,\nu^{\prime},\nu^{\prime\prime}}\left[\tilde{\mathbf{D}}_{\mathbf{k}+\frac{e}{\hbar}\mathbf{A}\left(t\right),\nu,\nu^{\prime\prime}}\tilde{T}_{\mathbf{k}+\frac{e}{\hbar}\mathbf{A}\left(t\right),\nu^{\prime\prime},\nu^{\prime}}\right]\tilde{c}_{\mathbf{k},\nu}^{\dagger}\tilde{c}_{\mathbf{k},\nu^{\prime}}\nonumber \\
 & -\mathrm{i}\frac{e}{\hbar}\sum_{\mathbf{k},\nu,\nu^{\prime},\nu^{\prime\prime}}\left[\tilde{T}_{\mathbf{k}+\frac{e}{\hbar}\mathbf{A}\left(t\right),\nu,\nu^{\prime\prime}}\tilde{\mathbf{D}}_{\mathbf{k}+\frac{e}{\hbar}\mathbf{A}\left(t\right),\nu^{\prime\prime},\nu^{\prime}}\right]\tilde{c}_{\mathbf{k},\nu}^{\dagger}\tilde{c}_{\mathbf{k},\nu^{\prime}},
\end{align}
where
\begin{align}
 & \tilde{T}_{\mathbf{k},\nu,\nu^{\prime}}=\frac{1}{M}\sum_{\mathbf{i},\mathbf{j}}\mathrm{e}^{-\mathrm{i}\mathbf{k}\bullet\mathbf{R}_{\mathbf{ij}}}\tilde{T}_{\mathbf{R}_{\mathbf{ij}},\nu,\nu^{\prime}},\label{eq:Tk_from_TR}\\
 & \tilde{\mathbf{D}}_{\mathbf{k},\nu,\nu^{\prime}}=\frac{1}{M}\sum_{\mathbf{i},\mathbf{j}}\mathrm{e}^{-\mathrm{i}\mathbf{k}\bullet\mathbf{R}_{\mathbf{ij}}}\tilde{\mathbf{D}}_{\mathbf{R}_{\mathbf{ij}},\nu,\nu^{\prime}},\label{eq:Dk_from_DR}
\end{align}
or equivalently
\begin{align}
 & \tilde{T}_{\mathbf{k},\nu,\nu^{\prime}}=\sum_{\mathbf{i}}\mathrm{e}^{-\mathrm{i}\mathbf{k}\bullet\mathbf{R}_{\mathbf{i}}}\tilde{T}_{\mathbf{R}_{\mathbf{i}},\nu,\nu^{\prime}},\label{eq:Tk_from_TR-1}\\
 & \tilde{\mathbf{D}}_{\mathbf{k},\nu,\nu^{\prime}}=\sum_{\mathbf{i}}\mathrm{e}^{-\mathrm{i}\mathbf{k}\bullet\mathbf{R}_{\mathbf{i}}}\tilde{\mathbf{D}}_{\mathbf{R}_{\mathbf{i}},\nu,\nu^{\prime}}.\label{eq:Dk_from_DR-1}
\end{align}

Again, it is worth noting that, in principle, for any \emph{real}
material, $\tilde{T}_{\mathbf{R}_{\mathbf{ij}},\nu,\nu^{\prime}}$
and $\tilde{\mathbf{D}}_{\mathbf{R}_{\mathbf{ij}},\nu,\nu^{\prime}}$
can be obtained as standard outputs of \emph{Wannier90} code~\citep{MOSTOFI20142309},
given its interfaces to a certain number of available DFT codes.

\section{Oscillations of the diagonal elements and multi-photon resonances\protect\label{app:Oscillations-of-diag}}

Let us study a simple two-level system with the equilibrium eigenstates
$\left|1\right\rangle ,\left|2\right\rangle $ and eigenenergies $\varepsilon_{1},\varepsilon_{2}$,
respectively. The gap frequency is $\omega_{g}=\left(\varepsilon_{2}-\varepsilon_{1}\right)/\hbar>0$.
The Hamiltonian of such a system under the application of a pump pulse,
has the following generic form:

\begin{equation}
\mathcal{H}\left(t\right)=\left(\begin{array}{cc}
\varepsilon_{1}\left(t\right) & \lambda_{12}\left(t\right)\\
\lambda_{21}\left(t\right) & \varepsilon_{2}\left(t\right)
\end{array}\right).
\end{equation}
The time dependent state of the system, $\left|\psi\left(t\right)\right\rangle $,
can be spanned in the basis of equilibrium eigenstates:

\begin{equation}
\left|\psi\left(t\right)\right\rangle =a_{1}\left(t\right)\left|1\right\rangle +a_{2}\left(t\right)\left|2\right\rangle ,
\end{equation}
and it is straightforward to show that,

\begin{equation}
\mathrm{i}\hbar\dot{a}_{\alpha}\left(t\right)=\varepsilon_{\alpha}\left(t\right)a_{\alpha}\left(t\right)+\lambda_{\alpha\bar{\alpha}}\left(t\right)a_{\bar{\alpha}}\left(t\right),
\end{equation}
where $\alpha,\bar{\alpha}\in\left\{ 1,2\right\} $ and $\bar{\alpha}=\left(\alpha\mod 2\right)+1$.

In this appendix, we consider a perfectly periodic pump, and for simplicity,
real off-diagonal terms, hence, $\lambda_{\alpha\bar{\alpha}}\left(t\right)=\lambda_{0}\sin\left(\omega t\right)$.

\subsection{Standard Rabi oscillation}

For the case of standard Rabi oscillations, one considers constant
diagonal terms in time, i.e., $\varepsilon_{\alpha}\left(t\right)=\varepsilon_{\alpha}$.
As it is quite well known, one can obtain the dynamics by simply defining
$a_{\alpha}\left(t\right)=a_{0,\alpha}\left(t\right)e^{-i\varepsilon_{\alpha}t/\hbar}$.
Considering the one photon resonant condition, $\omega=\omega_{g}$,
and the rotating wave approximation (RWA), one obtains the following
EOM for $a_{0,\alpha}\left(t\right)$,

\begin{equation}
\ddot{a}_{0,\alpha}\left(t\right)\approx-\left(\frac{\lambda_{0}}{2\hbar}\right)^{2}a_{0,\alpha}\left(t\right),\label{eq:perfect-Rabi}
\end{equation}
which results in a Rabi oscillation with the Rabi frequency $\omega_{R}=\frac{\lambda_{0}}{\hbar}$.
It is worth noticing that in this case, there is no way of getting
a multi-photon resonance.

\subsection{Oscillations of the diagonal elements}

In a more general case, which is very common in realistic setups (e.g.,
in lattice systems), the diagonal terms of the out-of-equilibrium
Hamiltonian also oscillate. One can write,

\begin{equation}
\varepsilon_{\alpha}\left(t\right)=\varepsilon_{\alpha}+\delta_{\alpha}\sin\left(\omega t+\phi_{\alpha}\right),
\end{equation}
where $\delta_{\alpha}$ is the amplitude of the diagonal oscillation
and $\phi_{\alpha}$ is its phase. Let's define,
\begin{equation}
a_{0,\alpha}\left(t\right)=a_{\alpha}\left(t\right)e^{\frac{i}{\hbar}\left(\varepsilon_{\alpha}t-\frac{\delta_{\alpha}}{\omega}\cos\left(\omega t+\phi_{\alpha}\right)\right)}.
\end{equation}
One can straightforwardly show that,
\begin{multline}
i\hbar\dot{a}_{0,\alpha}\left(t\right)=\lambda_{0}\sin\left(\omega t\right)e^{i\left(\varepsilon_{\alpha}-\varepsilon_{\bar{\alpha}}\right)t/\hbar}\\
e^{\frac{i}{\hbar\omega}\left(\delta_{\bar{\alpha}}\cos\left(\omega t+\phi_{\bar{\alpha}}\right)-\delta_{\alpha}\cos\left(\omega t+\phi_{\alpha}\right)\right)}a_{0,\bar{\alpha}}\left(t\right).
\end{multline}

Given that our aim is to understand the general effect of the oscillating
diagonal terms, we consider $\delta_{1}=\delta_{2}=\delta$, $\phi_{1}=\pi$
and $\phi_{2}=0$, so that $\varepsilon_{1/2}\left(t\right)=\varepsilon_{1/2}\mp\delta\sin\left(\omega t\right)$.
The EOM reduces to,
\begin{align}
i\hbar\dot{a}_{0,1/2}\left(t\right) & =\lambda_{0}\sin\left(\omega t\right)e^{\mp i\omega_{g}t}e^{\pm2i\frac{\delta}{\hbar\omega}\cos\left(\omega t\right)}a_{0,2/1}\left(t\right).
\end{align}
Expanding the term $e^{\pm2i\frac{\delta}{\hbar\omega}\cos\left(\omega t\right)}$
and performing some algebra, one can show that,

\begin{multline*}
\mathrm{i}\hbar\dot{a}_{0,1/2}\left(t\right)=a_{0,2/1}\left(t\right)\frac{\lambda_{0}}{2i}e^{\mp i\omega_{g}t}\\
\sum_{n=-\infty}^{\infty}\left(b_{n-1,\pm\delta}-b_{n+1,\pm\delta}\right)e^{in\omega t},
\end{multline*}
where
\begin{equation}
b_{n,\delta}=\sum_{r=0}^{\infty}\frac{1}{\left(2r+\left|n\right|\right)!}\left(i\frac{\delta}{\hbar\omega}\right)^{2r+\left|n\right|}\left(\begin{array}{c}
2r+\left|n\right|\\
r
\end{array}\right).
\end{equation}

In this case, one can verify the emergence of Rabi-like oscillations
also at multi-photon resonances. Consider a $n$-photon resonance
condition: $\omega_{g}=n\omega$, and apply RWA to remove all of the
fast oscillating terms. We remind the reader that here we do not want
to give an analytical solution to the EOM, but our goal is to investigate
the resonance conditions and the situations in which Rabi-like oscillations
emerge. The EOM reduces to,
\begin{align}
i\hbar\dot{a}_{0,1/2}\left(t\right) & \approx\frac{\lambda_{0}}{2i}\left(b_{\pm n-1,\pm\delta}-b_{\pm n+1,\pm\delta}\right)a_{0,2/1}\left(t\right).\label{eq:n-ph-Rabi}
\end{align}
Differentiating Eq. \ref{eq:n-ph-Rabi} with respect to time, we get
\begin{multline}
\ddot{a}_{0,\alpha}\left(t\right)\approx-\left(\frac{\lambda_{0}}{2\hbar}\right)^{2}\left|b_{n-1,\delta}-b_{n+1,\delta}\right|^{2}a_{0,\alpha}\left(t\right),
\end{multline}
which has the same form of Eq. \ref{eq:perfect-Rabi}. This shows
the emergence of a Rabi-like oscillatory behavior for a multi-photon
resonant case, with the Rabi frequency of $\omega_{R}^{\left(n\right)}=\frac{\lambda_{0}}{\hbar}\left|b_{n-1,\delta}-b_{n+1,\delta}\right|$
where $n\geq1$. As a conclusion, oscillating diagonal terms in the
Hamiltonian are a possible source of multi-photon resonances.

\begin{figure*}
\begin{centering}
\begin{tabular}{ccc}
\includegraphics[width=6cm]{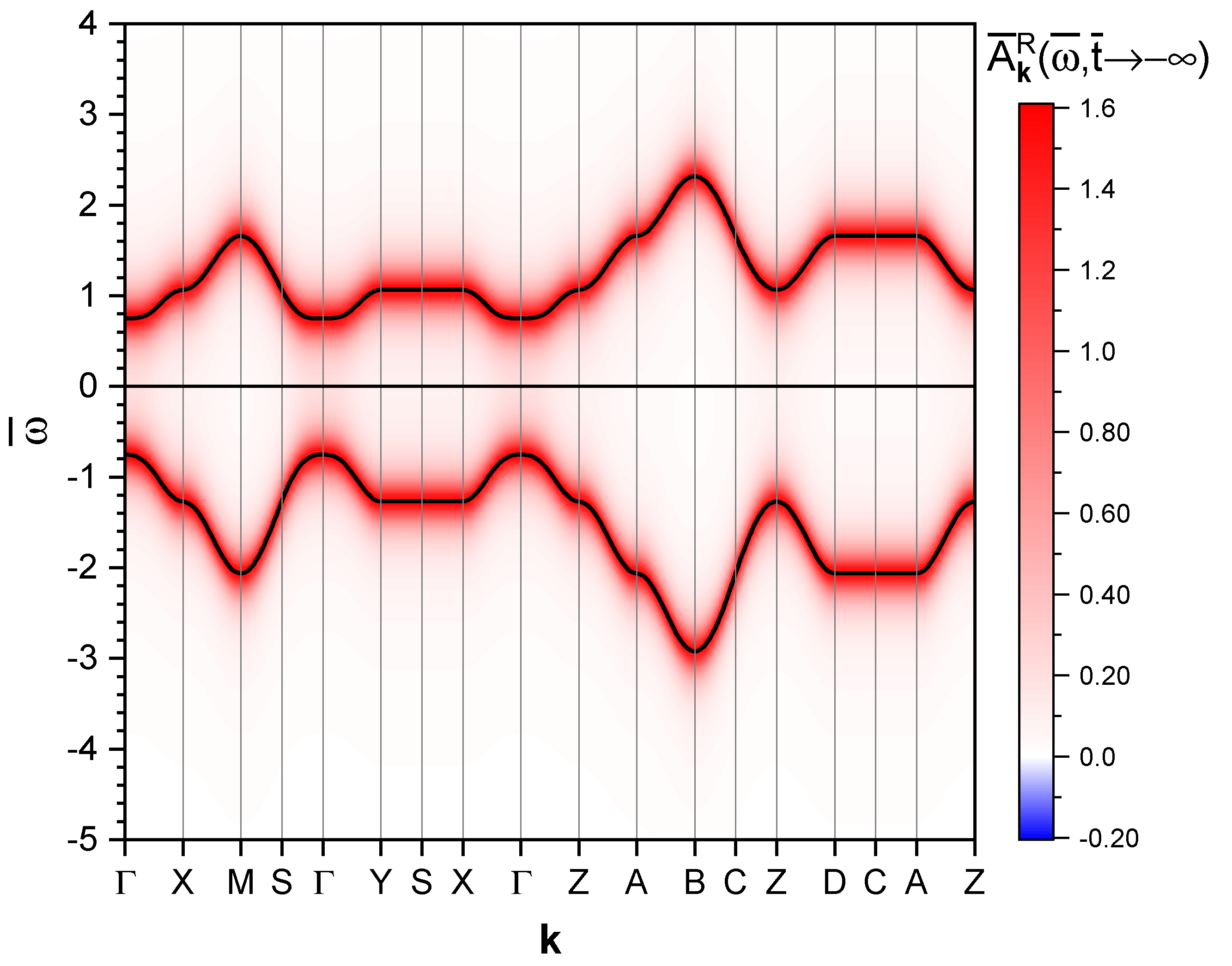} & \includegraphics[width=6cm]{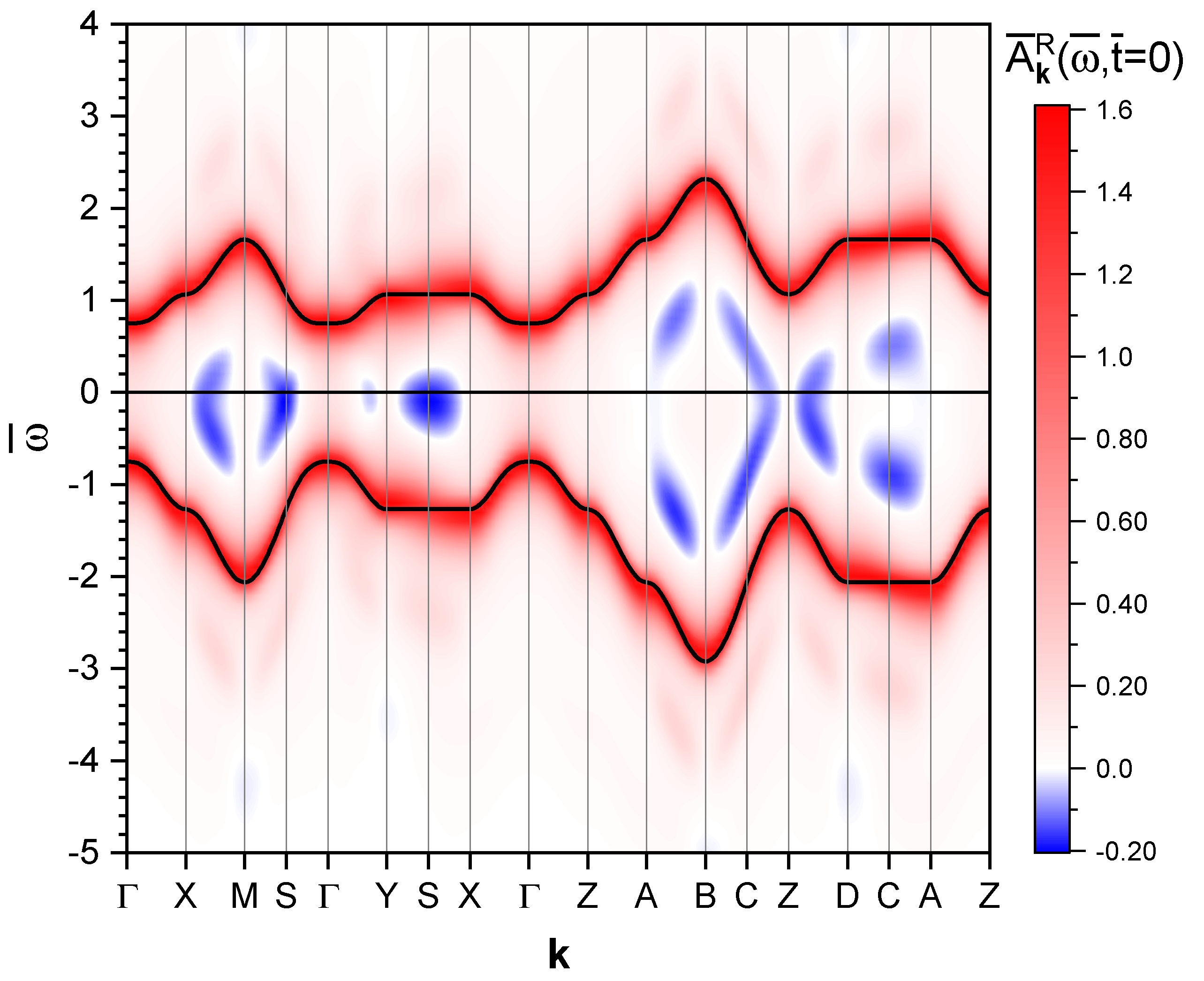} & \includegraphics[width=6cm]{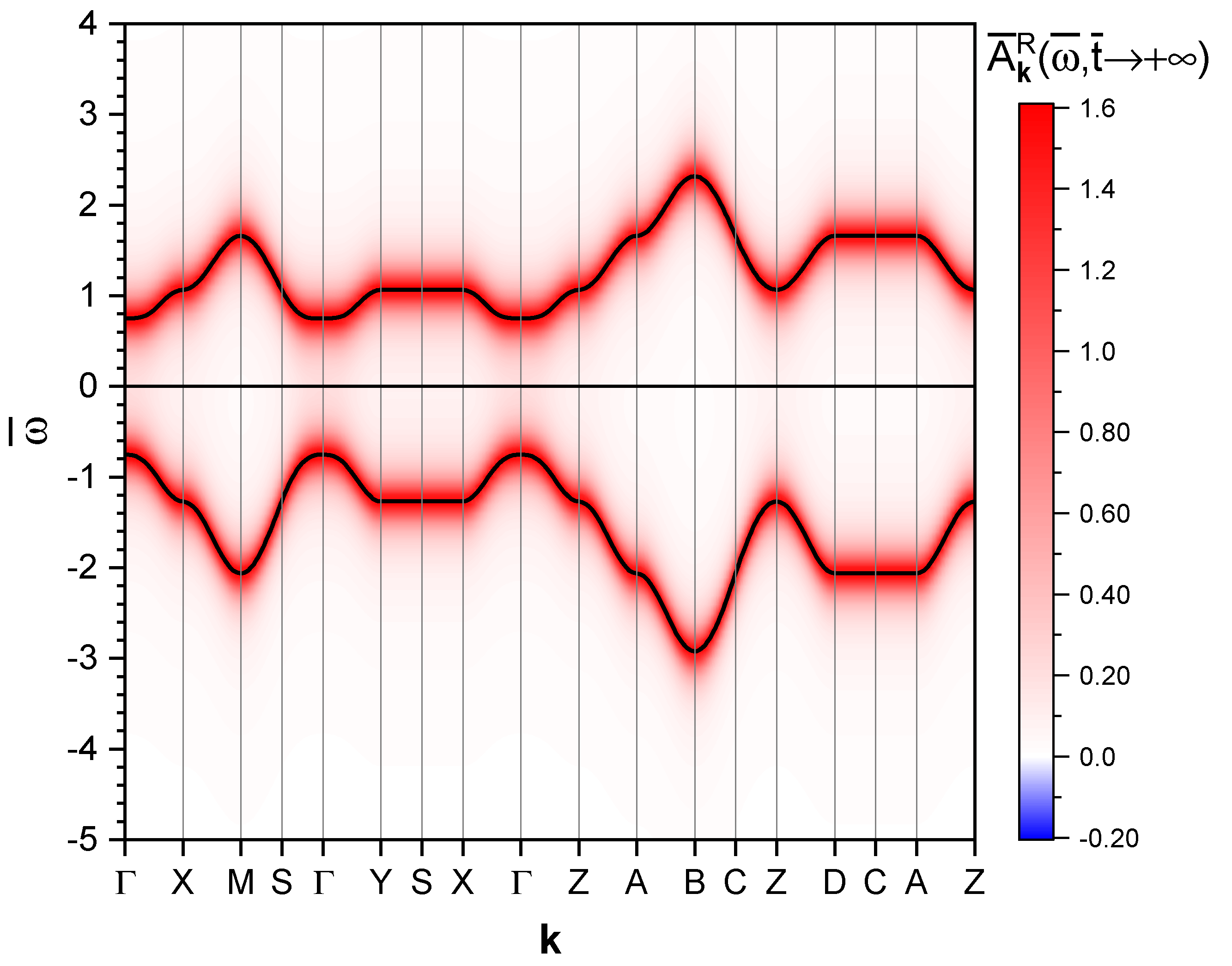}\tabularnewline
\includegraphics[width=6cm]{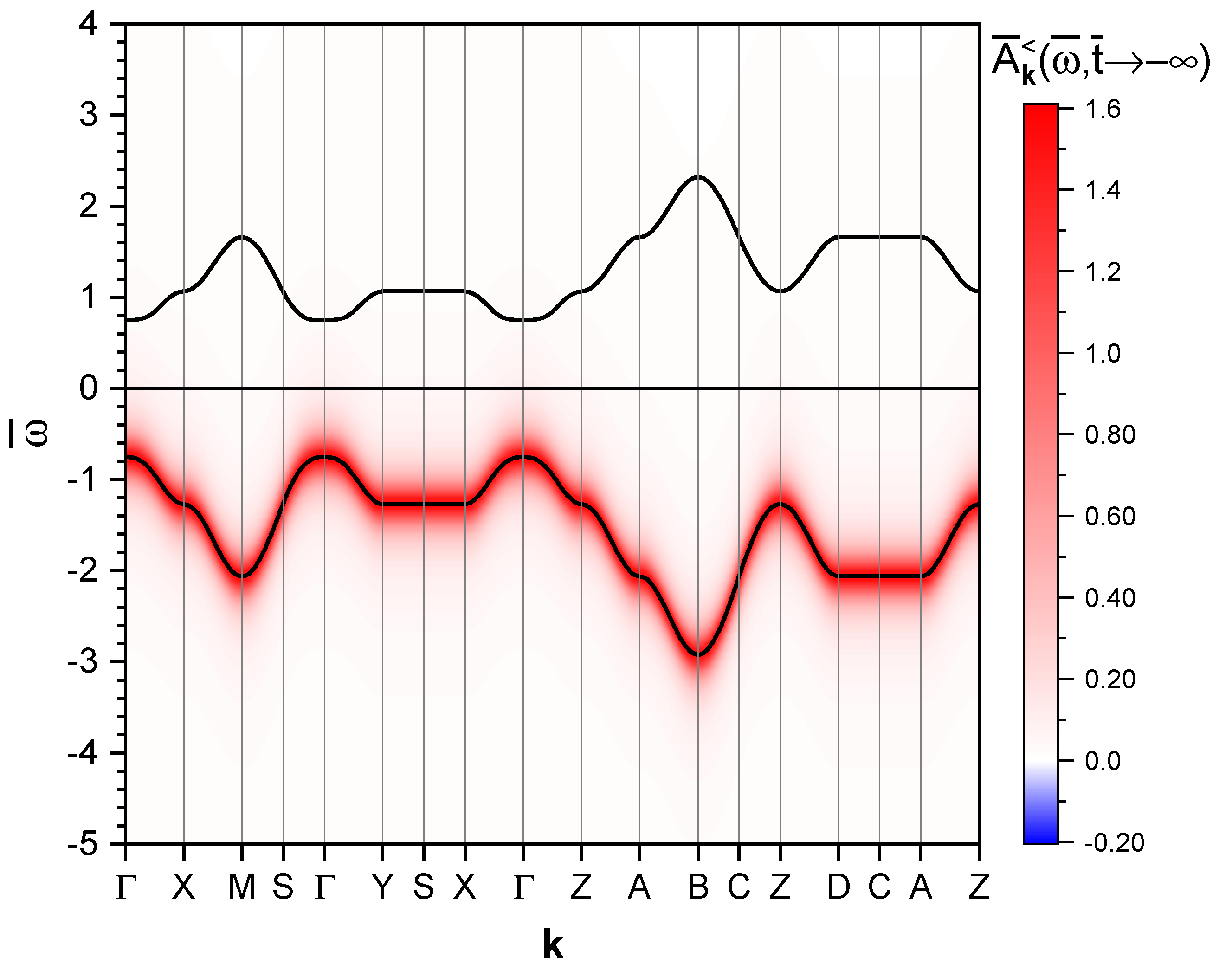} & \includegraphics[width=6cm]{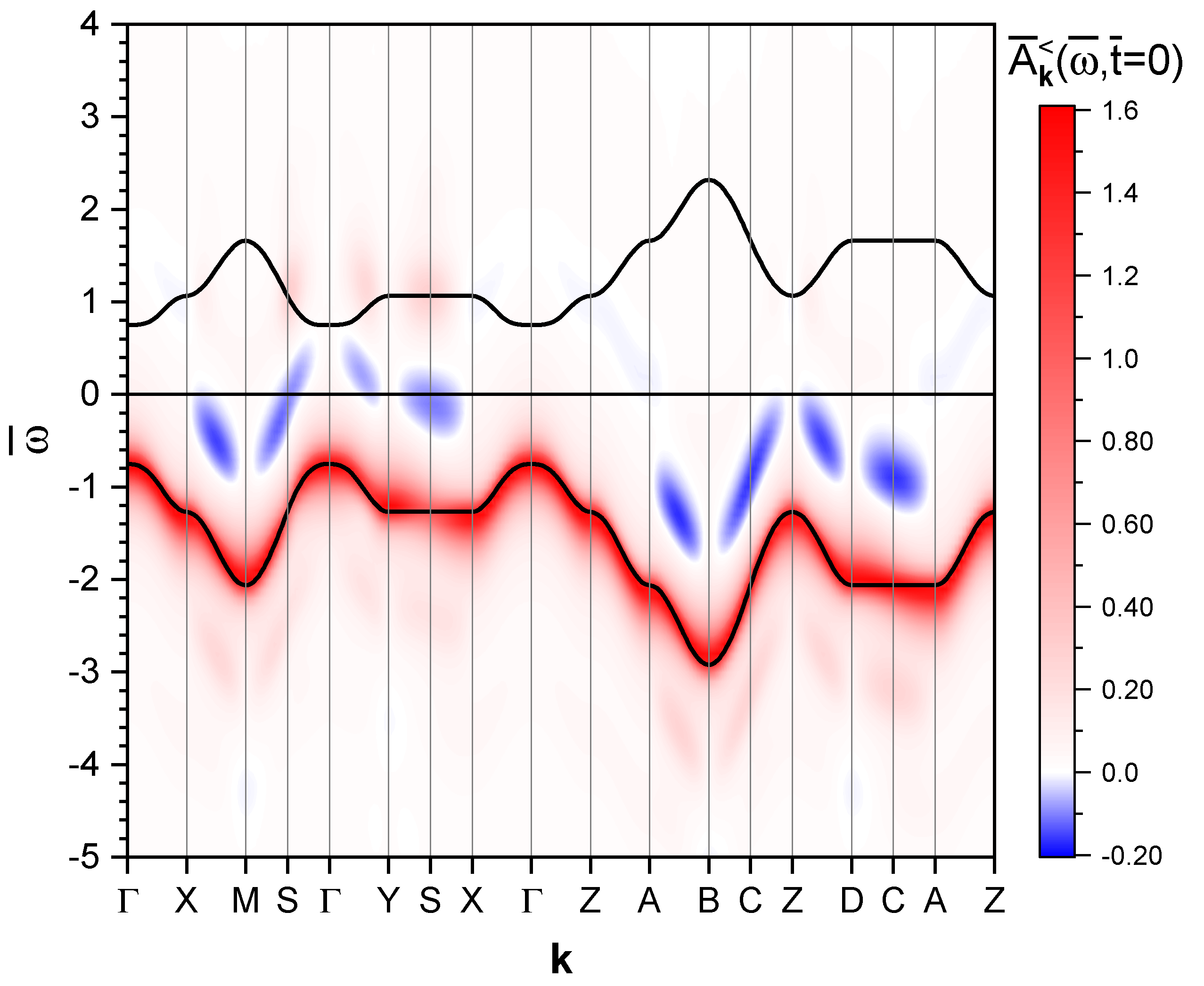} & \includegraphics[width=6cm]{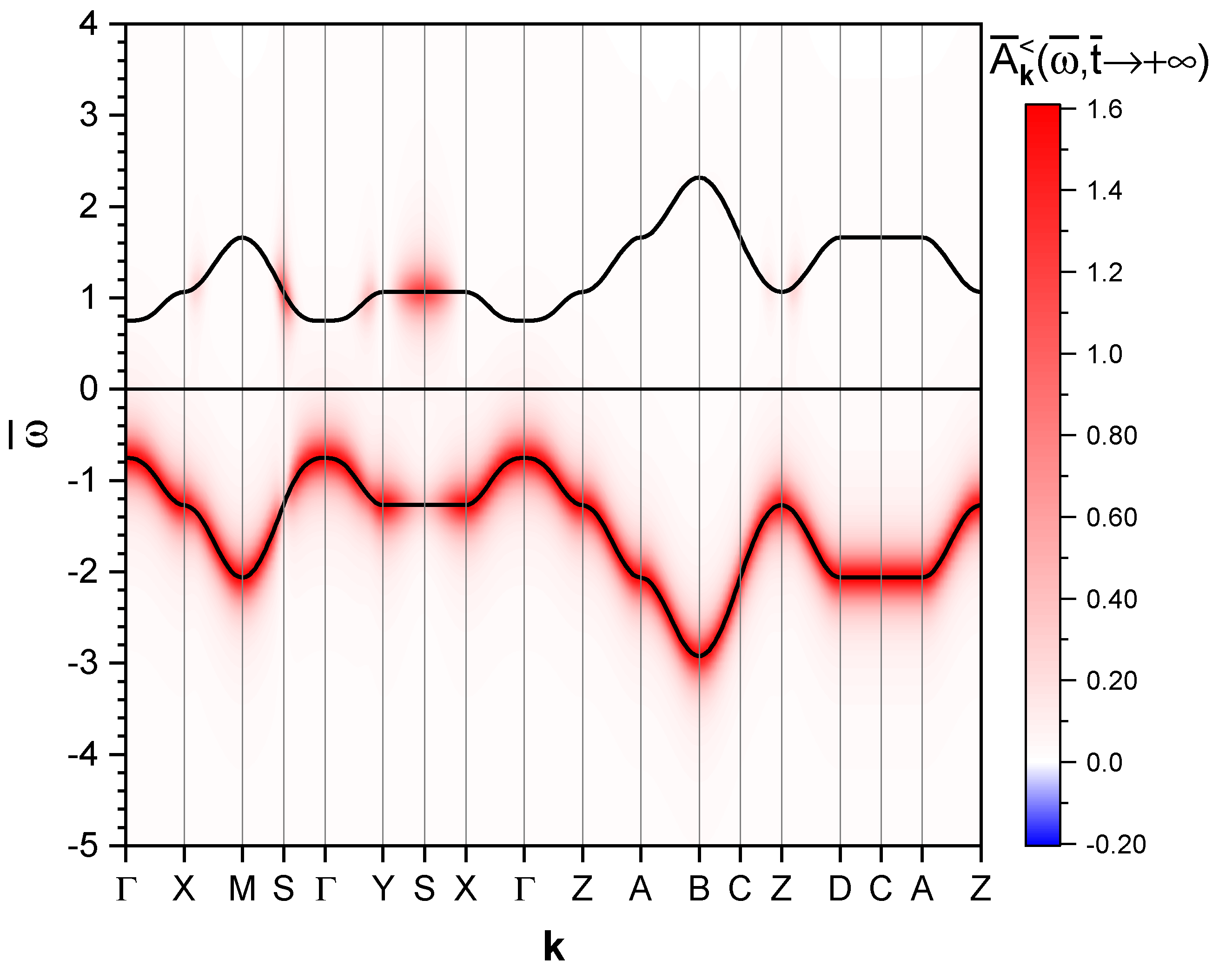}\tabularnewline
\end{tabular}
\par\end{centering}
\caption{Top and bottom panels: the dimensionless spectral function,$\bar{A}_{\mathbf{k}}^{R}\left(\bar{\omega},t\right)$,
and lesser spectral function, $\bar{A}_{\mathbf{k}}^{<}\left(\bar{\omega},t\right)$,
respectively, as function of $\omega$ along the \emph{main} path,
for the same pump-pulse and system parameters as the one of Fig.~\ref{fig:I-t0-NoD}.
In the left, middle and right panels, the time $t$ is chosen to be
well before the application of pump pulse, in the center of the pump
pulse ($t=0$), and well after the application of pump pulse, respectively.
The black solid curves show the equilibrium bands. The numerical value
of $0^{+}$ has been chosen to be $0.2\Delta/\hbar$.\protect\label{fig:SF-k}}
\end{figure*}

\section{The Houston approach\protect\label{app:Houston}}

One of the methods used to simulate the behavior of pumped semiconductors
is the Houston approach \citep{schlaepfer2018attosecond,sato2018role}.
Such an approach is usually formulated in the velocity gauge and first
quantization, just for the reason that will become clear in the following.
Let us start from the time-independent single-particle Hamiltonian
of the Bloch system under analysis,
\begin{equation}
\hat{H}_{0}=\frac{\hat{p}^{2}}{2m}+V\left(\hat{\mathbf{r}}\right),
\end{equation}
where $\hat{\mathbf{p}}$ and $\hat{\mathbf{r}}$ are the momentum
and position operators, respectively, $m$ is the electron mass, and
$V\left(\hat{\mathbf{r}}\right)$ is the periodic potential of the
system under analysis. The related time-independent Schrödinger equation
$\hat{H}_{0}\left|\psi_{\emph{\ensuremath{\boldsymbol{k}},n}}\right\rangle =\varepsilon_{\emph{\ensuremath{\boldsymbol{k}},n}}\left|\psi_{\emph{\ensuremath{\boldsymbol{k}},n}}\right\rangle $
is solved in terms of the Bloch bands $\varepsilon_{\emph{\ensuremath{\boldsymbol{k}},n}}$
and of the Bloch functions $\left|\psi_{\emph{\ensuremath{\boldsymbol{k}},n}}\right\rangle =\mathrm{e}^{-\mathrm{i}\boldsymbol{k}\bullet\hat{\mathbf{r}}}\left|u_{\emph{\ensuremath{\boldsymbol{k}},n}}\right\rangle $
where $\left|u_{\emph{\ensuremath{\boldsymbol{k}},n}}\right\rangle $
displays the same periodicity of the potential. Accordingly, we have
the following reduced equation $\hat{H}_{0,\boldsymbol{k}}\left|u_{\emph{\ensuremath{\boldsymbol{k}},n}}\right\rangle =\varepsilon_{\emph{\ensuremath{\boldsymbol{k}},n}}\left|u_{\emph{\ensuremath{\boldsymbol{k}},n}}\right\rangle $
where $\hat{H}_{0,\boldsymbol{k}}=\mathrm{e}^{\mathrm{i}\boldsymbol{k}\bullet\hat{\mathbf{r}}}\hat{H}_{0}\mathrm{e}^{-\mathrm{i}\boldsymbol{k}\bullet\hat{\mathbf{r}}}=\frac{\left(\hat{\mathbf{p}}-\hbar\boldsymbol{k}\right)^{2}}{2m}+V\left(\hat{\mathbf{r}}\right)$.
Now, if we have a pump pulse described by the vector potential $\mathbf{A}\left(t\right)$
impinging on the system, the related time-dependent minimal-coupling
Hamiltonian in the velocity gauge reads as
\begin{equation}
\hat{H}\left(t\right)=\frac{\left(\hat{\mathbf{p}}+e\mathbf{A}\left(t\right)\right)^{2}}{2m}+V\left(\hat{\mathbf{r}}\right),
\end{equation}
where $e>0$ is the electronic charge. It is straightforward to demonstrate
that the eigenfunctions and the eigenvalues of this Hamiltonian are
simply $\left|\varphi{}_{\emph{\ensuremath{\boldsymbol{k}},n}}\left(t\right)\right\rangle =\mathrm{e}^{-\mathrm{i}\boldsymbol{k}\bullet\hat{\mathbf{r}}}\left|u_{\emph{\ensuremath{\boldsymbol{k}}+\ensuremath{\frac{e}{\hbar}}\ensuremath{\mathbf{A}\left(t\right)},n}}\right\rangle $
and $\varepsilon_{\emph{\ensuremath{\boldsymbol{k}}+\ensuremath{\frac{e}{\hbar}}\ensuremath{\mathbf{A}\left(t\right)},n}}$,
respectively. The set of such eigenfunctions is usually named the
instantaneous or the adiabatic basis because these states would exactly
describe the behavior of the system only if the pump pulse would be
so slowly varying on the characteristic timescales (energies) of the
system to allow it to adjust to the pump pulse at each instant of
time (i.e., adiabatically). Accordingly, they do not solve the general
time-dependent Schrödinger equation $\hat{H}\left(t\right)\left|\phi{}_{\boldsymbol{k}}\left(t\right)\right\rangle =\mathrm{i}\hbar\frac{\partial}{\partial t}\left|\phi{}_{\boldsymbol{k}}\left(t\right)\right\rangle $,
but they can be used as a basis for expanding $\left|\phi{}_{\boldsymbol{k}}\left(t\right)\right\rangle =\sum_{n}\lambda{}_{\emph{\ensuremath{\boldsymbol{k}},n}}\left(t\right)\left|\varphi{}_{\emph{\ensuremath{\boldsymbol{k}},n}}\left(t\right)\right\rangle $.
The projection coefficients $\lambda{}_{\emph{\ensuremath{\boldsymbol{k}},n}}\left(t\right)$
are determined via the following equation of motion,
\begin{multline}
\mathrm{i}\hbar\frac{\partial}{\partial t}\lambda{}_{\emph{\ensuremath{\boldsymbol{k}},n}}\left(t\right)=\left(\varepsilon_{\emph{\ensuremath{\boldsymbol{k}}+\ensuremath{\frac{e}{\hbar}}\ensuremath{\mathbf{A}\left(t\right)},n}}-\theta{}_{\emph{\ensuremath{\boldsymbol{k}},n}}\left(t\right)\right)\lambda{}_{\emph{\ensuremath{\boldsymbol{k}},n}}\left(t\right)\\
+\mathrm{i}\hbar\frac{e}{m}\sum_{n^{\prime}\left(\neq n\right)}\frac{\mathbf{E}\left(t\right)\bullet\mathbf{p}{}_{\boldsymbol{k},n,n^{\prime}}\left(t\right)}{\Delta\varepsilon_{\boldsymbol{k},n,n^{\prime}}\left(t\right)}\lambda{}_{\boldsymbol{k},n^{\prime}}\left(t\right),\label{eq:Houston_coe}
\end{multline}
where $\theta{}_{\emph{\ensuremath{\boldsymbol{k}},n}}\left(t\right)=\left\langle \varphi{}_{\emph{\ensuremath{\boldsymbol{k}},n}}\left(t\right)\right|\mathrm{i}\hbar\frac{\partial}{\partial t}\left|\varphi{}_{\emph{\ensuremath{\boldsymbol{k}},n}}\left(t\right)\right\rangle $
is connected to the Berry phase of the system and can be neglected
if there is no degeneracy, $\mathbf{p}{}_{\boldsymbol{k},n,n^{\prime}}\left(t\right)=\left\langle \varphi{}_{\emph{\ensuremath{\boldsymbol{k}},n}}\left(t\right)\right|\hat{\mathbf{p}}\left|\varphi{}_{\boldsymbol{k},n^{\prime}}\left(t\right)\right\rangle $
is the matrix element of the momentum in the instantaneous basis,
$\Delta\varepsilon_{\boldsymbol{k},n,n^{\prime}}\left(t\right)=\varepsilon_{\emph{\ensuremath{\boldsymbol{k}}+\ensuremath{\frac{e}{\hbar}}\ensuremath{\mathbf{A}\left(t\right)},n}}-\varepsilon_{\boldsymbol{k}+\frac{e}{\hbar}\mathbf{A}\left(t\right),n^{\prime}}$
and $\mathbf{E}\left(t\right)=-\frac{\partial}{\partial t}\mathbf{A}\left(t\right)$
is the applied electric field. Defining new coefficients $\beta{}_{\boldsymbol{k},n}\left(t\right)$
such that $\lambda{}_{\emph{\ensuremath{\boldsymbol{k}},n}}\left(t\right)=\beta{}_{\emph{\ensuremath{\boldsymbol{k}},n}}\left(t\right)\mathrm{e}^{-\frac{\mathrm{i}}{\hbar}\int_{-\infty}^{t}\varepsilon_{\emph{\ensuremath{\boldsymbol{k}}+\ensuremath{\frac{e}{\hbar}}\ensuremath{\mathbf{A}\left(t^{\prime}\right)},n}}dt^{\prime}}$,
the projection-coefficient equation further simplifies to,
\begin{multline}
\mathrm{i}\hbar\frac{\partial}{\partial t}\beta{}_{\emph{\ensuremath{\boldsymbol{k}},n}}\left(t\right)=-\theta{}_{\emph{\ensuremath{\boldsymbol{k}},n}}\left(t\right)\beta{}_{\emph{\ensuremath{\boldsymbol{k}},n}}\left(t\right)\\
+\mathrm{i}\hbar\frac{e}{m}\sum_{n^{\prime}\left(\neq n\right)}\frac{\mathbf{E}\left(t\right)\bullet\mathbf{p}{}_{\boldsymbol{k},n,n^{\prime}}\left(t\right)}{\Delta\varepsilon_{\boldsymbol{k},n,n^{\prime}}\left(t\right)}\mathrm{e}^{\frac{\mathrm{i}}{\hbar}\int_{-\infty}^{t}\Delta\varepsilon_{\boldsymbol{k},n,n^{\prime}}\left(t'\right)dt^{\prime}}\beta{}_{\boldsymbol{k},n^{\prime}}\left(t\right)
\end{multline}
and the corresponding basis $\left\{ \mathrm{e}^{-\frac{\mathrm{i}}{\hbar}\int_{-\infty}^{t}\varepsilon_{\emph{\ensuremath{\boldsymbol{k}}+\ensuremath{\frac{e}{\hbar}}\ensuremath{\mathbf{A}\left(t^{\prime}\right)},n}}dt^{\prime}}\left|\varphi{}_{\emph{\ensuremath{\boldsymbol{k}},n}}\left(t\right)\right\rangle \right\} $
is the Houston basis, which differs just by a time-dependent phase
factor from the instantaneous basis (actually, both basis are often
dubbed in the literature as Houston basis).

The main appeal of such a procedure resides in the possibility of
obtaining sensible results even if one: (i) focuses only on few bands
(e.g., one valence, one conduction and, if needed, one core band),
(ii) supposes that $\mathbf{p}{}_{\boldsymbol{k},n,n^{\prime}}\left(t\right)$
is approximately $\boldsymbol{k}$ independent so that $\mathbf{p}{}_{\boldsymbol{k},n,n^{\prime}}\left(t\right)\approx\mathbf{p}{}_{n,n^{\prime}}$
(the time dependence gets cancelled as well), and (iii) uses the parabolic
approximation $\varepsilon_{\emph{\ensuremath{\boldsymbol{k}},n}}\approx\varepsilon_{n}+\frac{\hbar^{2}k^{2}}{2m_{n}}$,
that is, retains only relevant gaps $\left|\varepsilon_{n}-\varepsilon_{n^{\prime}}\right|$
and effective masses $m_{n}$ in the proximity of few selected $\boldsymbol{k}$
points.

\section{Out-of-Equilibrium Spectral Functions\protect\label{app:SF}}

To obtain the spectral functions, we need the Fourier transformation
of the GFs with respect to time, which we perform as follows,

\begin{align}
G_{\mathbf{k}}^{R,<}\left(\omega,t\right) & =\int_{-\infty}^{+\infty}d\tau e^{i\omega\tau-0^{+}\left|\tau\right|}G_{\mathbf{k}}^{R,<}\left(t+\frac{\tau}{2},t-\frac{\tau}{2}\right),
\end{align}
where $0^{+}$ is an infinitesimal convergence factor. Then, the (retarded)
spectral function is given by 
\begin{equation}
A_{\mathbf{k}}^{R}\left(\omega,t\right)=-\frac{1}{\pi}\Im\left[\Tr G_{\mathbf{k}}^{R}\left(\omega,t\right)\right],
\end{equation}
while the lesser spectral function is defined as
\begin{equation}
A_{\mathbf{k}}^{<}\left(\omega,t\right)=\frac{1}{2\pi}\Im\left[\Tr G_{\mathbf{k}}^{<}\left(\omega,t\right)\right].
\end{equation}
In Fig.~\ref{fig:SF-k} top and bottom panels, we report the dimensionless
retarded and lesser functions, $\bar{A}_{\mathbf{k}}^{R,<}\left(\bar{\omega},t\right)=A_{\mathbf{k}}^{R,<}\left(\bar{\omega},t\right)\Delta/\hbar$
, respectively, as function of $\bar{\omega}$ along the \emph{main}
path, for the same pump-pulse and system parameters as the ones of
Fig.~\ref{fig:I-t0-NoD}. In the left, middle and right panels, the
time $t$ is chosen to be well before the application of the pump
pulse, in the center of the pump pulse ($t=0$), and well after the
application of pump pulse, respectively. Clearly, during the application
of the pump pulse, the spectral functions, $\bar{A}_{\mathbf{k}}^{R,<}\left(\bar{\omega},t\right)$,
become negative and lose their original physical interpretations.

Obviously, in the absence of the pump pulse (both before and after
its application), the spectral function gives correct information
about the energy bands of the system. The left panel of Fig.~\ref{fig:SF-k}
can be directly compared to Fig.~\ref{fig:I_k-eq} and the only difference
to be acknowledged is that, in the former, the band broadening originates
from the finite numerical value of $0^{+}$, while, in the latter,
it originates from the finite FWHM of the probe pulse. After the application
of the pump pulse, the bands recover their equilibrium shape as it
can be seen by comparing the top-right panel of Fig.~\ref{fig:SF-k}
to its top-left panel. On the other hand, the lesser spectral function
(see bottom-right panel of Fig.~\ref{fig:SF-k}) shows that at some
\textbf{k} points we have residual excitations similarly to what is
reported in the top panel of Fig.~\ref{fig:I-res-NoD}.

\section{TR-ARPES signal in the band basis \protect\label{app:FD_derivition}}

In this appendix, we derive Eqs.~\ref{eq:FD_1} and \ref{eq:FD_2}
in the equilibrium band basis. Substituting Eq.~\ref{eq:GF_L_P}
in Eq.~\ref{eq:I_L} and using $\Im\left[z\right]=\frac{1}{2\mathrm{i}}\left(z-z^{\star}\right)$,
the lesser signal can be written as
\begin{multline}
I_{\mathbf{k}}^{<}\left(\omega,t_{\mathrm{pr}}\right)=\frac{\tau_{\mathrm{pr}}}{4\sqrt{2\pi\ln2}}\int_{-\infty}^{+\infty}dt_{1}\int_{-\infty}^{+\infty}dt_{2}\\
S_{\mathrm{pr}}\left(t_{1}-t_{\mathrm{pr}}\right)S_{\mathrm{pr}}\left(t_{2}-t_{\mathrm{pr}}\right)\\
\left[e^{\mathrm{i}\omega\left(t_{1}-t_{2}\right)}\sum_{n,n^{\prime}}f_{+}\left(\varepsilon_{\mathbf{k},n^{\prime}}\right)P_{\mathbf{k},n,n^{\prime}}\left(t_{1}\right)P_{\mathbf{k},n,n^{\prime}}^{\star}\left(t_{2}\right)\right.\\
\left.+e^{\mathrm{-i}\omega\left(t_{1}-t_{2}\right)}\sum_{n,n^{\prime}}f_{+}\left(\varepsilon_{\mathbf{k},n^{\prime}}\right)P_{\mathbf{k},n,n^{\prime}}^{\star}\left(t_{1}\right)P_{\mathbf{k},n,n^{\prime}}\left(t_{2}\right)\right].\label{eq:I_L_ex}
\end{multline}
Changing the dummy variable $t_{1}\longleftrightarrow t_{2}$, one
can verify that the 4th and 3rd lines of the Eq.~\ref{eq:I_L_ex}
are equal to each other. Rearranging the terms, one can simply show
that
\begin{multline}
I_{\mathbf{k}}^{<}\left(\omega,t_{\mathrm{pr}}\right)=\frac{\tau_{\mathrm{pr}}}{2\sqrt{2\pi\ln2}}\sum_{n,n^{\prime}}\\
f_{+}\left(\varepsilon_{\mathbf{k},n^{\prime}}\right)\left|Q_{\mathbf{k},n,n^{\prime}}\left(\omega,t_{\mathrm{pr}}\right)\right|^{2},
\end{multline}
where
\begin{equation}
Q_{\mathbf{k},n,n^{\prime}}\left(\omega,t_{\mathrm{pr}}\right)=\int_{-\infty}^{+\infty}dte^{\mathrm{i}\omega t}S_{\mathrm{pr}}\left(t-t_{\mathrm{pr}}\right)P_{\mathbf{k},n,n^{\prime}}\left(t\right).
\end{equation}
Defining $L_{\mathbf{k},n;n^{\prime}}\left(\omega,t_{\mathrm{pr}}\right)=\frac{\tau_{\mathrm{pr}}}{2\sqrt{2\pi\ln2}}\left|Q_{\mathbf{k},n,n^{\prime}}\left(\omega,t_{\mathrm{pr}}\right)\right|^{2}$
one obtains Eq.~\ref{eq:FD_1}.

For the retarded signal, substituting Eq.~\ref{eq:GF_R_P} in Eq.~\ref{eq:I_R}
and using $\Im\left[z\right]=\frac{1}{2\mathrm{i}}\left(z-z^{\star}\right)$,
we have, 
\begin{multline}
I_{\mathbf{k}}^{R}\left(\omega,t_{\mathrm{pr}}\right)=\frac{\tau_{\mathrm{pr}}}{2\sqrt{2\pi\ln2}}\int_{-\infty}^{+\infty}dt_{1}\int_{-\infty}^{+\infty}dt_{2}\\
S_{\mathrm{pr}}\left(t_{1}-t_{\mathrm{pr}}\right)S\left(t_{2}-t_{\mathrm{pr}}\right)\\
\left[\theta\left(t_{1}-t_{2}\right)e^{\mathrm{i}\omega\left(t_{1}-t_{2}\right)}\sum_{n,n^{\prime}}P_{\mathbf{k},n,n^{\prime}}\left(t_{1}\right)P_{\mathbf{k},n,n^{\prime}}^{\star}\left(t_{2}\right)\right.\\
\left.+\theta\left(t_{1}-t_{2}\right)e^{-\mathrm{i}\omega\left(t_{1}-t_{2}\right)}\sum_{n,n^{\prime}}P_{\mathbf{k},n,n^{\prime}}^{\star}\left(t_{1}\right)P_{\mathbf{k},n,n^{\prime}}\left(t_{2}\right)\right].
\end{multline}
changing the dummy variable $t_{1}\leftrightarrow t_{2}$ in the last
line of this equation and using $\theta\left(t_{1}-t_{2}\right)+\theta\left(t_{2}-t_{1}\right)=1$,
we get,
\begin{multline}
I_{\mathbf{k}}^{R}\left(\omega,t_{\mathrm{pr}}\right)=\frac{\tau_{\mathrm{pr}}}{2\sqrt{2\pi\ln2}}\int_{-\infty}^{+\infty}dt_{1}\int_{-\infty}^{+\infty}dt_{2}\\
S_{\mathrm{pr}}\left(t_{1}-t_{\mathrm{pr}}\right)S\left(t_{2}-t_{\mathrm{pr}}\right)\\
e^{\mathrm{i}\omega\left(t_{1}-t_{2}\right)}\sum_{n,n^{\prime}}P_{\mathbf{k},n,n^{\prime}}\left(t_{1}\right)P_{\mathbf{k},n,n^{\prime}}^{\star}\left(t_{2}\right),
\end{multline}
which simply results in,
\begin{multline}
I_{\mathbf{k}}^{R}\left(\omega,t_{\mathrm{pr}}\right)=\frac{\tau_{\mathrm{pr}}}{2\sqrt{2\pi\ln2}}\sum_{n,n^{\prime}}\left|Q_{\mathbf{k},n,n^{\prime}}\left(\omega,t_{\mathrm{pr}}\right)\right|^{2},
\end{multline}
and proves Eq.~\ref{eq:FD_2}.

\bibliographystyle{apsrev4-2}
\bibliography{biblio}

\end{document}